\documentclass[english]{article}
\usepackage[T1]{fontenc}
\usepackage[latin9]{inputenc}
\usepackage{geometry}
\usepackage{color}
\usepackage{amsmath}
\usepackage{amssymb}
\usepackage{graphicx}
\usepackage{esint}

\makeatletter

\newcommand{\lyxdot}{.}

\numberwithin{equation}{section}
\newcommand{\lyxaddress}[1]{
\par {\raggedright #1
\vspace{1.4em}
\noindent\par}
}

\@ifundefined{showcaptionsetup}{}{%
 \PassOptionsToPackage{caption=false}{subfig}}
\usepackage{subfig}
\makeatother

\usepackage{babel}
\begin{document}

\title{Partition-free theory of time-dependent current correlations in nanojunctions
in response to an arbitrary time-dependent bias}

\author{Michael Ridley$^{1}$, Angus MacKinnon$^{1}$, Lev Kantorovich$^{2}$}

\maketitle

\lyxaddress{$^{1}$Department of Physics, The Blackett Laboratory, Imperial College
London, South Kensington Campus, London SW7 2AZ, United Kingdom\\$^{2}$Department
of Physics, King's College London, Strand, London, WC2R 2LS, United
Kingdom}
\begin{abstract}
Working within the Nonequilibrium Green's Function (NEGF) formalism,
a formula for the two-time current correlation function is derived
for the case of transport through a nanojunction in response to an
arbitrary time-dependent bias. The one-particle Hamiltonian and the
Wide Band Limit Approximation (WBLA) are assumed, enabling us to extract
all necessary Green's functions and self energies for the system,
extending the analytic work presented previously {[}Ridley \textit{et
al. }Phys. Rev. B (2015){]}. We show that our new expression for the
two-time correlation function generalises the B{\"u}ttiker theory of shot
and thermal noise on the current through a nanojunction to the time-dependent
bias case including the transient regime following the switch-on.
Transient terms in the correlation function arise from an initial
state that does not assume (as is usually done) that the system is
initially uncoupled, i.e. our approach is partition-free. We show
that when the bias loses its time-dependence, the long time-limit
of the current correlation function depends on the time difference
only, as in this case an ideal steady state is reached. This enables
derivation of known results for the single frequency power spectrum
and for the zero frequency limit of this power spectrum. In addition,
we present a technique which for the first time facilitates fast calculations
of the transient quantum noise, valid for arbitrary temperature, time
and voltage scales. We apply this to the quantum dot and molecular
wire systems for both DC and AC biases, and find a novel signature
of the traversal time for electrons crossing the wire in the time-dependent
cross-lead current correlations.
\end{abstract}

\section{Introduction}

Electronic circuit components with nanoscale dimensions can now be
fabricated and tuned to form active circuit components \cite{Bergren2016}.
In addition to the speed-up in processing power that arises from sub-micrometre
size \cite{Shalf2015}, molecular junctions also enable a massive
speedup in device operation due to THz intramolecular transport processes
and fast electron traversal time \cite{Selzer2013}. Subsequent to
the initial proposal of molecular rectification in 1974 \cite{Aviram1974},
chemical fabrication techniques have led to the realization of many
interesting devices, including molecular wires \cite{Blum2003,Wohlgamuth2013},
single-electron transistors \cite{Rokhinson2000a}, frequency doublers
and detectors \cite{Swager1998,Iniguez-de-la-Torre2010} and switches
for fast memory storage \cite{JanvanderMolen2010,Liu2013}. In addition,
conductance properties of nanostructures subjected to strong time-dependent
external fields have been the subject of intense experimental research.
This research includes work on photon assisted tunneling \cite{Drexler1995,Covington2000}
and transport through AC biased carbon-based nanostructures in the
GHz-THz regime \cite{Li2004,Burke2004,Chaste2008,Zhang2014}. 

In contrast to classical electronics, the time-dependent current in
molecular structures may undergo fluctuations that have a comparable
magnitude to the current signal itself, so that a theory of time-dependent
fluctuations is essential for the design and control of these devices
\cite{Ratner2013}. Moreover, time-dependent current-current correlations
and their associated frequency-dependent noise spectra contain information
which is not present in the first moment of the current \cite{Blanter2000}.
This includes deviation from classical behaviour in the Fano factor
due to Pauli repulsion \cite{Reznikov1995,Kumar1996}, detection of
fractional charges for quantum Hall quasiparticles \cite{Saminadayar1997}
and the determination of transmission probabilities \cite{Djukic2006}.
When the external field driving the transport process depends upon
time, the transient current correlations provide information on intramolecular
`circular' currents that cannot be studied using the current alone
\cite{Ochoa2015}. Recent measurements of shot noise in graphene irradiated
by THz fields showed an enhancement of the shot noise due to the excitation
of electron-hole pairs in the sample \cite{Parmentier2016}.

In general, nanoelectronic devices possess noise spectra which are
nonlinear functions of frequency. When in equilibrium, there are two
regimes, namely the low $\omega$ regime, in which Johnson-Nyquist
noise is evident \cite{Nyquist1928,Johnson1928}, and the high $\omega$
scenario in which zero-point fluctuations dominate \cite{Aguado2000}.
When a bias is applied to the system, one observes in addition the
shot noise, which results from the discreteness of electronic charge
and the Pauli exclusion principle. At high frequencies, it was shown
that the correct noise spectra is an asymmetric function of the frequency
due to the dominance of zero-point photon fluctuations there \cite{Aguado2000,Gavish2000}.
Distinct negative and positive frequency components of the current
noise due to quasiparticle tunneling across a Josephson junction have
been measured experimentally \cite{Billangeon2006}, and may be physically
interpreted in terms of the transfer of energy quanta during the coresponding
absorption and emission processes \cite{Zamoum2015}. In the theoretical
literature, both symmetric \cite{Yang1992,Buttiker1992,Buttiker1992a,Wu2010}
and asymmetric \cite{Entin-Wohlman2007,Rothstein2009,Gabdank2011}
noise spectra have been classified and studied. 

The Landauer-B{\"u}ttiker (LB) theory of shot and thermal noise represents
a significant milestone in the development of the theory of current
fluctuations in nanoscale systems \cite{Buttiker1990,Buttiker1992,Buttiker1992a,Landauer1993,Blanter2000}.
Originally, it was developed within a scattering matrix approach to
coherent quantum transport, wherein one typically considers a molecular
junction as a subsystem coupled to macroscopic leads, which act as
heat and particle reservoirs. Electrons in the leads are treated as
independent plane waves, populated according to the Fermi distribution
function, and propagated onto the molecule, where they scatter. Experiments
have demonstrated a good agreement between experiment and the noise
spectra obtained from the scattering theory for both the low-frequency
noise \cite{VandenBrom1999,Dicarlo2008} and for power spectra that
depend upon the frequency of the measurement device \cite{Schoelkopf1997,Aguado2000,Gabelli2008}.
In these studies the scattering potential is chosen to be static,
but time-dependent scattering formalisms have been developed which
enable the calculation of current and current noise in response to
an AC potential in the leads \cite{Moskalets2004,Moskalets2007,Moskalets2012},
in both the adiabatic \cite{Moskalets2004} and non-adiabatic \cite{Dai2014,Zhu2015}
regimes. These approaches make use of the Floquet theorem, as do master
equation approaches, which expand scattering states into a harmonic
series \cite{Camalet2004,Wu2010}, generating functional approaches
to the full counting statistics (FCS) \cite{Lesovik1994} and reduced
density matrix methods that make a perturbative expansion in the lead-molecule
coupling \cite{Peskin2016}. The noise response to an AC field has
been shown to carry information on the production of electron-hole
pairs that does not appear in the noise response to a DC bias \cite{Vanevic2012}.
Moreover these electron-hole pairs are correlated and able to propagate
through the molecular junction into separate terminals \cite{Rychkov2005,Vanevic2008}.
In a generating functional approach to the full counting statistics
of an AC-driven system, it was proven that a periodic Lorentzian voltage
signal with quantized flux minimized the noise, i.e. it was reduced
to the DC level \cite{Ivanov1997,Keeling2006}. In recent experiments,
these quantized voltage pulses, known as \textit{levitons}, have been
experimentally realized \cite{Dubois2013} and approximated by a biharmonic
driving field \cite{Gabelli2013}. Even given the restriction of periodic
time-dependence, one can study a rich range of phenomena, such as
photon-assisted tunneling (PAT) \cite{Pedersen1998,Rychkov2005,Wu2010,Hammer2011},
quantum pumping \cite{Strass2005,Riwar2013}, and the interplay of
an external driving field parameters with Fabry-P{\'e}rot conductance
oscillations in graphene nanoribbon (GNR) and carbon nanotube (CNT)
systems \cite{Rocha2010}. 

The Nonequilibrium Green's Function (NEGF) or Keldysh method for the
calculation of dynamical quantum statistical averages can be used
to re-express time-dependent transmission functions, currents and
particle populations in terms of products of self-energies and Green's
Functions \cite{Konstantinov1961,Kadanoff1962,Keldysh1965,Stefanucci2013}.
The equivalence of this picture to the Landauer-B{\"u}ttiker theory in
the noninteracting case is well known \cite{Arrachea2006,DiVentra2008},
but it can also be extended to perturbative calculations of noise
in systems with a Coulombic interaction \cite{Joho2012,Riwar2013}.
Crucially for the present work, it involves the propagation of Green's
Function along a complex time contour that means the effects of the
equilibrium preparation of the system are automatically taken into
account in the dynamics resulting from the switch-on of a bias in
the leads \cite{Stefanucci2013}.

Many calculations of the time-dependent response of a nanojunction
to the switch-on of a bias across the junction make use of the \textit{partitioned}
approach, in which the leads and molecule are completely decoupled
prior to the switch on time $t_{0}$, and suddenly coupled simultaneously
with the addition of a time-dependent bias to the leads at $t_{0}$
\cite{Caroli1971,Caroli1971a,Wingreen1993,Jauho1994}. Partitioned
approaches often involve relegation of $t_{0}$ to the distant past,
because in noninteracting systems the Memory-loss Theorem \cite{Stefanucci2004}
guarantees that the initial condition does not affect the long-time
dynamics. However, transient dynamics was also studied within a partitioned
approach following an artificial quench that instantaneously couples
the molecule to the leads, as was recently done for phononic transport
\cite{Tuovinen2016} (assuming that such an experiment can be done
in practice). In the \textit{partition-free }framework, one includes
a coupling between the leads and molecule in the equilibrium Hamiltonian
which describes the preparation of the system prior to the switch-on.
Partition-free approaches to quantum transport have been implemented
within NEGF \cite{Cini1980,Stefanucci2004} and master equation \cite{Yang2015}
approaches. Recent calculations of transient noise characteristics
have made use of the partitioned approach \cite{Feng2008,Riwar2013,Yang2014}
and there are currently no published calculations of the transient
current noise arising from a partition-free switch on process. 

In recent years, partition-free generalizations of the LB formula
for the current and particle number response to the switch-on of a
static bias have been derived \cite{Perfetto2008,Tuovinen2013,Stefanucci2013,Tuovinen2014}.
This formalism makes use of the wide-band limit approximation (WBLA),
and enables fast calculation of the transport characteristics of realistic
systems at very low computational cost compared with other time-dependent
schemes \cite{Tuovinen2014,GomesdaRocha2015,Tuovinen2016b}. It was
then extended by the present authors to the current response to an
arbitrary time-dependent bias \cite{Ridley2015}, and a practical
scheme for implementation of this formula based upon the replacement
of all frequency integrals with special functions was then developed
\cite{Ridley,Tuovinen2016a}. In the static bias partition-free switch-on
approach pioneered in Refs. \cite{Tuovinen2013,Stefanucci2013,Tuovinen2014},
an analytic result for the \textit{equal time} lesser Green's function
$\mathbf{G}^{<}\left(t,t\right)$ was derived, from which the particle
number in the molecular region and current in the leads can be derived.
However, to calculate current-current correlations one needs an expression
for the lesser Green's function in the two-time plane, $\mathbf{G}^{<}\left(t_{1},t_{2}\right)$,
and the formalism presented in Refs. \cite{Ridley2015,Ridley,Ridley2016}
does this for the arbitrarily time-dependent bias. The ability to
deal with arbitrary time-dependence enables us to study a wider class
of switch-on problems, including those in which the bias is stochastic
in time \cite{Ridley2016}. In the present work, we will extend our
NEGF method further in order to develop an exact formalism enabling
the study of transient current correlations resulting from an \textit{arbitrary}
time dependent bias in the leads. This method does not involve any
assumption of adiabaticity or weak lead-molecule coupling, and neither
is there any limitation on the kind of time-dependence which can be
studied. This will be useful within the field of fast noise calculations
for real molecular junctions driven by ultrafast pulses \cite{Selzer2013,Gaury2014,Ochoa2015},
and to new physics arising from the time-resolved nanoelectronic response
to these pulses that includes the effects of the initial coupling. 

The paper is organized as follows. In Section \ref{sec:The-Model}
we introduce the partition-free time-dependent NEGF formalism developed
in Refs. \cite{Ridley2015,Ridley,Ridley2016}, and show how to obtain
generic formulas for the two-time current correlation function within
the WBLA. In Section \ref{sec:Recovery-of-Known} expressions are
derived for the long-time and static bias approximations in the frequency
domain, thereby confirming that our formalism agrees with other published
work. In Section \ref{sec:Numerics} we present the results of numerical
calculations of the two-time current-correlations in a two-terminal
nanojunction, based upon a fast algorithm that is based on an expansion
of the Fermi function with subsequent analytic removal of all frequency
integrals. We calculate the time-dependent cross-correlations for
single-site quantum dots and extended molecular wires of different
sizes. We identify finite-size effects in the transient current cross-correlations
which cannot be observed in single level systems. In particular, by
studying the competition between wire length, end-site coupling and
internal coupling on the molecule, we show that a resonant signature
of the time taken for electronic information to cross the system can
be seen in both the transient and steady state cross-lead correlations.

\section{\textcolor{black}{Partition-free Correlation Function\label{sec:The-Model}}}

\subsection{\textcolor{black}{Time-dependent NEGF}}

\textcolor{black}{In quantum transport processes, one is typically
concerned with the time-dependent electronic response through a junction
at measurement time $t$ to the switch-on of a bias at some initial
time $t_{0}$, which drives the system away from equilibrium. The
equations of motion for quantum statistical averages are evolved along
a complex time contour, consisting of an upper branch $C_{-}$ running
from $t_{0}+i0$ to $t+i0$, then along a lower branch $C_{+}$ running
back from $t-i0$ to $t_{0}-i0$, and finally along the imaginary
time branch $C_{M}$ from $t_{0}-i0$ to $t_{0}-i\beta$, where $\beta\equiv1/k_{B}T$
(it is adopted that $\hbar=1$ in the following). Real times on the
horizontal branches correspond to the nonequilibrium system, whereas
on the vertical branch the equilibrium system is represented. The
Hamiltonian we will use is formally identical to the one studied in
Ref. \cite{Ridley2016} and is parametrized by the variable $z$ which
denotes the contour `time' variable specifying positions on the Konstantinov-Perel'
contour $\gamma\equiv C_{-}\oplus C_{+}\oplus C_{M}$:}

\textcolor{black}{
\begin{gather}
\hat{H}\left(t\right)=\underset{k\alpha}{\sum}\varepsilon_{k\alpha}\left(z\right)\hat{d}_{k\alpha}^{\dagger}\hat{d}_{k\alpha}+\underset{mn}{\sum}H_{mn}\left(z\right)\hat{d}_{m}^{\dagger}\hat{d}_{n}+\underset{m,k\alpha}{\sum}\left[T_{mk\alpha}\left(z\right)\hat{d}_{m}^{\dagger}\hat{d}_{k\alpha}+T_{k\alpha m}\left(z\right)\hat{d}_{k\alpha}^{\dagger}\hat{d}_{m}\right]\label{eq:Hamiltonian}
\end{gather}
Here, $\hat{d}_{k\alpha}$, $\hat{d}_{m}$ and $\hat{d}_{k\alpha}^{\dagger}$,
$\hat{d}_{m}^{\dagger}$ are annihilation and creation operators of
leads and central system electronic states, where for simplicity spin
degrees of freedom are neglected. The first term is a Hamiltonian
of the lead states $k$ belonging to each lead $\alpha$, the second
is the Hamiltonian of the molecule sandwiched between the leads, describing
hopping within the molecular structure, and the third term describes
the coupling of the molecule to the leads. We collect elements of
this Hamiltonian into a matrix consisting of `blocks' corresponding
to each of the physical subsystems it describes. For example the $\alpha-C$
`block' is the matrix $\mathbf{h}_{\alpha C}\left(z\right)$ with
elements $V_{k\alpha,m}\left(z\right)$:}

\textcolor{black}{
\begin{equation}
\mathbf{h}\left(z\right)=\left(\begin{array}{cccc}
\mathbf{h}_{11}\left(z\right) & 0 & \cdots & \mathbf{h}_{1C}\left(z\right)\\
0 & \mathbf{h}_{22}\left(z\right) & \cdots & \mathbf{h}_{2C}\left(z\right)\\
\vdots & \vdots & \ddots & \vdots\\
\mathbf{h}_{C1}\left(z\right) & \mathbf{h}_{C2}\left(z\right) & \cdots & \mathbf{h}_{CC}\left(z\right)
\end{array}\right)\label{eq:hamiltonianmat}
\end{equation}
In the molecular basis, we also define the $i,j$-th component of
the one-particle Green's function on the Konstantinov-Perel' contour:}

\textcolor{black}{
\begin{equation}
G_{ij}\left(z_{1},z_{2}\right)=-i\frac{Tr\left[e^{-\beta\hat{H}^{M}}\hat{T}_{\gamma}\left[\hat{d}_{i,H}\left(z_{1}\right)\hat{d}_{j,H}^{\dagger}\left(z_{2}\right)\right]\right]}{Tr\left[e^{-\beta\hat{H}^{M}}\right]}\label{eq: (12)-2}
\end{equation}
The elements $G_{ij}$ of the Green's function form a matrix $\mathbf{G}$
defined on the whole space of orbitals of all leads and the central
region; correspondingly, one can introduce diagonal, $\mathbf{G}_{CC}$
and $\mathbf{G}_{\alpha\alpha}$, as well as non-diagonal, $\mathbf{G}_{C\alpha}$,
$\mathbf{G}_{\alpha C}$ and $\mathbf{G}_{\alpha\alpha^{\prime}}$,
blocks of this matrix:}

\textcolor{black}{
\begin{equation}
\mathbf{G}\left(z_{1},z_{2}\right)=\left(\begin{array}{cccc}
\mathbf{G}_{11}\left(z_{1},z_{2}\right) & \mathbf{G}_{12}\left(z_{1},z_{2}\right) & \cdots & \mathbf{G}_{1C}\left(z_{1},z_{2}\right)\\
\mathbf{G}_{21}\left(z_{1},z_{2}\right) & \mathbf{G}_{22}\left(z_{1},z_{2}\right) & \cdots & \mathbf{G}_{2C}\left(z_{1},z_{2}\right)\\
\vdots & \vdots & \ddots & \vdots\\
\mathbf{G}_{C1}\left(z_{1},z_{2}\right) & \mathbf{G}_{C2}\left(z_{1},z_{2}\right) & \cdots & \mathbf{G}_{CC}\left(z_{1},z_{2}\right)
\end{array}\right)\label{eq:GFmat}
\end{equation}
The Green's function $\mathbf{G}_{CC}$ for the central region is
obtained by projecting the general equation of motion onto the $CC$}
matrix block:\textcolor{black}{
\begin{equation}
\left[i\frac{d}{dz_{1}}-\mathbf{h}_{CC}\left(z_{1}\right)\right]\mathbf{G}_{CC}\left(z_{1},z_{2}\right)=\mathbf{1}_{CC}\delta\left(z_{1},z_{2}\right)+\int_{\gamma}d\bar{z}\,\mathbf{\Sigma}_{CC}\left(z_{1},\bar{z}\right)\mathbf{G}_{CC}\left(\bar{z},z_{2}\right)\label{eq:EoM-for-G_CC}
\end{equation}
where $\mathbf{1}_{CC}$ is the unit matrix in the $C$ subspace,
and 
\begin{equation}
\mathbf{\Sigma}_{CC}\left(z_{1},z_{2}\right)=\underset{\alpha}{\sum}\mathbf{h}_{C\alpha}\left(z_{1}\right)\mathbf{g}_{\alpha\alpha}\left(z_{1},z_{2}\right)\mathbf{h}_{\alpha C}\left(z_{2}\right)\label{eq:self-energy-Def}
\end{equation}
is the matrix of the embedding self-energy, where $\mathbf{g}_{\alpha\alpha}\left(z_{1},z_{2}\right)$
is the isolated lead Green's function, whose evolution is governed
solely by the $\alpha\alpha$ block of the Hamiltonian matrix, Eq.
(\ref{eq:hamiltonianmat}). The non-diagonal matrix blocks of the
Green's function are given by Eqs. (\ref{eq: 2}) and (\ref{eq: 3})
in Appendix (\ref{sec:Appendix-B}). The blocks in Eq. (\ref{eq:GFmat})
can then be further subdivided into subspaces defined by regions of
the complex time plane. For example, the `left' Green's function $\mathbf{G}^{\ulcorner}$
is obtained by choosing $z_{1}\in C_{M}$ and $z_{2}\in C_{\mp}$,
and one can obtain its equation of motion using the Langreth rules
\cite{Langreth1972,Langreth1976}:}

\textcolor{black}{
\begin{equation}
\mathbf{G}_{CC}^{\ulcorner}\left(\tau_{1},t_{2}\right)\left[-i\overleftarrow{\frac{d}{dt_{2}}}-\mathbf{h}_{CC}\left(t_{2}\right)\right]=\left[\mathbf{G}_{CC}^{\ulcorner}.\mathbf{\Sigma}_{CC}^{a}+\mathbf{G}_{CC}^{M}\star\mathbf{\Sigma}_{CC}^{\ulcorner}\right]_{\left(\tau_{1},t_{2}\right)}\label{eq:leftEoM}
\end{equation}
where the differential operator in the left hand side acts on the
left. One also defines the `right' Green's function $\mathbf{G}^{\urcorner}$
by choosing $z_{1}\in C_{\mp}$ and $z_{2}\in C_{M}$, the `lesser'
and `greater' Green's functions $\mathbf{G}^{\lessgtr}$ with, e.g.,
$z_{1}\in C_{-},\, z_{2}\in C_{+}$ and $z_{1}\in C_{+},\, z_{2}\in C_{-}$,
respectively, and the Matsubara Green's function $\mathbf{G}^{M}$
with $z_{1},z_{2}\in C_{M}$. In addition, `retarded' and `advanced'
Green's functions are stipulated with a definite real-time ordering:}

\textcolor{black}{
\begin{equation}
\mathbf{G}^{r}\left(t_{1},t_{2}\right)=\theta\left(t_{1},t_{2}\right)\left[\mathbf{G}^{>}\left(t_{1},t_{2}\right)-\mathbf{G}^{<}\left(t_{1},t_{2}\right)\right]
\end{equation}
}

\textcolor{black}{
\begin{equation}
\mathbf{G}^{a}\left(t_{1},t_{2}\right)=-\theta\left(t_{2},t_{1}\right)\left[\mathbf{G}^{>}\left(t_{1},t_{2}\right)-\mathbf{G}^{<}\left(t_{1},t_{2}\right)\right]
\end{equation}
The equations obtained by projecting Eq. (\ref{eq:EoM-for-G_CC})
and its complex conjugate onto these subregions of the complex time
plane are known as the Kadanoff-Baym equations, see, e.g., Ref. \cite{Stefanucci2013}. }

\subsection{Generalized Expression from Wick's Theorem\label{sub:Wick}}

The current in lead $\alpha$ can be obtained as the thermal average
of the time derivative of the average charge in that\textcolor{black}{{}
lead, $I_{\alpha}\left(t\right)\equiv q\left\langle \frac{d\hat{N}_{\alpha}\left(t\right)}{dt}\right\rangle $
(where the spin-degenerate particle number is }$\hat{N}_{\alpha}=2\underset{k}{\sum}\hat{d}_{k\alpha}^{\dagger}\hat{d}_{k\alpha}$\textcolor{black}{).
In all numerical calculations that follow, the electron charge will
be set to $q=-1$. Given the noninteracting Hamiltonian in Eq. (\ref{eq:Hamiltonian}),
the current operator has the form:}

\begin{eqnarray}
\hat{I}_{\alpha}\left(t\right) & = & 2iq\underset{k,m}{\sum}\left[T_{mk\alpha}\hat{d}_{m}^{\dagger}\left(t\right)\hat{d}_{k\alpha}\left(t\right)-T_{mk\alpha}^{*}\hat{d}_{k\alpha}^{\dagger}\left(t\right)\hat{d}_{m}\left(t\right)\right]\label{eq:currentoperator}
\end{eqnarray}
We define the current deviation operator with a mean value of zero:

\begin{eqnarray}
\triangle\hat{I}_{\alpha}\left(t\right) & = & 2iq\underset{k,m}{\sum}\left[T_{mk\alpha}\left(t\right)\left(\hat{d}_{m}^{\dagger}\left(t\right)\hat{d}_{k\alpha}\left(t\right)-\left\langle \hat{d}_{m}^{\dagger}\left(t\right)\hat{d}_{k\alpha}\left(t\right)\right\rangle \right)-T_{mk\alpha}^{*}\left(t\right)\left(\hat{d}_{k\alpha}^{\dagger}\left(t\right)\hat{d}_{m}\left(t\right)-\left\langle \hat{d}_{k\alpha}^{\dagger}\left(t\right)\hat{d}_{m}\left(t\right)\right\rangle \right)\right]\label{eq:currentdeviation}
\end{eqnarray}
The two-time current correlator between leads $\alpha$ and $\beta$
is defined as:

\begin{equation}
C_{\alpha\beta}\left(t_{1},t_{2}\right)\equiv\left\langle \triangle\hat{I}_{\alpha}\left(t_{1}\right)\triangle\hat{I}_{\beta}\left(t_{2}\right)\right\rangle \label{eq:currcorr}
\end{equation}
This correlator obviously satisfies the symmetry property:

\begin{equation}
C_{\alpha\beta}\left(t_{1},t_{2}\right)^{*}=C_{\beta\alpha}\left(t_{2},t_{1}\right)\label{eq:csymmetry}
\end{equation}
Since $\triangle\hat{I}_{\alpha}\left(t_{1}\right)$ and $\triangle\hat{I}_{\beta}\left(t_{2}\right)$
do not commute in general, $C_{\alpha\beta}\left(t_{1},t_{2}\right)$
is not guaranteed to be real and so in several studies the symmetrized
correlation function is preferred \cite{Buttiker1992,Buttiker1992a}:
\begin{equation}
P_{\alpha\beta}\left(t_{1},t_{2}\right)\equiv\frac{1}{2}\left\langle \triangle\hat{I}_{\alpha}\left(t_{1}\right)\triangle\hat{I}_{\beta}\left(t_{2}\right)+\triangle\hat{I}_{\beta}\left(t_{2}\right)\triangle\hat{I}_{\alpha}\left(t_{1}\right)\right\rangle =\textrm{Re}\left[C_{\alpha\beta}\left(t_{1},t_{2}\right)\right]\label{eq:twotimecorr}
\end{equation}
Since $P_{\alpha\beta}\left(t_{1},t_{2}\right)$ is just the real
part of $C_{\alpha\beta}\left(t_{1},t_{2}\right)$, knowledge of the
latter object is sufficient for a full characterization of the symmetric
noise properties of the junction. The lack of two-particle interactions
in the Hamiltonian (\ref{eq:Hamiltonian}) means we can simplify the
non-symmetrized correlator using Wick's Theorem, which is valid for
a noninteracting Hamiltonian with arbitrary time-dependence \cite{Stefanucci2013}:

\[
C_{\alpha\beta}\left(t_{1},t_{2}\right)=-4q^{2}\underset{k,k',m,m'}{\sum}\left[T_{mk\alpha}\left(t_{1}\right)T_{m'k'\beta}\left(t_{2}\right)\left\langle \hat{d}_{m}^{\dagger}\left(t_{1}\right)\hat{d}_{k'\beta}\left(t_{2}\right)\right\rangle \left\langle \hat{d}_{k\alpha}\left(t_{1}\right)\hat{d}_{m'}^{\dagger}\left(t_{2}\right)\right\rangle \right.
\]

\[
-T_{mk\alpha}\left(t_{1}\right)T_{m'k'\beta}^{*}\left(t_{2}\right)\left\langle \hat{d}_{m}^{\dagger}\left(t_{1}\right)\hat{d}_{m'}\left(t_{2}\right)\right\rangle \left\langle \hat{d}_{k\alpha}\left(t_{1}\right)\hat{d}_{k'\beta}^{\dagger}\left(t_{2}\right)\right\rangle 
\]
\[
-T_{mk\alpha}^{*}\left(t_{1}\right)T_{m'k'\beta}\left(t_{2}\right)\left\langle \hat{d}_{k\alpha}^{\dagger}\left(t_{1}\right)\hat{d}_{k'\beta}\left(t_{2}\right)\right\rangle \left\langle \hat{d}_{m}\left(t_{1}\right)\hat{d}_{m'}^{\dagger}\left(t_{2}\right)\right\rangle 
\]

\begin{equation}
\left.+T_{mk\alpha}^{*}\left(t_{1}\right)-T_{m'k'\beta}^{*}\left(t_{2}\right)\left\langle \hat{d}_{k\alpha}^{\dagger}\left(t_{1}\right)\hat{d}_{m'}\left(t_{2}\right)\right\rangle \left\langle \hat{d}_{m}\left(t_{1}\right)\hat{d}_{k'\beta}^{\dagger}\left(t_{2}\right)\right\rangle \right]
\end{equation}

One identifies the following Green's functions in this expression:

\begin{equation}
\left[\mathbf{G}_{AB}^{>}\left(t_{1},t_{2}\right)\right]_{kk'}=-i\left\langle \hat{d}_{kA}\left(t_{1}\right)\hat{d}_{k'B}^{\dagger}\left(t_{2}\right)\right\rangle 
\end{equation}

\begin{equation}
\left[\mathbf{G}_{AB}^{<}\left(t_{1},t_{2}\right)\right]_{kk'}=i\left\langle \hat{d}_{k'B}^{\dagger}\left(t_{2}\right)\hat{d}_{kA}\left(t_{1}\right)\right\rangle 
\end{equation}
where $A$ and $B$ correspond to either the lead or central molecule
regions. It is then possible to rewrite Eq. (\ref{eq:twotimecorr})
in the compact analytic form:

\[
C_{\alpha\beta}\left(t_{1},t_{2}\right)=-4q^{2}\textrm{Tr}_{C}\left[\mathbf{h}_{C\alpha}\left(t_{1}\right)\mathbf{G}_{\alpha C}^{>}\left(t_{1},t_{2}\right)\mathbf{h}_{C\beta}\left(t_{2}\right)\mathbf{G}_{\beta C}^{<}\left(t_{2},t_{1}\right)-\mathbf{h}_{C\alpha}\left(t_{1}\right)\mathbf{G}_{\alpha\beta}^{>}\left(t_{1},t_{2}\right)\mathbf{h}_{\beta C}\left(t_{2}\right)\mathbf{G}_{CC}^{<}\left(t_{2},t_{1}\right)\right.
\]

\begin{equation}
\left.-\mathbf{G}_{CC}^{>}\left(t_{1},t_{2}\right)\mathbf{h}_{C\beta}\left(t_{2}\right)\mathbf{G}_{\beta\alpha}^{<}\left(t_{2},t_{1}\right)\mathbf{h}_{\alpha C}\left(t_{1}\right)+\mathbf{G}_{C\beta}^{>}\left(t_{1},t_{2}\right)\mathbf{h}_{\beta C}\left(t_{2}\right)\mathbf{G}_{C\alpha}^{<}\left(t_{2},t_{1}\right)\mathbf{h}_{\alpha C}\left(t_{1}\right)\right]\label{eq: 1}
\end{equation}
The expression (\ref{eq: 1}) is structurally identical to current
correlation functions in Refs. \cite{Yang2014,Ochoa2015}, but we
emphasize that here the two time Green's functions appearing in Eq.
(\ref{eq: 1}) evolve in response to the switch-on of an arbitrary
time-dependent bias in the partition-free approach, i.e. they contain
convolution integrals taken along the vertical part of the Konstantinov-Perel'
contour as well. Notice that, in addition to correlation functions
describing particle hopping events between the leads and the molecule,
Eq. (\ref{eq: 1}) also contains information on lead-lead hopping
events and on `circular' \cite{Ochoa2015} currents involving electronic
transport processes within the molecular structure. In some work on
the time-dependent noise, the two-time correlator was given as a function
of a single time \cite{Feng2008}, but we emphasize that we need to
solve the Kadanoff-Baym equations for all Green's functions `blocks'
in Eq. (\ref{eq:GFmat}) in the two-time plane for a complete picture
of current fluctuations. We present the main steps of this derivation
in Appendix \ref{sec:Appendix-B}, and the derived Green's functions
are inserted into Eq. (\ref{eq: 1}), resulting in a sum of terms
involving only self-energy components and components of the $CC$
region Green's function: 

\[
C_{\alpha\beta}\left(t_{1},t_{2}\right)=4q^{2}\textrm{Tr}_{C}\left[\left(\mathbf{\Sigma}_{\alpha}^{>}\left(t_{1},t_{2}\right)\delta_{\alpha\beta}+\left(\left(\mathbf{\Sigma}_{\alpha}^{>}\cdot\mathbf{G}_{CC}^{a}+\mathbf{\Sigma}_{\alpha}^{r}\cdot\mathbf{G}_{CC}^{>}+\mathbf{\Sigma}_{\alpha}^{\urcorner}\star\mathbf{G}_{CC}^{\ulcorner}\right)\cdot\mathbf{\Sigma}_{\beta}^{a}\right.\right.\right.
\]

\[
\left.\left.+\mathbf{\Sigma}_{\alpha}^{r}\cdot\left(\mathbf{G}_{CC}^{r}.\mathbf{\Sigma}_{\beta}^{>}+\mathbf{G}_{CC}^{\urcorner}\star\mathbf{\Sigma}_{\beta}^{\ulcorner}\right)\right)_{\left(t_{1}^{+},t_{2}^{-}\right)}\right)\cdot\mathbf{G}_{CC}^{<}\left(t_{2},t_{1}\right)
\]

\[
+\mathbf{G}_{CC}^{>}\left(t_{1},t_{2}\right)\left(\mathbf{\Sigma}_{\alpha}^{<}\left(t_{2},t_{1}\right)\delta_{\alpha\beta}+\left(\left(\mathbf{\Sigma}_{\beta}^{<}\cdot\mathbf{G}_{CC}^{a}+\mathbf{\Sigma}_{\beta}^{r}\cdot\mathbf{G}_{CC}^{<}+\mathbf{\Sigma}_{\beta}^{\urcorner}\star\mathbf{G}_{CC}^{\ulcorner}\right)\cdot\mathbf{\Sigma}_{\alpha}^{a}\right.\right.
\]

\[
\left.\left.+\mathbf{\Sigma}_{\beta}^{r}\cdot\left(\mathbf{G}_{CC}^{r}\cdot\mathbf{\Sigma}_{\alpha}^{<}+\mathbf{G}_{CC}^{\urcorner}\star\mathbf{\Sigma}_{\alpha}^{\ulcorner}\right)\right)_{\left(t_{2}^{-},t_{1}^{+}\right)}\right)
\]

\[
-\left(\mathbf{\Sigma}_{\alpha}^{>}\cdot\mathbf{G}_{CC}^{a}+\mathbf{\Sigma}_{\alpha}^{r}\cdot\mathbf{G}_{CC}^{>}+\mathbf{\Sigma}_{\alpha}^{\urcorner}\star\mathbf{G}_{CC}^{\ulcorner}\right)_{\left(t_{1}^{+},t_{2}^{-}\right)}\left(\mathbf{\Sigma}_{\beta}^{<}\cdot\mathbf{G}_{CC}^{a}+\mathbf{\Sigma}_{\beta}^{r}\cdot\mathbf{G}_{CC}^{<}+\mathbf{\Sigma}_{\beta}^{\urcorner}\star\mathbf{G}_{CC}^{\ulcorner}\right)_{\left(t_{2}^{-},t_{1}^{+}\right)}
\]

\begin{equation}
\left.\left.-\left(\mathbf{G}_{CC}^{>}\cdot\mathbf{\Sigma}_{\beta}^{a}+\mathbf{G}_{CC}^{r}\cdot\mathbf{\Sigma}_{\beta}^{>}+\mathbf{G}_{CC}^{\urcorner}\star\mathbf{\Sigma}_{\beta}^{\ulcorner}\right)_{\left(t_{1}^{+},t_{2}^{-}\right)}\left(\mathbf{G}_{CC}^{<}\cdot\mathbf{\Sigma}_{\alpha}^{a}+\mathbf{G}_{CC}^{r}\cdot\mathbf{\Sigma}_{\alpha}^{<}+\mathbf{G}_{CC}^{\urcorner}\star\mathbf{\Sigma}_{\alpha}^{\ulcorner}\right)_{\left(t_{2}^{-},t_{1}^{+}\right)}\right)\right]\label{eq:power1}
\end{equation}
Here the sign superscripts indicate the contour position of each time
variable. So far, no assumptions have been made on the system Hamiltonian,
i.e. we have not yet stated which regions are subject to a time-dependent
perturbation, and neither have we made assumptions about the nature
of the lead-molecule coupling. Up to this point, the derivation is
completely algebraic, and so for noninteracting systems Eq. (\ref{eq:power1})
is completely general.

\subsection{Time-dependent Model and the WBLA\label{sub:Time-dependent-Model-and}}

\textcolor{black}{In this section we make assumptions on the model
that enable us to solve the Kadanoff-Baym equations analytically.
We assume that, prior to $t_{0}$, the Hamiltonian $\hat{H}_{0}\equiv\hat{H}\left(z\in C_{M}\right)$
is given by Eq. (\ref{eq:Hamiltonian}) with time-independent energies
$\varepsilon_{k\alpha}\left(z\in C_{M}\right)=\varepsilon_{k\alpha}$
and molecular site and hopping integrals $H_{mn}\left(z\in C_{M}\right)=h_{mn}$.
The lead-molecule couplings $T_{m,k\alpha}\left(z\in\gamma\right)=T_{m,k\alpha}$
are assumed to be present in equilibrium in the partition-free approach
and unchanged by the switch-on process. As all subsystems are coupled
during their equilibration, they all possess the same initial temperature
$T$ and chemical potential $\mu$, which means the system is initially
described by the density operator $\widehat{\rho}_{0}=Z^{-1}e^{-\beta\left(\widehat{H}_{0}-\mu\widehat{N}\right)}$
(where $Z$ is the partition function and $\hat{N}$ is the number
operator for the entire coupled system). Following Ref. \cite{Ridley2016},
we add an arbitrary spatially homogeneous time-dependent shift to
the lead energies as their bias. To the molecular Hamiltonian, we
add a static correction $\mathbf{u}_{CC}=\underset{mn}{\sum}u_{mn}\hat{d}_{m}^{\dagger}\hat{d}_{n}$
\cite{Tuovinen2014}, and a time-dependent shift that scales the particle
number operator $\hat{N}_{C}=\underset{mn}{\sum}\hat{d}_{m}^{\dagger}\hat{d}_{n}$
\cite{Ridley2016}:}

\textcolor{black}{
\begin{eqnarray}
\varepsilon_{k\alpha}\left(z\in C_{\mp}\right) & = & \varepsilon_{k\alpha}+V_{\alpha}\left(t\right)\\
H_{mn}\left(z\in C_{\mp}\right) & = & h_{mn}+u_{mn}+\delta_{mn}V_{C}\left(t\right)
\end{eqnarray}
Now we assume that the leads satisfy the WBLA, i.e. we neglect the
energy dependence of the lead-molecule coupling. As described in Ref.
\cite{Ridley2015}, this assumption enables us to write down all components
of the effective embedding self-energy in terms of the level-width
matrix $\mathbf{\Gamma}_{\alpha}$, defined as: 
\begin{equation}
\Gamma_{\alpha,mn}=2\pi\sum_{k}T_{m,k\alpha}T_{k\alpha,n}\delta\left(\varepsilon_{\alpha}^{F}-\varepsilon_{k\alpha}\right)\label{eq:gamWBLA}
\end{equation}
where $\varepsilon_{\alpha}^{F}$ is the equilibrium Fermi energy
of lead $\alpha$. The self-energy components for this problem are
collected together in Eqs. (\ref{eq:retardedself})-(\ref{eq:lessgreatself})
of Appendix (\ref{sec:Appendix-A}), where the time-dependence of
the lead states is contained in phase factors of the form: 
\begin{equation}
\psi_{\alpha}\left(t_{1},t_{2}\right)\equiv\underset{t_{2}}{\overset{t_{1}}{\int}}d\tau\, V_{\alpha}\left(\tau\right)\label{eq:phasealpha}
\end{equation}
Within the WBLA, the KB equations \cite{Kadanoff1962} for the different
components of $\mathbf{G}_{CC}$ are linearized in terms of the effective
Hamiltonian $\widetilde{\mathbf{h}}_{CC}^{eff}\equiv\widetilde{\mathbf{h}}_{CC}-\frac{i}{2}\underset{\alpha}{\sum}\mathbf{\Gamma}_{\alpha}$
of the central region, where $\widetilde{\mathbf{h}}_{CC}=\mathbf{h}_{CC}+\mathbf{u}_{CC}$.
The derivation of these components was published in Refs. \cite{Ridley2015,Ridley2016},
and leads to the following compact formula for the greater and lesser
Green's functions:}

\textcolor{black}{
\begin{equation}
\mathbf{G}_{CC}^{\gtrless}\left(t_{1},t_{2}\right)=\mp i\int\frac{d\omega}{2\pi}f\left(\mp\left(\omega-\mu\right)\right)\underset{\gamma}{\sum}\mathbf{S}_{\gamma}\left(t_{1},t_{0};\omega\right)\mathbf{\Gamma}_{\gamma}\mathbf{S}_{\gamma}^{\dagger}\left(t_{2},t_{0};\omega\right)\label{eq:greater/lesserGF}
\end{equation}
where we introduce the matrix}

\textcolor{black}{
\begin{equation}
\mathbf{S}_{\alpha}\left(t,t_{0};\omega\right)\equiv e^{-i\mathbf{\tilde{h}}_{CC}^{eff}\left(t-t_{0}\right)}e^{-i\varphi_{C}\left(t,t_{0}\right)}\left[\mathbf{G}_{CC}^{r}\left(\omega\right)-i\intop_{t_{0}}^{t}d\bar{t}e^{-i\left(\omega\mathbf{1}-\mathbf{\widetilde{h}}_{CC}^{eff}\right)\left(\bar{t}-t_{0}\right)}e^{i\left(\varphi_{C}-\psi_{\alpha}\right)\left(\bar{t},t_{0}\right)}\right]\label{eq:smatrix}
\end{equation}
defined in terms of $\mathbf{G}_{CC}^{r}\left(\omega\right)=\left(\omega\mathbf{I}-\mathbf{h}_{CC}^{eff}\right)^{-1}$
(i.e. defined without the tilde on the effective Hamiltonian), and
the phase factor associated with the molecular time-dependence:}

\textcolor{black}{
\begin{equation}
\varphi_{C}\left(t_{1},t_{2}\right)\equiv\underset{t_{2}}{\overset{t_{1}}{\int}}d\tau V_{C}\left(\tau\right)\label{eq:phiC}
\end{equation}
All other components of the GF can be explicitly calculated in the
time domain \cite{Ridley2015,Ridley2016}, and are listed in Appendix
(\ref{sec:Appendix-A}). The quantum statistical expectation value
of the current operator (\ref{eq:currentoperator}) can also be reformulated
as a sum of convolution integrals on the Konstantinov-Perel' contour,
which may be evaluated exactly within the WBLA. Setting the electronic
charge $q=-1$, the current may be expressed in terms of the $\mathbf{S}_{\alpha}$
as \cite{Ridley}:}

\begin{equation}
I_{\alpha}\left(t\right)=\frac{1}{\pi}\int d\omega f\left(\omega-\mu\right)\,\mbox{Tr}_{C}\left[2\mbox{Re}\left[i\mathbf{\Gamma}_{\alpha}e^{i\omega\left(t-t_{0}\right)}e^{i\psi_{\alpha}\left(t,t_{0}\right)}\mathbf{S}_{\alpha}\left(t,t_{0};\omega\right)\right]-\mathbf{\Gamma}_{\alpha}\underset{\gamma}{\sum}\mathbf{S}_{\gamma}\left(t,t_{0};\omega\right)\mathbf{\Gamma}_{\gamma}\mathbf{S}_{\gamma}^{\dagger}\left(t,t_{0};\omega\right)\right]\label{eq:compactcurrent}
\end{equation}
The WBLA enables us to derive a closed form for the current correlation
function. We substitute the equations (\ref{eq:retardedself}) and
(\ref{eq:advancedself}) for the retarded/advanced self energies into
Eq. (\ref{eq:power1}), which then reduces to a rather compact form:

\begin{eqnarray*}
C_{\alpha\beta}\left(t_{1},t_{2}\right) & = & 4q^{2}\mathrm{Tr}_{C}\left[\delta_{\alpha\beta}\left(\mathbf{\Sigma}_{\alpha}^{>}\left(t_{1},t_{2}\right)\mathbf{G}_{CC}^{<}\left(t_{2},t_{1}\right)+\mathbf{G}_{CC}^{>}\left(t_{1},t_{2}\right)\mathbf{\Sigma}_{\alpha}^{<}\left(t_{2},t_{1}\right)\right)\right.
\end{eqnarray*}
\[
+\mathbf{\Gamma}_{\alpha}\mathbf{G}_{CC}^{>}\left(t_{1},t_{2}\right)\mathbf{\Gamma}_{\beta}\mathbf{G}_{CC}^{<}\left(t_{2},t_{1}\right)
\]

\[
+i\mathbf{G}_{CC}^{>}\left(t_{1},t_{2}\right)\left[\mathbf{\Lambda}_{\beta}^{+}\left(t_{2},t_{1}\right)\mathbf{\Gamma}_{\alpha}+\mathbf{\Gamma}_{\beta}\left(\mathbf{\Lambda}_{\alpha}^{+}\right)^{\dagger}\left(t_{1},t_{2}\right)\right]
\]
\[
+i\left[\mathbf{\Lambda}_{\alpha}^{-}\left(t_{1},t_{2}\right)\mathbf{\Gamma}_{\beta}+\mathbf{\Gamma}_{\alpha}\left(\mathbf{\Lambda}_{\beta}^{-}\right)^{\dagger}\left(t_{2},t_{1}\right)\right]\mathbf{G}_{CC}^{<}\left(t_{2},t_{1}\right)
\]

\begin{equation}
\left.-\mathbf{\Lambda}_{\beta}^{+}\left(t_{2},t_{1}\right)\mathbf{\Lambda}_{\alpha}^{-}\left(t_{1},t_{2}\right)-\left(\mathbf{\Lambda}_{\alpha}^{+}\right)^{\dagger}\left(t_{1},t_{2}\right)\left(\mathbf{\Lambda}_{\beta}^{-}\right)^{\dagger}\left(t_{2},t_{1}\right)\right]\label{eq:corrcompact}
\end{equation}
Here we have collected convolution integrals on the Konstantinov-Perel'
contour into the objects $\Lambda_{\alpha}^{\pm}\left(t_{1},t_{2}\right)$:

\begin{eqnarray}
\mathbf{\Lambda}_{\beta}^{+}\left(t_{2},t_{1}\right) & \equiv & \left(\mathbf{\Sigma}_{\beta}^{<}.\mathbf{G}_{CC}^{a}+\mathbf{\Sigma}_{\beta}^{\urcorner}\star\mathbf{G}_{CC}^{\ulcorner}\right)_{\left(t_{2}^{-},t_{1}^{+}\right)}\label{eq:lambdabetaplus}\\
\left(\mathbf{\Lambda}_{\alpha}^{+}\right)^{\dagger}\left(t_{1},t_{2}\right) & \equiv & -\left(\mathbf{G}_{CC}^{r}.\mathbf{\Sigma}_{\alpha}^{<}+\mathbf{G}_{CC}^{\urcorner}\star\mathbf{\Sigma}_{\alpha}^{\ulcorner}\right)_{\left(t_{2}^{-},t_{1}^{+}\right)}\\
\mathbf{\Lambda}_{\alpha}^{-}\left(t_{1},t_{2}\right) & \equiv & \left(\mathbf{\Sigma}_{\alpha}^{>}.\mathbf{G}_{CC}^{a}+\mathbf{\Sigma}_{\alpha}^{\urcorner}\star\mathbf{G}_{CC}^{\ulcorner}\right)_{\left(t_{1}^{+},t_{2}^{-}\right)}\\
\left(\mathbf{\Lambda}_{\beta}^{-}\right)^{\dagger}\left(t_{2},t_{1}\right) & \equiv & -\left(\mathbf{G}_{CC}^{r}.\mathbf{\Sigma}_{\beta}^{>}+\mathbf{G}_{CC}^{\urcorner}\star\mathbf{\Sigma}_{\beta}^{\ulcorner}\right)_{\left(t_{1}^{+},t_{2}^{-}\right)}\label{eq:lambdabetaminusconj}
\end{eqnarray}
We may now perform the convolution integrals in Eqs. (\ref{eq:lambdabetaplus})-(\ref{eq:lambdabetaminusconj})
using the formulas obtained for the self energies and GFs of the $CC$
region in Appendix \ref{sec:Appendix-A}. The convolution integrals
in $\mathbf{\Lambda}^{\pm}$ are evaluated using the methods of Refs.
\cite{Ridley2015,Ridley2016}, where the transformation from Matsubara
summations to frequency integrals \cite{Stefanucci2013} is done taking
account of the ordering of time-variables on the contour. This guarantees
the linearity of each term in the fermion/hole distribution function
$f\left(\pm\left(\omega-\mu\right)\right)$, and results in the following
pair of functional identities:

\begin{eqnarray}
\mathbf{\Lambda}_{\beta}^{+}\left(t_{2},t_{1}\right) & = & ie^{-i\psi_{\beta}\left(t_{2},t_{0}\right)}\int\frac{d\omega}{2\pi}f\left(\omega-\mu\right)e^{-i\omega\left(t_{2}-t_{0}\right)}\mathbf{\Gamma}_{\beta}\mathbf{S}_{\beta}^{\dagger}\left(t_{1},t_{0};\omega\right)\label{eq:amatrix}
\end{eqnarray}

\begin{equation}
\mathbf{\Lambda}_{\alpha}^{-}\left(t_{1},t_{2}\right)=-ie^{-i\psi_{\alpha}\left(t_{1},t_{0}\right)}\int\frac{d\omega}{2\pi}\left(1-f\left(\omega-\mu\right)\right)e^{-i\omega\left(t_{1}-t_{0}\right)}\mathbf{\Gamma}_{\alpha}\mathbf{S}_{\alpha}^{\dagger}\left(t_{2},t_{0};\omega\right)\label{eq:lambda2}
\end{equation}
Here we have expressed $\mathbf{\Lambda}_{\alpha}^{\pm}$ in terms
of the matrix $\mathbf{S}_{\alpha}$ defined in Eq. (\ref{eq:smatrix}).
Notice on the first line of Eq. (\ref{eq:corrcompact}) the presence
of the object $4q^{2}\textrm{Tr}_{C}\left[\mathbf{\Gamma}_{\alpha}\mathbf{G}_{CC}^{>}\left(t_{1},t_{2}\right)\mathbf{\Gamma}_{\beta}\mathbf{G}_{CC}^{<}\left(t_{2},t_{1}\right)\right]$.
In the single level case, all the objects inside the trace are replaced
by scalars, and this object is equal to $q^{2}\Gamma_{\alpha}\Gamma_{\beta}N_{C}\left(t\right)\left(1-N_{C}\left(t\right)\right)$,
where the particle number on the molecular region is defined by $N_{C}\left(t\right)=-2i\textrm{Tr}_{C}\left[\mathbf{G}_{CC}^{<}\left(t,t\right)\right]$
. The time-dependence of this object is thus entirely due to the internal
dynamics of electron and hole populations on sites of the molecule.
The lead dependent matrices $\mathbf{\Lambda}_{\beta}^{+}$ and $\mathbf{\Lambda}_{\alpha}^{-}$
correspond physically to electrons propagating from lead $\beta$
and positively charged holes propagating from lead $\alpha$, respectively.
We therefore interpret the two terms appearing on the second line
of Eq. (\ref{eq:corrcompact}) as describing processes in which electrons
in the leads interfere with holes in the molecular region, or holes
in the leads interefere with electrons in the molecule. The terms
on the third line of Eq. (\ref{eq:corrcompact}) are interpreted as
cross-lead particle-hole interference terms. 

In \cite{Ridley2015,Ridley2016}, the greater and lesser Green's functions
were expressed in terms of the $\mathbf{S}_{\alpha}$ matrices following
a line-integral of the Kadanoff Baym equations in the two-time plane,
and these are given in Eq. (\ref{eq:greater/lesserGF}). We thus have
explicit formulas for all terms which appear in the two-time correlation
function, which may be evaluated numerically in the $\left(t_{1},t_{2}\right)$
plane as follows:

\begin{equation}
C_{\alpha\beta}\left(t_{1},t_{2}\right)=4q^{2}\int\frac{d\omega}{2\pi}\frac{d\omega'}{2\pi}\left(1-f\left(\omega-\mu\right)\right)f\left(\omega'-\mu\right)\textrm{Tr}{}_{C}\left\{ \delta_{\alpha\beta}\underset{\gamma}{\sum}\left(\mathbf{\Gamma}_{\alpha}e^{-i\psi_{\alpha}\left(t_{1},t_{2}\right)}e^{-i\omega\left(t_{1}-t_{2}\right)}\mathbf{S}_{\gamma}\left(t_{2},t_{0};\omega'\right)\mathbf{\Gamma}_{\gamma}\mathbf{S}_{\gamma}^{\dagger}\left(t_{1},t_{0};\omega'\right)+h.c.\right)\right.\label{eq:correlationnonsymmetric}
\end{equation}

\[
+\underset{\gamma,\gamma'}{\sum}\mathbf{\Gamma}_{\alpha}\mathbf{S}_{\gamma}\left(t_{1},t_{0};\omega\right)\mathbf{\Gamma}_{\gamma}\mathbf{S}_{\gamma}^{\dagger}\left(t_{2},t_{0};\omega\right)\mathbf{\Gamma}_{\beta}\mathbf{S}_{\gamma'}\left(t_{2},t_{0};\omega'\right)\mathbf{\Gamma}_{\gamma'}\mathbf{S}_{\gamma'}^{\dagger}\left(t_{1},t_{0};\omega'\right)
\]

\[
+i\underset{\gamma}{\sum}\left(\mathbf{\Gamma}_{\alpha}\mathbf{S}_{\gamma}\left(t_{1},t_{0};\omega\right)\mathbf{\Gamma}_{\gamma}\mathbf{S}_{\gamma}^{\dagger}\left(t_{2},t_{0};\omega\right)\mathbf{\Gamma}_{\beta}\left(e^{-i\psi_{\beta}\left(t_{2},t_{0}\right)}e^{-i\omega'\left(t_{2}-t_{0}\right)}\mathbf{S}_{\beta}^{\dagger}\left(t_{1},t_{0};\omega'\right)-e^{i\psi_{\alpha}\left(t_{1},t_{0}\right)}e^{i\omega'\left(t_{1}-t_{0}\right)}\mathbf{S}_{\alpha}\left(t_{2},t_{0};\omega'\right)\right)\right)
\]

\[
-\left(e^{-i\psi_{\beta}\left(t_{2},t_{0}\right)}e^{-i\omega\left(t_{2}-t_{0}\right)}\mathbf{\Gamma}_{\beta}\mathbf{S}_{\beta}^{\dagger}\left(t_{1},t_{0};\omega\right)e^{-i\psi_{\alpha}\left(t_{1},t_{0}\right)}e^{-i\omega'\left(t_{1}-t_{0}\right)}\mathbf{\Gamma}_{\alpha}\mathbf{S}_{\alpha}^{\dagger}\left(t_{2},t_{0};\omega'\right)\right.
\]

\[
\left.\left.+e^{i\psi_{\alpha}\left(t_{1},t_{0}\right)}e^{i\omega\left(t_{1}-t_{0}\right)}\mathbf{S}_{\alpha}\left(t_{2},t_{0};\omega\right)\mathbf{\Gamma}_{\alpha}e^{i\psi_{\beta}\left(t_{2},t_{0}\right)}e^{i\omega'\left(t_{2}-t_{0}\right)}\mathbf{S}_{\beta}\left(t_{1},t_{0};\omega'\right)\mathbf{\Gamma}_{\beta}\right)\right\} 
\]
This expression contains a great deal of information, and it is the
central result of this paper. It is the two-time correlation function
for a molecular junction connected to an arbitrary number of leads,
through which time-dependent voltages are passed. It contains transient
parts which decay as $t_{1},t_{2}\rightarrow\infty$, while $\tau\equiv t_{1}-t_{2}$
remains finite. It automatically enables evaluation of cross-correlation
functions between different leads when $\alpha\neq\beta$, and the
correlation between currents through the same lead when $\alpha=\beta$.
It describes the noise on the current signal due to non-zero temperatures
(the thermal noise), and due to a non-zero bias (the shot noise),
as will be made clearer in the next section. The leads are assumed
to satisfy \textcolor{black}{the WBLA, and the additive contribution
of the voltage to the lead state energies is assumed, but the approach
is otherwise exact for electrons interacting at the mean field level
\cite{Gabdank2011}. Under close inspection, using the definition
(\ref{eq:smatrix}), we find that the explicit time-dependence enters
into (\ref{eq:correlationnonsymmetric}) only within structures of
the form $e^{i\left(\psi_{\alpha}-\varphi_{C}\right)\left(t,t_{0}\right)}$,
so that the noise does not distinguish between external fields that
bias all leads identically ($V_{\alpha}\left(t\right)=V\left(t\right)$,
for all $\alpha$) or a gate voltage which moves energies in the negative
energy direction ($V_{C}\left(t\right)=-V\left(t\right)$). }This
is also true for the current \cite{Ridley2016}. The expression (\ref{eq:correlationnonsymmetric})
will be used for the proof of analytic identities in Section (\ref{sec:Recovery-of-Known}),
but it is not entirely convenient for numerical evaluation. Instead,
we describe in Section (\ref{sec:Numerics}) and Appendix (\ref{sec:Appendix-D})
how to evaluate Eq. (\ref{eq:corrcompact}) directly.

\section{Recovery of Known Results For a Static Bias\label{sec:Recovery-of-Known}}

To parameterize our system with experimentally relevant variables,
we work in the relative time coordinate system so that $t_{1}=\tau+t$
and $t_{2}=t$, where $\tau\equiv t_{1}-t_{2}$ is the relative time
that we wish to take a Fourier transform with respect to $\tau$.
Note that, to make the mapping to the Fourier space associated with
$\tau$, one needs $\tau$ to take on negative values. However, since
both $t_{1}$ and $t_{2}$ must be times greater than $t_{0}$, this
means that $\tau$ is restricted to lie in the range $\left[-\left(t-t_{0}\right),t-t_{0}\right]$,
as was done in Ref. \cite{Joho2012}. We define the Fourier-transform
of the correlation with respect to the \textit{relative time} $\tau\equiv t_{1}-t_{2}$,
as a function of a single frequency $\Omega$ and the \textit{measurement
time} $t$:

\begin{equation}
P_{\alpha\beta}\left(\Omega,t\right)\equiv\underset{-t+t_{0}}{\overset{t-t_{0}}{\int}}d\tau e^{i\Omega\tau}P_{\alpha\beta}\left(t+\tau,t\right)=\frac{1}{2}\left(C_{\alpha\beta}\left(\Omega,t\right)+C_{\alpha\beta}^{*}\left(-\Omega,t\right)\right)\label{eq:PSomega}
\end{equation}
where $C_{\alpha\beta}\left(\Omega,t\right)$ is the Fourier transform
of $C_{\alpha\beta}\left(t+\tau,t\right)$ with respect to $\tau$.
Note that the relation 
\begin{eqnarray}
P_{\alpha\beta}^{*}\left(\Omega,t\right) & = & P_{\alpha\beta}\left(-\Omega,t\right)\label{eq:psym}
\end{eqnarray}
immediately follows. In Section \ref{sub:Wick} we remarked that it
is sufficient for knowledge of $P_{\alpha\beta}\left(t_{1},t_{2}\right)$
to know the non-symmetrized function $C_{\alpha\beta}\left(t_{1},t_{2}\right)$. 

In addition to the power spectrum, one can calculate several other
useful quantities in terms of the $C_{\alpha\beta}$. For instance,
in a two-lead junction, one may focus on the net current,

\begin{equation}
\hat{I}_{LR}^{\left(-\right)}\left(t\right)=\frac{1}{2}\left(\hat{I}_{L}\left(t\right)-\hat{I}_{R}\left(t\right)\right)\label{eq:inet}
\end{equation}
or on the sum of currents, which by the continuity equation is proportional
to the rate of change of charge in the molecule \cite{Ridley}:

\begin{equation}
\hat{I}_{LR}^{\left(+\right)}\left(t\right)=\frac{1}{2}\left(\hat{I}_{L}\left(t\right)+\hat{I}_{R}\left(t\right)\right)\label{eq:isum}
\end{equation}
The time-dependent noise spectra of these objects can be written:

\begin{equation}
C^{\left(-\right)}\left(\Omega,t\right)=\int d\tau e^{i\Omega\tau}\left\langle \triangle\hat{I}_{LR}^{\left(-\right)}\left(t+\tau\right)\triangle\hat{I}_{LR}^{\left(-\right)}\left(t\right)\right\rangle =\frac{1}{2}\left(C^{\left(auto\right)}\left(\Omega,t\right)-C^{\left(\times\right)}\left(\Omega,t\right)\right)\label{eq:corrnet}
\end{equation}

\begin{equation}
C^{\left(+\right)}\left(\Omega,t\right)=\int d\tau e^{i\Omega\tau}\left\langle \triangle\hat{I}_{LR}^{\left(+\right)}\left(t+\tau\right)\triangle\hat{I}_{LR}^{\left(+\right)}\left(t\right)\right\rangle =\frac{1}{2}\left(C^{\left(auto\right)}\left(\Omega,t\right)+C^{\left(\times\right)}\left(\Omega,t\right)\right)\label{eq:corrsum}
\end{equation}
where we have defined Fourier transforms of the \textit{\textcolor{black}{average}}
autocorrelation and cross-correlations:

\begin{equation}
C^{\left(auto\right)}\left(t+\tau,t\right)\equiv\frac{1}{2}\left(C_{LL}\left(t+\tau,t\right)+C_{RR}\left(t+\tau,t\right)\right)\label{eq:Cauto}
\end{equation}

\begin{equation}
C^{\left(\times\right)}\left(t+\tau,t\right)\equiv\frac{1}{2}\left(C_{LR}\left(t+\tau,t\right)+C_{RL}\left(t+\tau,t\right)\right)\label{eq:Ccross}
\end{equation}
\textcolor{black}{In general, $C^{\left(auto\right)}$ and $C^{\left(\times\right)}$
are complex quantities and so cannot be observed. However, due to
the symmetry property (\ref{eq:csymmetry}), they are both real at
the equal observation time point $\tau=0$. This fact was exploited
in Ref. \cite{Feng2008}, where the equal time autocorrelation in
the left lead, $C_{LL}\left(t,t\right)$, was studied in the time
domain. Using the identity (\ref{eq:csymmetry}), one can show that
the real parts of these functions are always }\textit{\textcolor{black}{symmetric}}\textcolor{black}{{}
in the $\tau=0$ line:}

\begin{equation}
\textrm{Re}\left[C^{\left(auto/\times\right)}\left(t+\tau,t\right)\right]=\textrm{Re}\left[C^{\left(auto/\times\right)}\left(t,t+\tau\right)\right]\label{eq:Csymreal1}
\end{equation}
, whereas the imaginary parts are always \textit{antisymmetric} about
this line:

\begin{equation}
\textrm{Im}\left[C^{\left(auto/\times\right)}\left(t+\tau,t\right)\right]=-\textrm{Im}\left[C^{\left(auto/\times\right)}\left(t,t+\tau\right)\right]\label{eq:Csymimag1}
\end{equation}

\textcolor{black}{To check the validity of our theory, we must confirm
that it reduces to known expressions in the long time and static bias
limits, as was already demonstrated for the current in Ref. \cite{Ridley2015}.
We shall assume that the bias is applied only to the leads ($\varphi_{C}\left(t_{1},t_{2}\right)\equiv0$),
that the equilibrium and nonequilibrium effective molecular Hamiltonians
are identical ($\widetilde{\mathbf{h}}_{CC}^{eff}=\mathbf{h}_{CC}^{eff}$),
and that $V_{\alpha}\left(t\right)=V_{\alpha}$ is constant in time
($t>t_{0}$). In th}is case the $\mathbf{S}_{\alpha}$ defined in
Eq. (\ref{eq:smatrix}) can be evaluated explicitly, and in the $t_{0}\rightarrow-\infty$
limit we obtain 

\begin{gather}
\mathbf{S}_{\gamma}\left(t_{1},t_{0};\omega\right)\mathbf{\Gamma}_{\gamma}\mathbf{S}_{\gamma}^{\dagger}\left(t_{2},t_{0};\omega\right)\underset{t_{0}\rightarrow-\infty}{\longrightarrow}e^{-i\left(\omega+V_{\gamma}\right)\left(t_{1}-t_{2}\right)}\mathbf{A}_{\gamma}\left(\omega+V_{\gamma}\right)\label{eq:longtimesSS}
\end{gather}
where $\mathbf{A}_{\gamma}\left(\omega\right)\equiv\mathbf{G}_{CC}^{r}\left(\omega\right)\mathbf{\Gamma}_{\gamma}\mathbf{G}_{CC}^{a}\left(\omega\right)$.
Other expressions appearing in the generalized two-time correlation
function can be worked out in a similar way, for instance:

\begin{equation}
e^{-i\psi_{\beta}\left(t_{2},t_{0}\right)}e^{-i\omega\left(t_{2}-t_{0}\right)}\mathbf{S}_{\beta}^{\dagger}\left(t_{1},t_{0};\omega\right)\underset{t_{0}\rightarrow-\infty}{\longrightarrow}\mathbf{G}_{CC}^{a}\left(\omega+V_{\beta}\right)e^{i\left(\omega+V_{\beta}\right)\tau}\label{eq:longtimesdagg}
\end{equation}
The $\mathbf{S}_{\alpha}$ matrices enter into the general expression
(\ref{eq:correlationnonsymmetric}) only in the form of structures
like (\ref{eq:longtimesSS}) and (\ref{eq:longtimesdagg}), so we
easily conclude that the correlation function $C_{\alpha\beta}\left(t_{1},t_{2}\right)$
depends only on the time difference $\tau$, the power spectrum does
not depend on time $t$. Hence the current becomes a stationary stochastic
process under the conditions that the bias is static and that the
switch-on time is relegated to the distant past. This is implied by
the fact that the current itself is simply the steady-state LB formula
in this case, as it was proven in \cite{Ridley2015} that all terms
arising from the initial conditions (vertical contour convolutions)
vanish in the long-time limit. With the exception of the initial condition
term, every vanishing term includes a convolution with a left or right
self-energy. In the partitioned approach to the transport problem,
these quantities vanish, as one can see from the definition (\ref{eq:self-energy-Def})
and the fact that $\mathbf{h}_{C\alpha}\left(z\right)=0$ for all
$z\in C_{M}$. It also enables us to write down a generic formula
for the lesser/greater Green's functions and $\mathbf{\Lambda}^{\pm}$
matrices that can be chosen either partitioned or partition-free,
for all times, by premultiplying all terms arising from a vertical
contour convolution integral by the \textit{\textcolor{black}{partitioning
parameter}}
\begin{equation}
\xi_{p}=\begin{cases}
\begin{array}{c}
1,\,\textrm{partition-free}\\
0,\,\textrm{partitioned}
\end{array}\end{cases}\label{eq:partitioningparameter}
\end{equation}
We include this parameter in the formulas for numerical implementation
in Appendix \ref{sec:Appendix-D}, which will enable us to directly
compare the noise resulting from a partitioned and partition-free
switch-on in Section \ref{sec:Numerics} of this paper. In studies
of high-frequency shot noise, the interesting physical observable
is usually the static non-symmetrized power spectrum \cite{Aguado2000,Zamoum2015},
which is the regular Fourier transform (denoted via $\mathcal{F}$
hereafter) of $C_{\alpha\beta}\left(\tau\right)\equiv\underset{t_{0}\rightarrow-\infty}{\lim}C_{\alpha\beta}\left(t+\tau,t\right)$:
\begin{equation}
C_{\alpha\beta}\left(\Omega\right)\equiv\underset{t_{0}\rightarrow-\infty}{\lim}C_{\alpha\beta}\left(\Omega,t\right)=\underset{-\infty}{\overset{\infty}{\int}}d\tau e^{i\Omega\tau}C_{\alpha\beta}\left(\tau\right)\equiv\mathcal{F}\left[C_{\alpha\beta}\left(\tau\right);\Omega\right]\label{eq:longPS}
\end{equation}
Note that infinite limits are possible here as $t_{0}\rightarrow-\infty$.
The above quantity satisfies the relation $C_{\alpha\beta}^{*}\left(\Omega\right)=C_{\beta\alpha}\left(\Omega\right)$
\cite{Engel2004}. For those experiments which do distinguish between
absorption and emission processes, the quantity of interest is most
often $C_{\alpha\alpha}\left(\Omega\right)$, which in general satisfies
the inequality $C_{\alpha\alpha}\left(\Omega\right)\neq C_{\alpha\alpha}\left(-\Omega\right)$.
$C_{\alpha\alpha}\left(\Omega\right)$ can therefore be used to describe
measurements in which a quanta of energy $\hbar\Omega$ is transferred
from the measuring device to the system. By contrast, the symmetrized
spectrum obeys $P_{\alpha\alpha}\left(\Omega,t\right)=P_{\alpha\alpha}\left(-\Omega,t\right)$,
i.e. it does not distinguish between emission and absorption processes.
Moreover, in recently published work \cite{Yang2014}, a master equation
formalism was used to derive an exact formula for the frequency-dependent
autocorrelation and cross-lead current correlations in a nanojunction
composed of a quantum dot coupled to two leads, which were treated
within the WBLA. In Appendix \ref{sec:Appendix-C}, we derive an explicit
formula for $C_{\alpha\beta}\left(\Omega\right)$. Here we simply
note that, if the discussion is restricted to a molecule coupled to
left (L) and right (R) leads, we find that the non-symmetrized autocorrelation
associated with a single lead is given by

\[
C_{\alpha\alpha}\left(\Omega\right)=4q^{2}\int\frac{d\omega}{2\pi}\textrm{Tr}_{C}\left[\left(1-f_{\alpha}\left(\omega+\Omega-\mu\right)\right)f_{\bar{\alpha}}\left(\omega-\mu\right)\mathbf{T}_{CC}^{\left(\alpha\bar{\alpha}\right)}\left(\omega\right)\mathbf{T}_{CC}^{\dagger\left(\alpha\bar{\alpha}\right)}\left(\omega\right)\right.
\]

\[
+\left(1-f_{\bar{\alpha}}\left(\omega+\Omega-\mu\right)\right)f_{\alpha}\left(\omega-\mu\right)\mathbf{T}_{CC}^{\left(\alpha\bar{\alpha}\right)}\left(\omega+\Omega\right)\mathbf{T}_{CC}^{\dagger\left(\alpha\bar{\alpha}\right)}\left(\omega+\Omega\right)
\]

\[
-\left(f_{\alpha}\left(\omega-\mu\right)-f_{\bar{\alpha}}\left(\omega-\mu\right)\right)\left(f_{\alpha}\left(\omega+\Omega-\mu\right)-f_{\bar{\alpha}}\left(\omega+\Omega-\mu\right)\right)\mathbf{T}_{CC}^{\left(\alpha\bar{\alpha}\right)}\left(\omega\right)\mathbf{T}_{CC}^{\dagger\left(\alpha\bar{\alpha}\right)}\left(\omega\right)\mathbf{T}_{CC}^{\left(\alpha\bar{\alpha}\right)}\left(\omega+\Omega\right)\mathbf{T}_{CC}^{\dagger\left(\alpha\bar{\alpha}\right)}\left(\omega+\Omega\right)
\]

\begin{equation}
\left.+\Omega^{2}\left(1-f_{\alpha}\left(\omega+\Omega-\mu\right)\right)f_{\alpha}\left(\omega-\mu\right)\mathbf{\Gamma}_{\alpha}\mathbf{G}^{r}\left(\omega\right)\mathbf{A}_{\alpha}\left(\omega+\Omega\right)\mathbf{G}^{a}\left(\omega\right)\right]\label{eq:CautoZhang}
\end{equation}
where $\bar{\alpha}\neq\alpha$, and we have defined the \textit{transmission
matrices} in the standard way \cite{DiVentra2008}:

\begin{eqnarray}
\mathbf{T}_{CC}^{\left(\alpha\beta\right)}\left(\omega\right) & \equiv & \left[\mathbf{\Gamma}_{\alpha}\right]^{\frac{1}{2}}\mathbf{G}_{CC}^{r}\left(\omega\right)\left[\mathbf{\Gamma}_{\beta}\right]^{\frac{1}{2}}\label{eq:transmission1}
\end{eqnarray}

\begin{equation}
\mathbf{T}_{CC}^{\dagger\left(\alpha\beta\right)}\left(\omega\right)\equiv\left[\mathbf{\Gamma}_{\beta}\right]^{\frac{1}{2}}\mathbf{G}_{CC}^{a}\left(\omega\right)\left[\mathbf{\Gamma}_{\alpha}\right]^{\frac{1}{2}}\label{eq:transmission2}
\end{equation}

Physically, the eigenvalues of $\mathbf{T}_{CC}^{\left(\alpha\beta\right)}\left(\omega\right)$
may be interpreted as probability amplitudes for electron scattering
events between the $\alpha$ and $\beta$\textcolor{black}{{} leads.
Eq. (\ref{eq:CautoZhang}) gives the analytic behaviour of a function
which should be accessible to the experimentalist - it is a power
spectrum for current measurements carried out with arbitrary detection
frequency, taken at long times after the switch-on of a constant bias.
It is expressed in terms of the transmission matrices, which depend
on the molecular Hamiltonian and on the coupling of the molecule to
the leads.} We remark that if one restricts the $CC$ region to a
single energy level, and replaces $\Omega\rightarrow-\Omega$ on the
right-hand side, then Eq. (\ref{eq:CautoZhang}) is exactly equivalent
to the expression found in Ref. \cite{Yang2014} (there the Fourier
transform was taken with a phase of $-i\Omega\tau$). We note that
Ref. \cite{Yang2014} also included a numerical scheme for moving
beyond the WBLA, and is in this sense more general than the formalism
presented in the current paper. Indeed, the self-energy in Eq. (\ref{eq:selflessgreat})
contains a singularity at $t_{1}=t_{2}$, and appears in Eq. (\ref{eq:corrcompact})
multiplied by $\delta_{\alpha\beta}$, so that the autocorrelation
function is singular in the two-time plane for $t_{1}=t_{2}$ whereas
the cross-correlation function is finite. This singularity in the
autocorrelation is an artefact of the WBLA, and does not exist when
the band-width of the leads is taken to be finite \cite{Feng2008,Yang2014}.
However our scheme can be used for rapid calculations on extended
molecules with a far larger spectrum than a quantum dot, and in such
molecules the WBLA is an increasingly accurate approximation \cite{Verzijl2012,Verzijl2013}.
To remove this singularity in the current autocorrelations, one may
leave the observation time representation and instead compute the
noise in a time-averaged sense. We defer this to a future work and
will instead perform calculations of the average cross-correlation
$C^{\left(\times\right)}\left(t,t\right)$ in the $t$ domain in Section
(\ref{sec:Numerics}) as this quantity is free from any singularities. 

\textcolor{black}{From Eqs. (\ref{eq:correlationnonsymmetric}) and
(\ref{eq:twotimecorr}) one obtains the symmetrized two-time correlation
function}:

\[
P_{\alpha\beta}\left(t_{1},t_{2}\right)=2q^{2}\int\frac{d\omega}{2\pi}\frac{d\omega'}{2\pi}F\left(\omega,\omega'\right)\textrm{Tr}{}_{C}\left\{ \delta_{\alpha\beta}\underset{\gamma}{\sum}\left(\mathbf{\Gamma}_{\alpha}e^{-i\psi_{\alpha}\left(t_{1},t_{2}\right)}e^{-i\omega\left(t_{1}-t_{2}\right)}\mathbf{S}_{\gamma}\left(t_{2},t_{0};\omega'\right)\mathbf{\Gamma}_{\gamma}\mathbf{S}_{\gamma}^{\dagger}\left(t_{1},t_{0};\omega'\right)+h.c\right)\right.
\]

\[
+\underset{\gamma,\gamma'}{\sum}\mathbf{\Gamma}_{\alpha}\mathbf{S}_{\gamma}\left(t_{1},t_{0};\omega\right)\mathbf{\Gamma}_{\gamma}\mathbf{S}_{\gamma}^{\dagger}\left(t_{2},t_{0};\omega\right)\mathbf{\Gamma}_{\beta}\mathbf{S}_{\gamma'}\left(t_{2},t_{0};\omega'\right)\mathbf{\Gamma}_{\gamma'}\mathbf{S}_{\gamma'}^{\dagger}\left(t_{1},t_{0};\omega'\right)
\]

\[
+i\underset{\gamma}{\sum}\left(\mathbf{\Gamma}_{\alpha}\mathbf{S}_{\gamma}\left(t_{1},t_{0};\omega\right)\mathbf{\Gamma}_{\gamma}\mathbf{S}_{\gamma}^{\dagger}\left(t_{2},t_{0};\omega\right)\mathbf{\Gamma}_{\beta}\left(e^{-i\psi_{\beta}\left(t_{2},t_{0}\right)}e^{-i\omega'\left(t_{2}-t_{0}\right)}\mathbf{S}_{\beta}^{\dagger}\left(t_{1},t_{0};\omega'\right)-e^{i\psi_{\alpha}\left(t_{1},t_{0}\right)}e^{i\omega'\left(t_{1}-t_{0}\right)}\mathbf{S}_{\alpha}\left(t_{2},t_{0};\omega'\right)\right)+h.c\right)
\]

\begin{equation}
\left.\left.-\left(e^{-i\psi_{\beta}\left(t_{2},t_{0}\right)}e^{-i\omega\left(t_{2}-t_{0}\right)}\mathbf{\Gamma}_{\beta}\mathbf{S}_{\beta}^{\dagger}\left(t_{1},t_{0};\omega\right)e^{-i\psi_{\alpha}\left(t_{1},t_{0}\right)}e^{-i\omega'\left(t_{1}-t_{0}\right)}\mathbf{\Gamma}_{\alpha}\mathbf{S}_{\alpha}^{\dagger}\left(t_{2},t_{0};\omega'\right)+h.c\right)\right)\right\} \label{eq:correlationcomplete}
\end{equation}
where we define the combination of electron-hole distribution functions:

\begin{equation}
F\left(\omega,\omega'\right)\equiv\left(1-f\left(\omega-\mu\right)\right)f\left(\omega'-\mu\right)+\left(1-f\left(\omega'-\mu\right)\right)f\left(\omega-\mu\right)\label{eq:fermcomb1}
\end{equation}
The steady-state symmetrized power spectrum can then be obtained either
by substituting the long time formulas (\ref{eq:longtimesSS}) and
(\ref{eq:longtimesdagg}) into Eq. (\ref{eq:correlationcomplete})
and taking the Fourier Transform, or simply by substituting the expression
for $C_{\alpha\beta}\left(\Omega\right)$ into (\ref{eq:PSomega}):

\[
P_{\alpha\beta}\left(\Omega\right)\equiv\underset{t_{0}\rightarrow-\infty}{\lim}P_{\alpha\beta}\left(\Omega,t\right)=2q^{2}\int\frac{d\omega}{2\pi}\left\{ \delta_{\alpha\beta}\underset{\gamma}{\sum}\textrm{Tr}_{C}\left[\mathbf{T}_{CC}^{\left(\alpha\gamma\right)}\left(\omega\right)\mathbf{T}_{CC}^{\dagger\left(\alpha\gamma\right)}\left(\omega\right)\right]\left(F_{\alpha\gamma}\left(\omega+\Omega,\omega\right)+F_{\alpha\gamma}\left(\omega-\Omega,\omega\right)\right)\right.
\]

\[
+\underset{\gamma,\gamma'}{\sum}F_{\gamma\gamma'}\left(\omega,\omega-\Omega\right)\textrm{Tr}_{C}\left[\mathbf{T}_{CC}^{\left(\alpha\gamma\right)}\left(\omega\right)\mathbf{T}_{CC}^{\dagger\left(\beta\gamma\right)}\left(\omega\right)\mathbf{T}_{CC}^{\left(\beta\gamma'\right)}\left(\omega-\Omega\right)\mathbf{T}_{CC}^{\dagger\left(\alpha\gamma'\right)}\left(\omega-\Omega\right)\right]
\]

\[
+i\underset{\gamma}{\sum}\textrm{Tr}_{C}\left[\mathbf{T}_{CC}^{\left(\alpha\gamma\right)}\left(\omega\right)\mathbf{T}_{CC}^{\dagger\left(\beta\gamma\right)}\left(\omega\right)\left(F_{\gamma\beta}\left(\omega,\omega-\Omega\right)\mathbf{T}_{CC}^{\dagger\left(\alpha\beta\right)}\left(\omega-\Omega\right)-F_{\gamma\alpha}\left(\omega,\omega-\Omega\right)\mathbf{T}_{CC}^{\left(\beta\alpha\right)}\left(\omega-\Omega\right)\right)\right.
\]

\[
\left.+\mathbf{T}_{CC}^{\left(\beta\gamma\right)}\left(\omega\right)\mathbf{T}_{CC}^{\dagger\left(\alpha\gamma\right)}\left(\omega\right)\left(F_{\gamma\alpha}\left(\omega,\omega+\Omega\right)\mathbf{T}_{CC}^{\dagger\left(\beta\alpha\right)}\left(\omega+\Omega\right)-F_{\gamma\beta}\left(\omega,\omega+\Omega\right)\mathbf{T}_{CC}^{\left(\alpha\beta\right)}\left(\omega+\Omega\right)\right)\right]
\]

\begin{equation}
\left.-\textrm{Tr}_{C}\left[F_{\alpha\beta}\left(\omega,\omega-\Omega\right)\mathbf{T}_{CC}^{\dagger\left(\alpha\beta\right)}\left(\omega-\Omega\right)\mathbf{T}_{CC}^{\dagger\left(\beta\alpha\right)}\left(\omega\right)+F_{\alpha\beta}\left(\omega,\omega+\Omega\right)\mathbf{T}_{CC}^{\left(\alpha\beta\right)}\left(\omega+\Omega\right)\mathbf{T}_{CC}^{\left(\beta\alpha\right)}\left(\omega\right)\right]\right\} \label{eq: 4}
\end{equation}
where we introduce $f_{\alpha}\left(x\right)\equiv f\left(x-V_{\alpha}\right)$,
and make the definition:

\[
F_{\alpha\beta}\left(\omega,\omega'\right)\equiv\left(1-f_{\alpha}\left(\omega-\mu\right)\right)f_{\beta}\left(\omega'-\mu\right)+\left(1-f_{\beta}\left(\omega'-\mu\right)\right)f_{\alpha}\left(\omega-\mu\right)=F_{\beta\alpha}\left(\omega',\omega\right)
\]
It is instructive to compare this formula with the finite frequency
power spectrum derived by B�ttiker and Yang \cite{Buttiker1992,Buttiker1992a,Yang1992,Moskalets2012}
within their S-matrix approach. In particular, when one assumes the
$C$ region to be a single level, and there is a Breit-Wigner resonance
in the scattering matrix amplitudes of their approach with energy-independent
resonance widths \cite{Entin-Wohlman2007,Rothstein2009,Hammer2011},
the LB formalism is exactly equivalent to ours. In many experiments,
the quantity (\ref{eq: 4}) is measured in the \textit{zero-frequency
limit} i.e. when the time separating measurements is much longer than
the timescale over which the current fluctuates \cite{Gabelli2008}.
Taking this limit, and using the identity:

\begin{equation}
\mathbf{G}_{CC}^{r}\left(\omega\right)=\mathbf{G}_{CC}^{a}\left(\omega\right)-i\underset{\gamma}{\sum}\mathbf{G}_{CC}^{r}\left(\omega\right)\mathbf{\Gamma}_{\gamma}\mathbf{G}_{CC}^{a}\left(\omega\right)\label{eq: 6}
\end{equation}
we obtain the zero-frequency power spectrum as:

\[
\underset{\Omega\rightarrow0}{\lim}P_{\alpha\beta}\left(\Omega\right)=2q^{2}\int\frac{d\omega}{2\pi}\left\{ 2\delta_{\alpha\beta}\underset{\gamma}{\sum}F_{\alpha\gamma}\left(\omega,\omega\right)Tr_{C}\left[\mathbf{T}_{CC}^{\left(\alpha\gamma\right)}\left(\omega\right)\mathbf{T}_{CC}^{\dagger\left(\alpha\gamma\right)}\left(\omega\right)\right]\right.
\]

\[
-F_{\alpha\beta}\left(\omega,\omega\right)Tr_{C}\left[\mathbf{T}_{CC}^{\left(\alpha\beta\right)}\left(\omega\right)\mathbf{T}_{CC}^{\dagger\left(\alpha\beta\right)}\left(\omega\right)+\mathbf{T}_{CC}^{\left(\beta\alpha\right)}\left(\omega\right)\mathbf{T}_{CC}^{\dagger\left(\beta\alpha\right)}\left(\omega\right)\right]
\]

\[
+\underset{\gamma,\gamma'}{\sum}\left[F_{\gamma\gamma'}\left(\omega,\omega\right)+F_{\alpha\beta}\left(\omega,\omega\right)\right]Tr_{C}\left[\mathbf{T}_{CC}^{\left(\alpha\gamma\right)}\left(\omega\right)\mathbf{T}_{CC}^{\dagger\left(\beta\gamma\right)}\left(\omega\right)\mathbf{T}_{CC}^{\left(\beta\gamma'\right)}\left(\omega\right)\mathbf{T}_{CC}^{\dagger\left(\alpha\gamma'\right)}\left(\omega\right)\right]
\]

\[
+i\underset{\gamma}{\sum}Tr_{C}\left[\mathbf{T}_{CC}^{\left(\alpha\gamma\right)}\left(\omega\right)\mathbf{T}_{CC}^{\dagger\left(\beta\gamma\right)}\left(\omega\right)\left(F_{\gamma\beta}\left(\omega,\omega\right)\mathbf{T}_{CC}^{\dagger\left(\alpha\beta\right)}\left(\omega\right)-F_{\gamma\alpha}\left(\omega,\omega\right)\mathbf{T}_{CC}^{\left(\beta\alpha\right)}\left(\omega\right)\right)\right.
\]

\begin{equation}
\left.\left.+\mathbf{T}_{CC}^{\left(\beta\gamma\right)}\left(\omega\right)\mathbf{T}_{CC}^{\dagger\left(\alpha\gamma\right)}\left(\omega\right)\left(F_{\gamma\alpha}\left(\omega,\omega\right)\mathbf{T}_{CC}^{\dagger\left(\beta\alpha\right)}\left(\omega\right)-F_{\gamma\beta}\left(\omega,\omega\right)\mathbf{T}_{CC}^{\left(\alpha\beta\right)}\left(\omega\right)\right)\right]\right\} \label{eq: 7}
\end{equation}
To better understand the content of the expression (\ref{eq: 7}),
we consider the special case where $\alpha=\beta$. In this case,
use of Eq. (\ref{eq: 6}) enables us to replace the single summation
with a double sum, and we can use the identity:

\begin{equation}
F_{\alpha\gamma}\left(\omega,\omega\right)=\frac{1}{2}\left[F_{\alpha\alpha}\left(\omega,\omega\right)+F_{\gamma\gamma}\left(\omega,\omega\right)+2\left(f_{\alpha}\left(\omega-\mu\right)-f_{\gamma}\left(\omega-\mu\right)\right)^{2}\right]\label{eq:Fagam}
\end{equation}
to give

\begin{equation}
\underset{\Omega\rightarrow0}{\lim}P_{\alpha\alpha}\left(\Omega\right)=P_{\alpha\alpha}^{\left(thermal\right)}\left(\Omega\right)+P_{\alpha\alpha}^{\left(shot\right)}\left(\Omega\right)\label{eq: 9}
\end{equation}
where we identify both the \textit{generalized thermal noise}, which
vanishes when the temperature $T=0$, and the \textit{generalized
shot noise}, which vanishes when $V_{\gamma}=0$ for all $\gamma$:

\begin{equation}
\underset{\Omega\rightarrow0}{\lim}P_{\alpha\alpha}^{\left(thermal\right)}\left(\Omega\right)=2q^{2}\int\frac{d\omega}{2\pi}\underset{\gamma\neq\alpha}{\sum}\left[F_{\alpha\alpha}\left(\omega,\omega\right)+F_{\gamma\gamma}\left(\omega,\omega\right)\right]\textrm{Tr}_{C}\left[\mathbf{T}_{CC}^{\left(\alpha\gamma\right)}\left(\omega\right)\mathbf{T}_{CC}^{\dagger\left(\alpha\gamma\right)}\left(\omega\right)\right]
\end{equation}

\[
\underset{\Omega\rightarrow\infty}{\lim}P_{\alpha\alpha}^{\left(shot\right)}\left(\Omega\right)=2q^{2}\int\frac{d\omega}{2\pi}\left\{ 2\underset{\gamma\neq\alpha}{\sum}\left(f_{\alpha}\left(\omega-\mu\right)-f_{\gamma}\left(\omega-\mu\right)\right)^{2}\textrm{Tr}_{C}\left[\mathbf{T}_{CC}^{\left(\alpha\gamma\right)}\left(\omega\right)\mathbf{T}_{CC}^{\dagger\left(\alpha\gamma\right)}\left(\omega\right)\left(1-\underset{\gamma'}{\sum}\mathbf{T}_{CC}^{\left(\alpha\gamma'\right)}\left(\omega\right)\mathbf{T}_{CC}^{\dagger\left(\alpha\gamma'\right)}\left(\omega\right)\right)\right]\right.
\]

\begin{equation}
\left.+\underset{\gamma,\gamma'}{\sum}\left(f_{\gamma}\left(\omega-\mu\right)-f_{\gamma'}\left(\omega-\mu\right)\right)^{2}\textrm{Tr}_{C}\left[\mathbf{T}_{CC}^{\left(\alpha\gamma\right)}\left(\omega\right)\mathbf{T}_{CC}^{\dagger\left(\alpha\gamma\right)}\left(\omega\right)\mathbf{T}_{CC}^{\left(\alpha\gamma'\right)}\left(\omega\right)\mathbf{T}_{CC}^{\dagger\left(\alpha\gamma'\right)}\left(\omega\right)\right]\right\} \label{eq:shotgen}
\end{equation}
If we now specialize this discussion to the case of a two-lead junction,
i.e. a junction in which $\alpha$ may be one of two indices L, R,
we recover the following well-known results for the thermal and shot
noise, respectively:

\begin{equation}
\underset{\Omega\rightarrow0}{\lim}P_{LL}^{\left(thermal\right)}\left(\Omega\right)=4q^{2}\int\frac{d\omega}{2\pi}\left(\left(1-f_{L}\left(\omega-\mu\right)\right)f_{L}\left(\omega-\mu\right)+\left(1-f_{R}\left(\omega-\mu\right)\right)f_{R}\left(\omega-\mu\right)\right)\textrm{Tr}_{C}\left[\mathbf{T}_{CC}^{\left(LR\right)}\left(\omega\right)\mathbf{T}_{CC}^{\dagger\left(LR\right)}\left(\omega\right)\right]
\end{equation}

\begin{equation}
\underset{\Omega\rightarrow0}{\lim}P_{LL}^{\left(shot\right)}\left(\Omega\right)=4q^{2}\int\frac{d\omega}{2\pi}\left(f_{L}\left(\omega-\mu\right)-f_{R}\left(\omega-\mu\right)\right)^{2}\textrm{Tr}_{C}\left[\mathbf{T}_{CC}^{\left(LR\right)}\left(\omega\right)\mathbf{T}_{CC}^{\dagger\left(LR\right)}\left(\omega\right)\left(1-\mathbf{T}_{CC}^{\left(LR\right)}\left(\omega\right)\mathbf{T}_{CC}^{\dagger\left(LR\right)}\left(\omega\right)\right)\right]
\end{equation}

Finally, we note that it is common practice \cite{Yang1992,DiVentra2008}
to neglect the frequency-dependence of the transmission functions,
$\mathbf{T}_{CC}\left(\omega\right)\sim\mathbf{T}_{CC}$ in Eq. (\ref{eq: 4}),
which allows for the trivial removal of all frequency integrals. It
is then simple to show that the $LL$ component of Eq. (\ref{eq: 4})
reduces to the well-known expression:

\[
P_{LL}\left(\Omega\right)=\frac{q^{2}}{\pi}\left\{ \textrm{Tr}_{C}\left[\mathbf{T}_{CC}^{\left(LR\right)}\mathbf{T}_{CC}^{\dagger\left(LR\right)}\mathbf{T}_{CC}^{\left(LR\right)}\mathbf{T}_{CC}^{\dagger\left(LR\right)}\right]2\Omega\coth\left(\frac{\Omega}{2k_{B}T}\right)\right.
\]
\[
+\textrm{Tr}_{C}\left[\mathbf{T}_{CC}^{\left(LR\right)}\mathbf{T}_{CC}^{\dagger\left(LR\right)}\left(1-\mathbf{T}_{CC}^{\left(LR\right)}\mathbf{T}_{CC}^{\dagger\left(LR\right)}\right)\right]
\]

\begin{equation}
\times\left.\left(\left(V_{L}-V_{R}-\Omega\right)\coth\left(\frac{V_{L}-V_{R}-\Omega}{2k_{B}T}\right)+\left(V_{L}-V_{R}+\Omega\right)\coth\left(\frac{V_{L}-V_{R}+\Omega}{2k_{B}T}\right)\right)\right\} \label{eq:SRYangconfirm}
\end{equation}
This formula expresses the interplay of the shot noise, Nyquist noise
and quantum vacuum fluctuations in a conductor, and moreover has been
verified experimentally for a wide range of mesoscale conductors \cite{Schoelkopf1997,Zakka-Bajjani2007}.

\section{Numerics\label{sec:Numerics}}

\subsection{Calculations on a Single Site}

In Appendix \ref{sec:Appendix-D} we present an efficient technique
for evaluating each term in (\ref{eq:corrcompact}) based on the analytical
removal of all frequency integrals in these expressions, as was done
for the current in Ref. \cite{Ridley2016}. Other schemes in the literature
perform the frequency integrals in the transient noise numerically
\cite{Feng2008,Yang2014}, so we acquire a significant computational
speed-up in comparison to those works, as well as access to the noise
response to an explicit time-dependent driving. Many cases of interest
can be studied by inserting into these formulas the following \textit{biharmonic}
bias, consisting of a constant shift $V_{\alpha}$ and two harmonic
modes: 
\begin{equation}
V_{\alpha}\left(t\right)=V_{\alpha}+A_{\alpha}^{\left(1\right)}\cos\left(p_{1}\Omega_{\alpha}\left(t-t_{0}\right)+\phi_{\alpha}\right)+A_{\alpha}^{\left(2\right)}\cos\left(p_{2}\Omega_{\alpha}\left(t-t_{0}\right)\right)\label{eq:biasdouble}
\end{equation}
Here $p_{1}$, $p_{2}$ are any integers and $\phi_{\alpha}$ is a
lead-dependent phase shift that breaks the dynamical symmetry of the
system under time reversal (TR), $t\rightarrow2t_{0}-t$. According
to a well-known Bessel function identity, this choice of bias leads
to the following representation of the exponential phase factor appearing
in Eq. (\ref{eq:ggreatlessexpanded-1}):

\begin{equation}
e^{i\psi_{\alpha}\left(t_{1},t_{2}\right)}=e^{iV_{\alpha}\left(t_{1}-t_{2}\right)}\underset{r,r',s,s'}{\sum}J_{r}\left(\frac{A_{\alpha}^{\left(1\right)}}{p_{1}\Omega_{\alpha}}\right)J_{r'}\left(\frac{A_{\alpha}^{\left(1\right)}}{p_{1}\Omega_{\alpha}}\right)J_{s}\left(\frac{A_{\alpha}^{\left(2\right)}}{p_{2}\Omega_{\alpha}}\right)J_{s'}\left(\frac{A_{\alpha}^{\left(2\right)}}{p_{2}\Omega_{\alpha}}\right)e^{i\left(r-r'\right)\phi_{\alpha}}e^{i\Omega_{\alpha}\left(p_{1}r+p_{2}s\right)\left(t_{1}-t_{0}\right)}e^{-i\Omega_{\alpha}\left(p_{1}r'+p_{2}s'\right)\left(t_{2}-t_{0}\right)}\label{eq:BESSELphase}
\end{equation}
In Appendix \ref{sec:FormulasBiharmonic}, we include explicit formulas
for $\mathbf{G}^{\lessgtr}\left(t_{1},t_{2}\right)$ and the $\mathbf{\Lambda}^{\pm}$
matrices within this biharmonic model, with all time integrals explicitly
removed by hand. In calculations presented here, we will only consider
the case of a single harmonic in order to reflect numerical work carried
out elsewhere \cite{Ridley2015,Ridley}. However, the equations presented
in this work will be valid for the general biharmonic case studied
experimentally in Ref. \cite{Gabelli2013}. 

\begin{figure}
\subfloat[]{\includegraphics[scale=0.5]{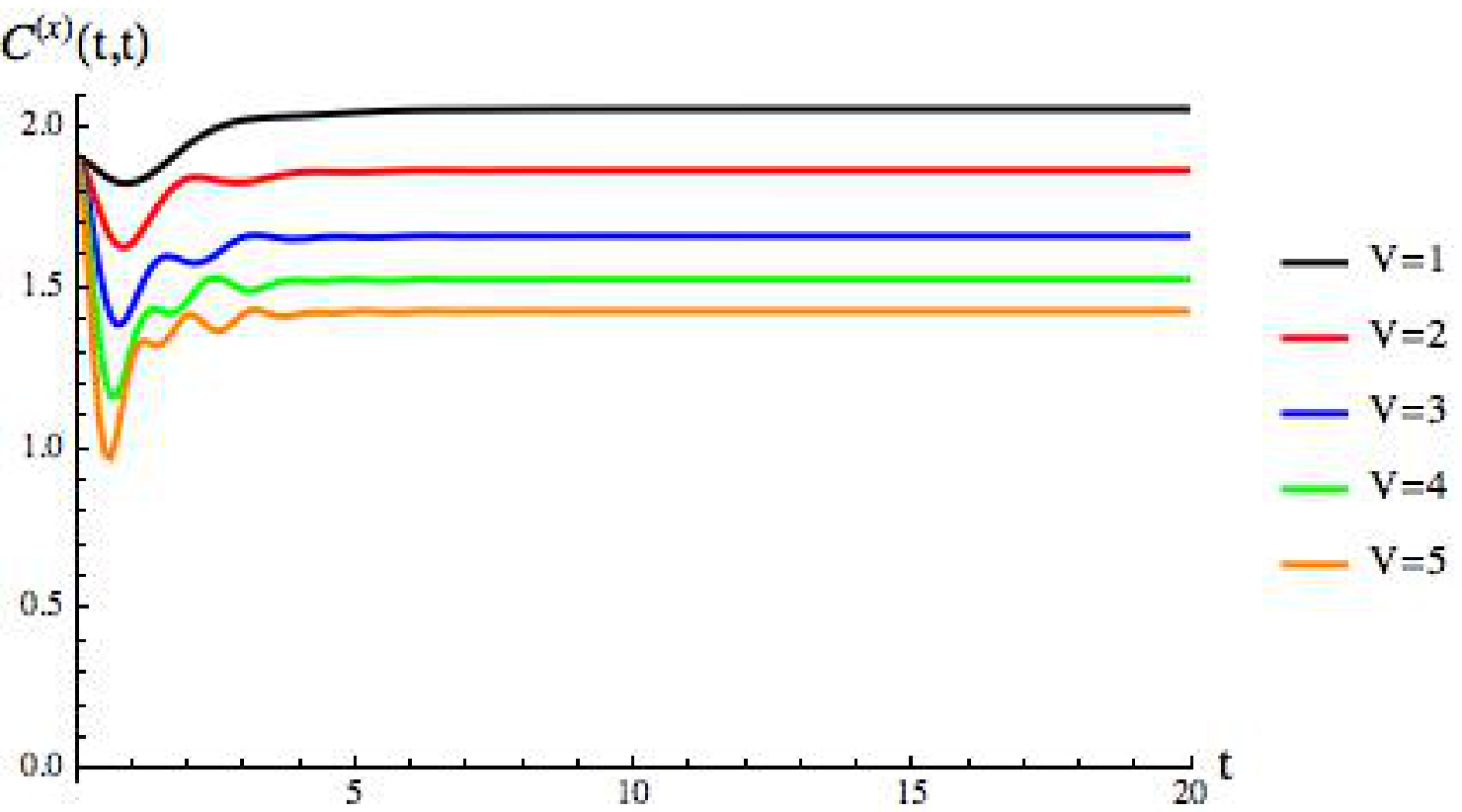}}\subfloat[]{\includegraphics[scale=0.5]{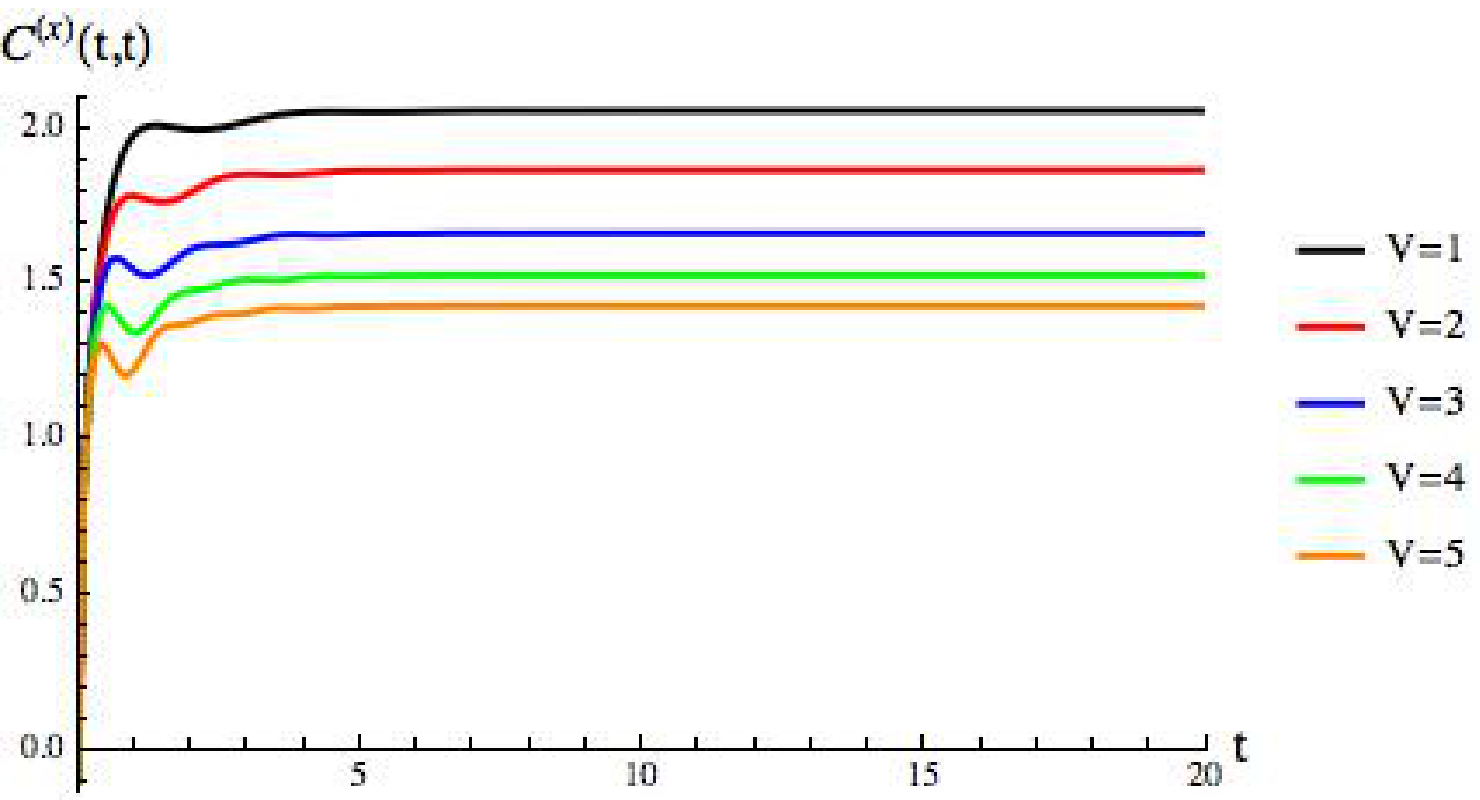}}\caption{\label{fig:dotcalculations}Cross correlation $C^{\left(\times\right)}\left(t,t\right)$
for different values of $V=V_{L}=-V_{R}$, arising from (a) the partition-free
($\zeta_{p}=1$) and (b) the partitioned ($\zeta_{p}=0$) switch-on
processes.}
\end{figure}

As a point of reference for the kind of physics to expect from this
formalism, we first perform some calculations for the simple case
of a single level quantum dot coupled symmetrically to left and right
leads. In Fig. \ref{fig:dotcalculations} the equal-time average cross
correlation function $C^{\left(\times\right)}\left(t,t\right)$ is
plotted as a function of the observation time $t$ for the quantum
dot ($\mathbf{h}_{CC}=\varepsilon_{0}$), where the level-width matrices
are scalars given by $\Gamma_{L}=\Gamma/2=\Gamma_{R}$. In Figs. \ref{fig:dotcalculations}
(a) and (b) the bias applied to the leads $V_{\alpha}\left(t\right)=V_{\alpha}$
is a constant, and we choose $V_{L}=V=-V_{R}$. In Fig. \ref{fig:dotcalculations}
(a) the cross-correlation response to a partition-free switch-on is
shown, for different values of $V$. This is compared with a switch-on
in the partitioned process, i.e. with $\zeta_{p}=0$, shown in Fig.
\ref{fig:dotcalculations} (b). In all cases, the two approaches give
the same steady-state value, satisfying the Memory-loss Theorem of
Ref. \cite{Stefanucci2004}. However, the transient correlations differ
significantly due to the system preparation. For the partition-free
switch-on process, the subsystems of the lead-molecule-lead nanojunction
have already equilibrated prior to the switch-on time $t_{0}$ (the
zero time in the Figures), so one would expect the initial cross-correlations
between the signal response in the $L$ and $R$ leads to be stronger
in this case, as this initial transient corresponds to the time delay
between switching on the bias and establishing the correlations. This
situation is indeed the case in Fig. \ref{fig:dotcalculations}, where
partitioning the system prior to the switch-on is found to strongly
reduce the cross-correlations in the transient regime of $t\in\left[0,2/\Gamma\right]$.
We also observe more `ringing' oscillations in the partition-free
transient with a higher frequency compared to the partitioned case;
this is due to the additional transient modes that are included in
the partition-free switch-on. 

\begin{figure}
\subfloat[]{\includegraphics[scale=0.5]{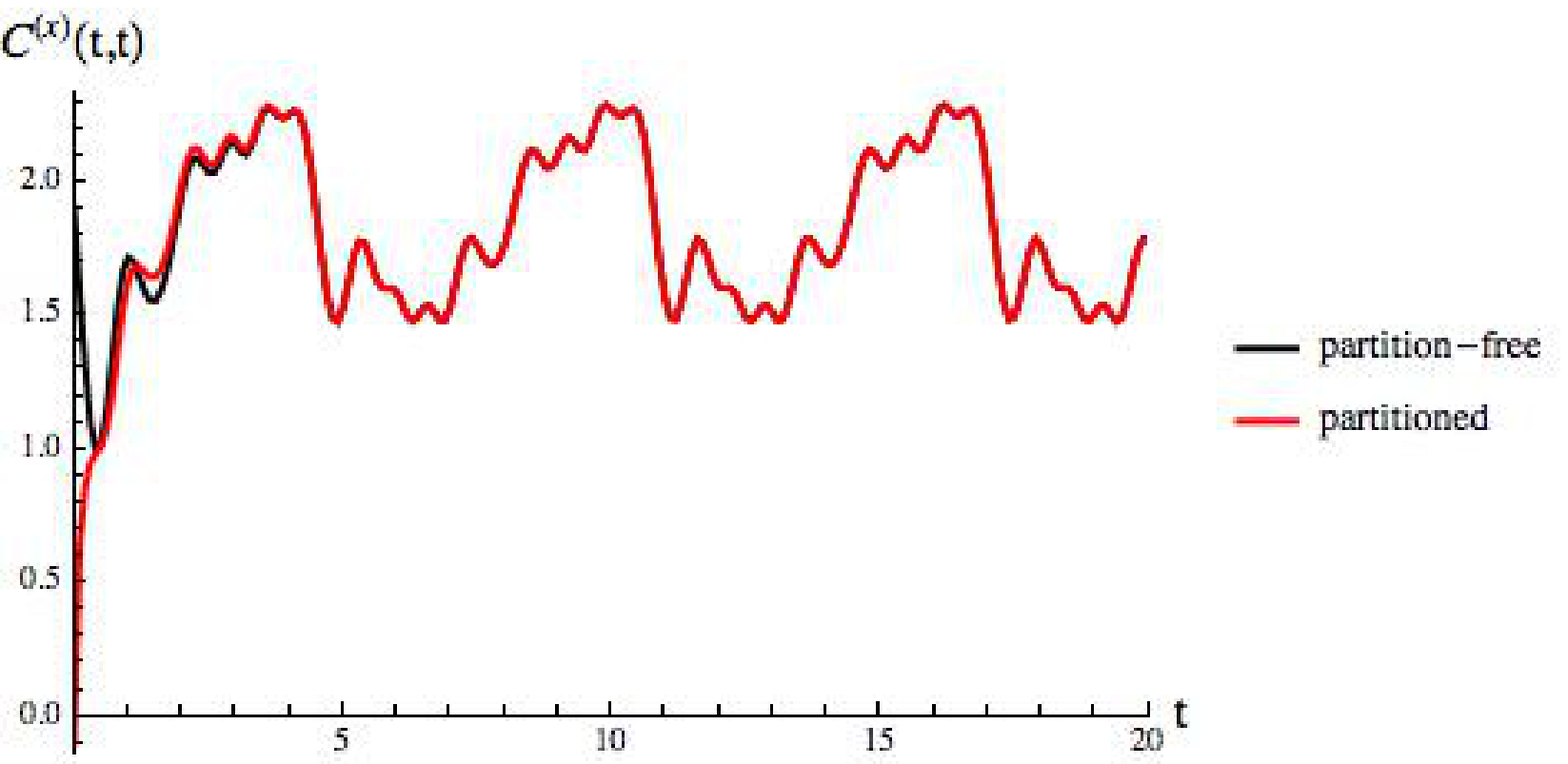}

}\subfloat[]{\includegraphics[scale=0.5]{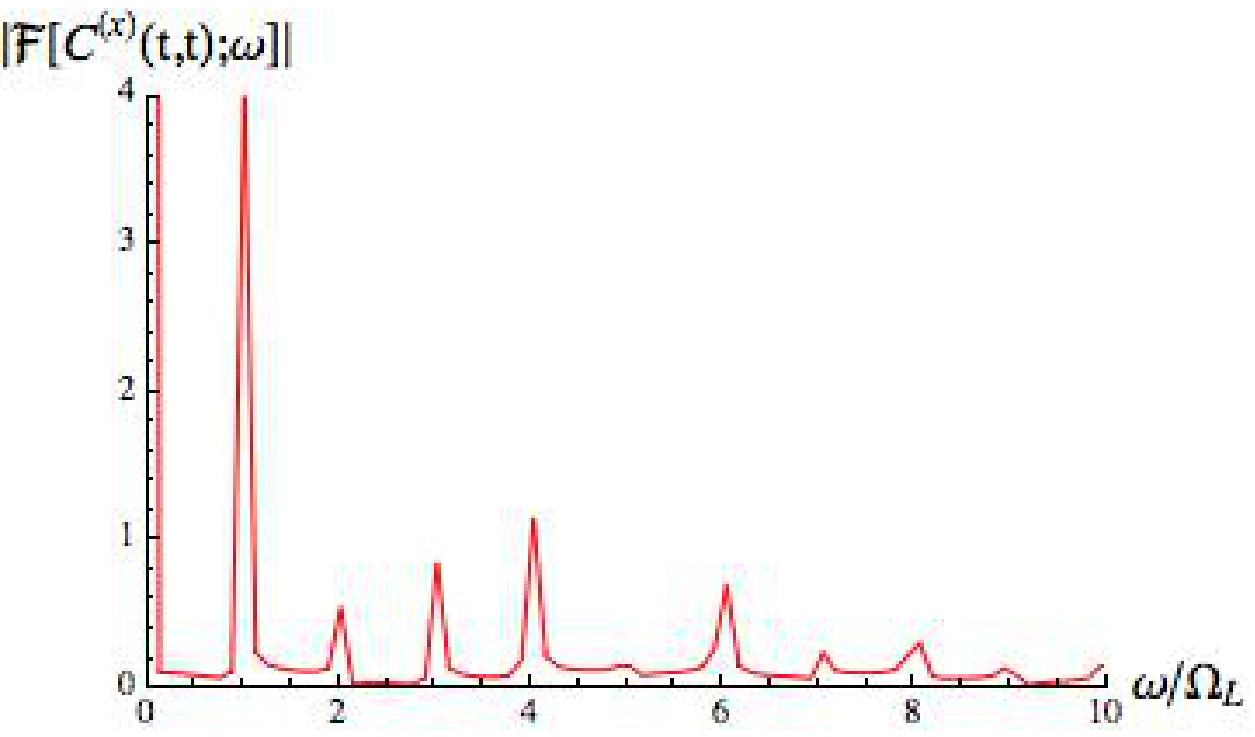}}

\caption{\label{fig:CLRdotsin} (a) The average cross-correlation $C^{\left(\times\right)}\left(t,t\right)$
between the $L$ and $R$ leads is plotted in the partition-free (black
line) and partitioned (red line) switch-on processes with system parameters
$\varepsilon_{0}=1,\,\Gamma=1,\, k_{B}T=0.1,\,\mu=0,\, V_{L}=5,\, V_{R}=0,\, A_{L}^{\left(1\right)}=4,\, A_{R}^{\left(1\right)}=A_{L}^{\left(2\right)}=A_{R}^{\left(2\right)}=0,\,\Omega_{L}=1$.
(b) Numerical Fourier Transform of the partition-free cross-correlation
shown in (a).}
\end{figure}

Next, we study the same driven single-site system that was considered
in calculations of the left lead current $I_{L}\left(t\right)$ in
Ref. \cite{Ridley2015}, i.e. we set $V_{L}=5$, $V_{R}=0$, $A_{L}^{\left(1\right)}=4,\, A_{R}^{\left(1\right)}=A_{L}^{\left(2\right)}=A_{R}^{\left(2\right)}=0$,
$\Omega_{L}=1$, with all other system parameters unchanged from the
static bias calculations of Fig. \ref{fig:dotcalculations}. Fig.
\ref{fig:CLRdotsin} (a) shows the time-dependence of the cross-correlation
for both the partition-free and partitioned switch-on processes. Once
again, the initial cross-correlations are stronger in the partition-free
approach, and both approaches relax to the same signal repeating with
a period of $2\pi/\Omega_{L}$. In Fig. \ref{fig:CLRdotsin} (b) we
plot the absolute value of the discrete Fourier Transform of the partition-free
$C^{\left(\times\right)}\left(t,t\right)$ curve in units of $\Omega_{L}$.
In addition to the dominant peak at the fundamental driving frequency
$\omega=\Omega_{L}$, the frequency spectrum of cross-correlations
exhibits smaller peaks at integer multiples $\omega=n\Omega_{L}$
of this number, which is a signature of PAT \cite{Dayem1962}. Electrons
which have been excited by higher multiples of $\Omega_{L}$ cause
the short-time `ringing' oscillations occurring on top of the signal
with the fundamental frequency in Fig. \ref{fig:CLRdotsin} (a), and
the difference in resonance heights in Fig. \ref{fig:CLRdotsin} (b)
is due to multiphoton processes that interfere with each other \cite{Wang2003}.

\begin{figure}
\subfloat[]{\includegraphics[scale=0.5]{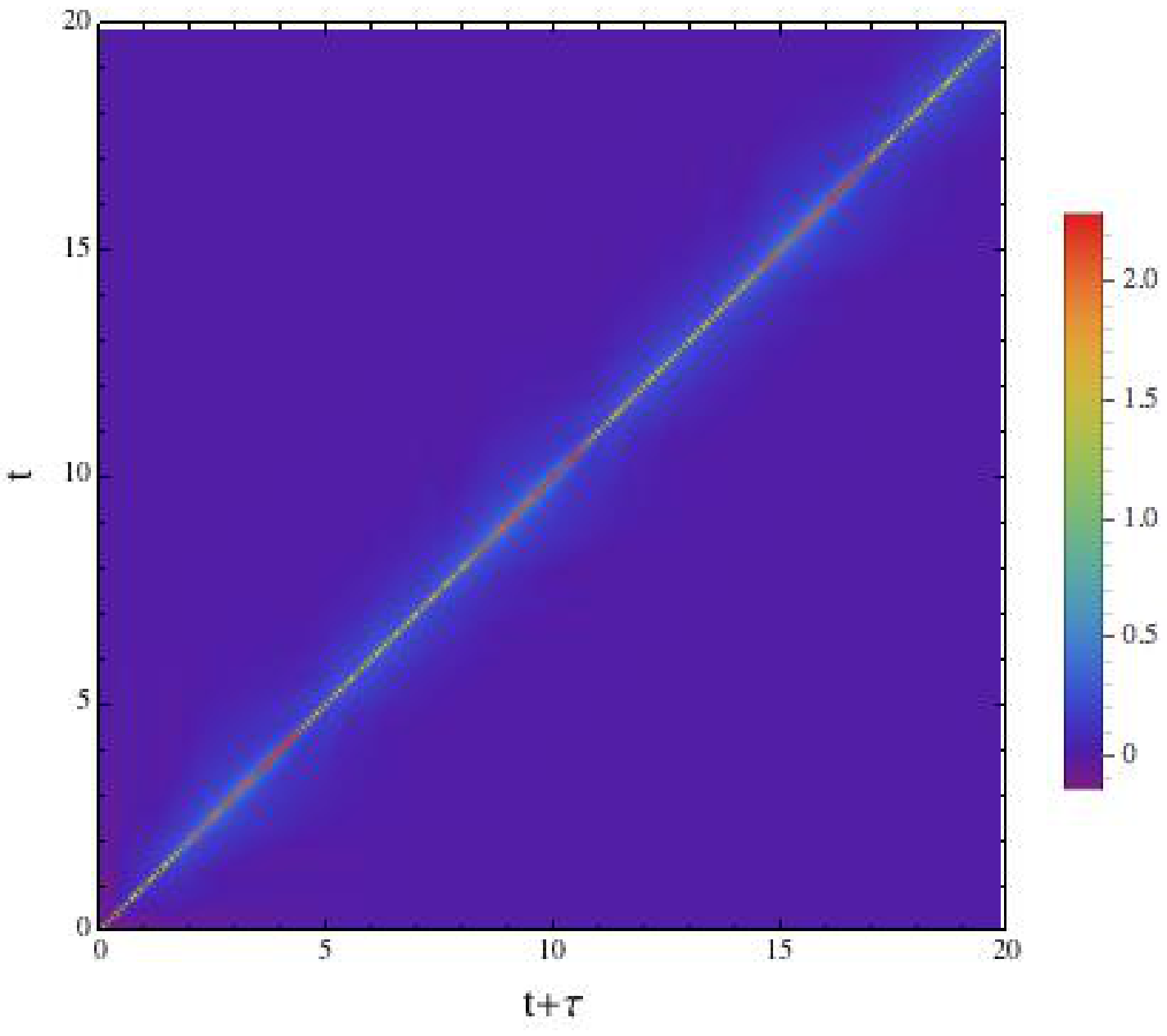}}\subfloat[]{\includegraphics[scale=0.5]{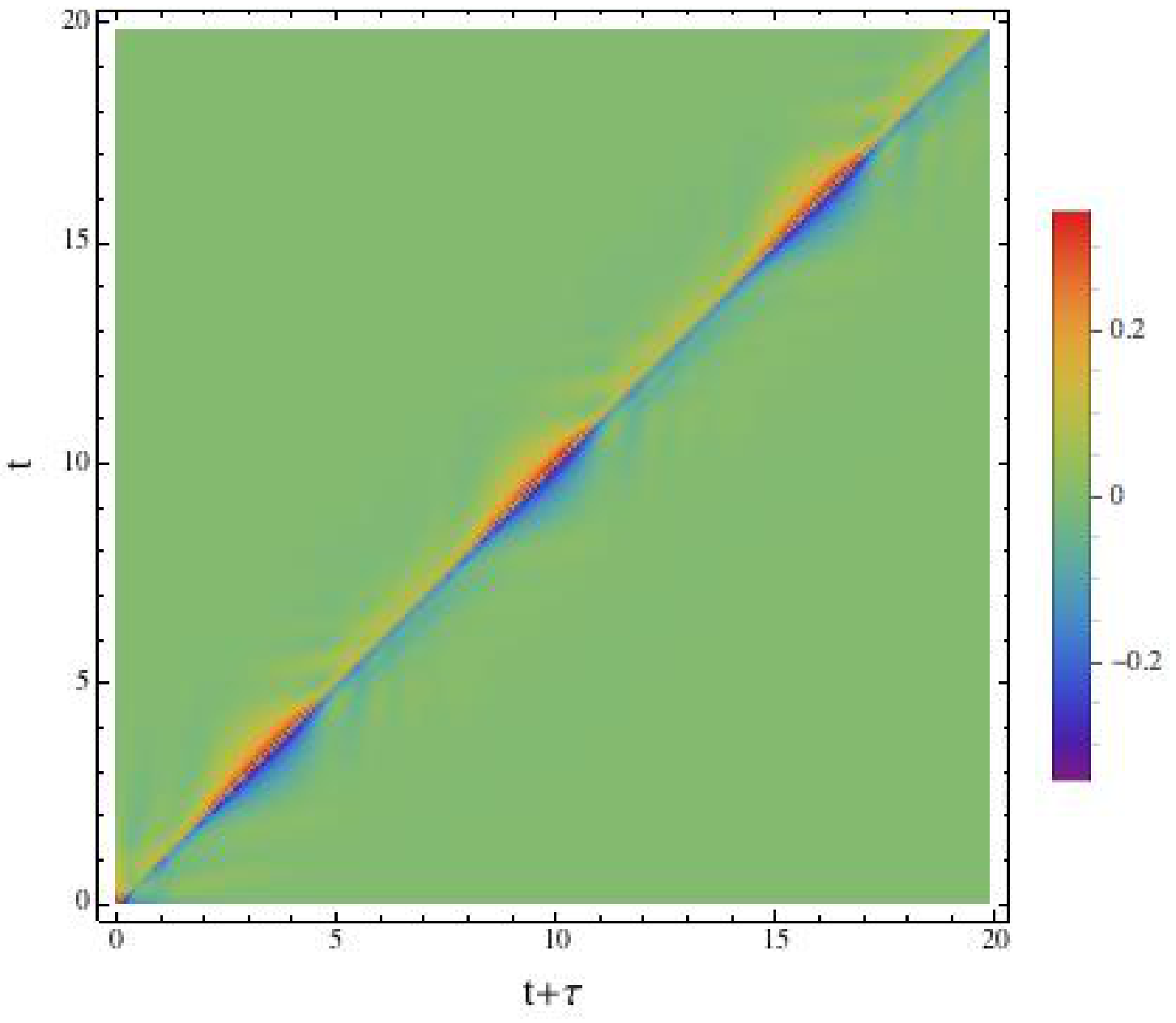}

}\caption{\label{fig:CLRdotsincontour}For the same driven system considered
in Fig. \ref{fig:CLRdotsin}, with the dot level set at $\varepsilon_{0}=1$,
we plot (a) $\textrm{Re}\left[C^{\left(\times\right)}\left(t+\tau,t\right)\right]$,
and (b) $\textrm{Im}\left[C^{\left(\times\right)}\left(t+\tau,t\right)\right]$.
In these plots we set $\zeta_{p}=1$, i.e. we take the approach to
be partition-free.}
\end{figure}

In Fig. \ref{fig:CLRdotsincontour} we present contour plots of the
real and imaginary parts of the full two-time cross correlation function
$C^{\left(\times\right)}\left(t+\tau,t\right)$ for the same single
level system as was considered in Fig. \ref{fig:CLRdotsin} for the
partition-free case. Fig. \ref{fig:CLRdotsincontour} (a) clearly
demonstrates the oscillating value of $C^{\left(\times\right)}\left(t,t\right)$
along the $\tau=0$ diagonal, and also numerically demonstrates the
reflection symmetry of $\textrm{Re}\left[C^{\left(\times\right)}\left(t+\tau,t\right)\right]$
in this line, a fact which was proven analytically and stated in Eq.
(\ref{eq:Csymreal1}). In Fig. \ref{fig:CLRdotsincontour} (b), we
notice that strong oscillations are also seen in the value of $\textrm{Im}\left[C^{\left(\times\right)}\left(t+\tau,t\right)\right]$
with a period of $2\pi/\Omega_{L}$. Although this component of the
cross-correlation is zero along the main diagonal, it satisfies the
reflection antisymmetry property $\textrm{Im}\left[C^{\left(\times\right)}\left(t+\tau,t\right)\right]=-\textrm{Im}\left[C^{\left(\times\right)}\left(t,t+\tau\right)\right]$,
as stated in Eq. (\ref{eq:Csymimag1}).

\subsection{Application to the Molecular Wire}

\begin{figure}
\centering{}\subfloat[]{\begin{centering}
\includegraphics[scale=0.3]{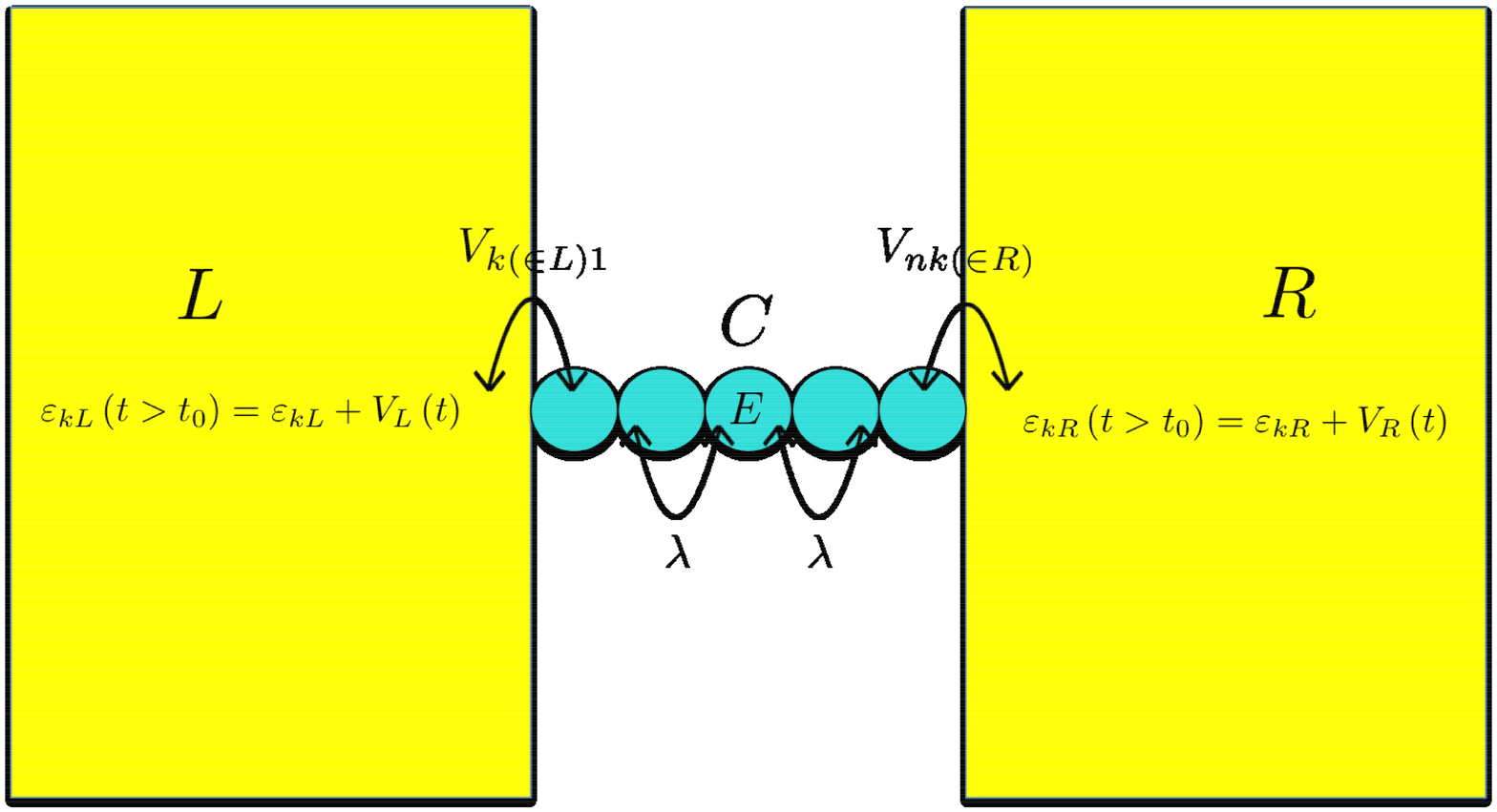}
\par\end{centering}

}\subfloat[]{\includegraphics[scale=0.5]{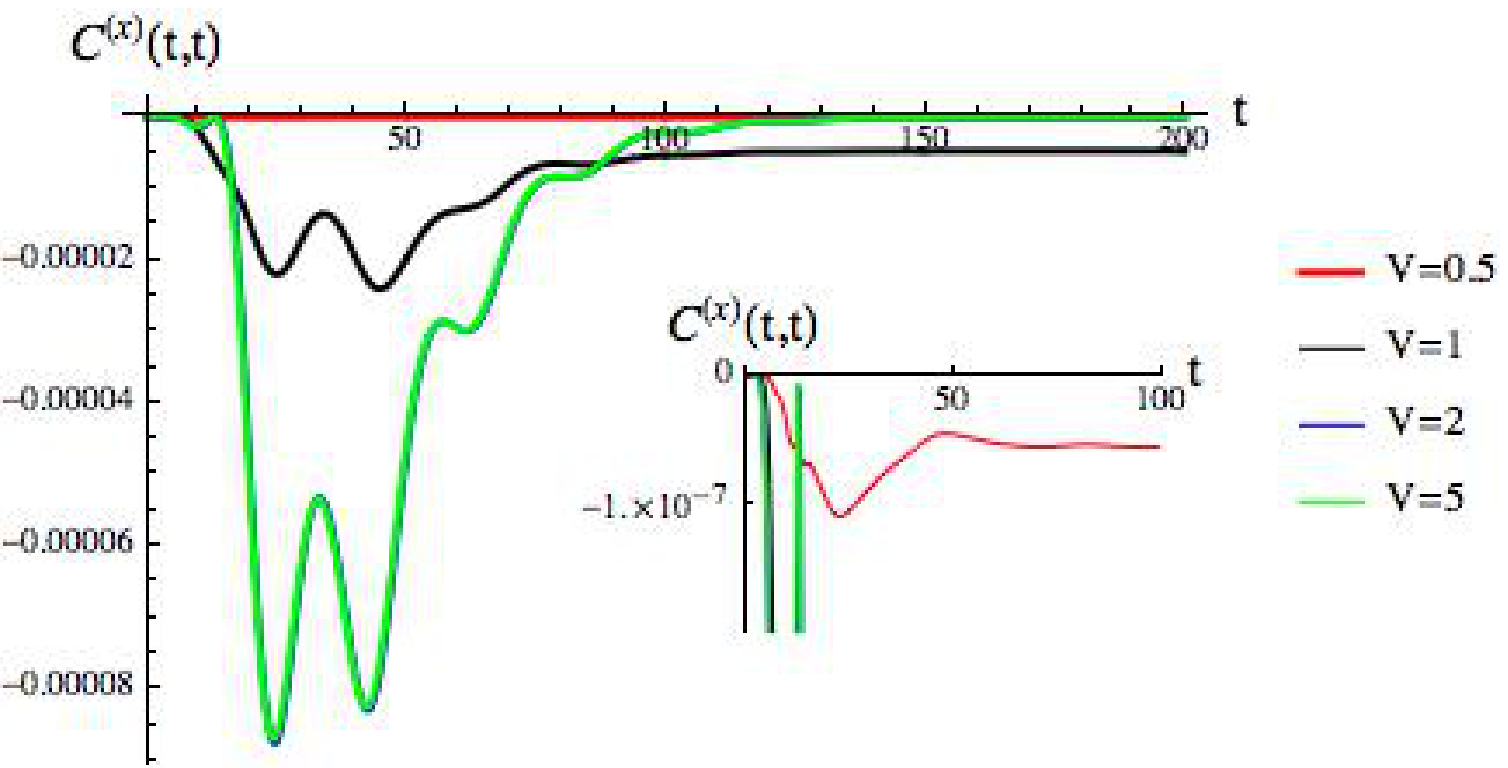}}\caption{\label{fig:n=00003D5Chain}(a) Schematic of a typical two-lead molecular
junction consisting of the left ($L$) and right ($R$) leads bridged
by a molecular system $C$, which in this case is chosen to be a molecular
wire composed of $N_{s}=5$ atomic sites with nearest neighbour hopping.
(b) Plot of $C^{\left(\times\right)}\left(t,t\right)$ for the switch-on
of a constant bias $V=V_{L}=-V_{R}$, for the parameter choice $E=1$,
$\Gamma=0.5$, $\lambda=0.1$.}
\end{figure}

Now we shall apply our formalism to the transport properties of the
molecular wire, using the tight-binding model of a one-dimensional
wire with nearest neighbor hopping from Refs. \cite{Mujica1994,Mujica1994a}.
\textcolor{black}{We previously studied the current response in this
system for sinusoidal \cite{Ridley} and stochastic \cite{Ridley2016}
biases in the leads. We assume that each site corresponds to a single
energy level, which may have a maximum occupation of $2$ due to spin
degeneracy, so that our model is equivalent to a wire of coupled quantum
dots \cite{Lehmann2002,Chen2013}. The Hamiltonian describing this
molecular wire is assumed to have on-site energies equal to the constant
value $\left[\mathbf{h}_{CC}\right]_{k,k}\equiv E$ and hopping elements
all given by $\left[\mathbf{h}_{CC}\right]_{k,k+1}=\left[\mathbf{h}_{CC}\right]_{k+1,k}\equiv\lambda$.
All quantities will be given in arbitrary units, and we choose the
chemical potential $\mu=0$ as the zero of energy. We model the perpendicular
orientation of the wire between the leads. In Fig. \ref{fig:n=00003D5Chain}
(a), we illustrate the configuration in which only the end sites of
the wire are coupled to their neighboring lead, in which case the
only non-zero elements of the level width are $\Gamma_{L,11}$ and
$\Gamma_{R,N_{s}N_{s}}$, where $N_{s}$ is the number of molecular
sites. For simplicity, we will assume that sites are symmetrically
coupled to the left and right leads, $\Gamma_{L,11}=\Gamma/2=\Gamma_{R,N_{s}N_{s}}$.
We will now study the response of the cross-correlation in this system
to the switch-on of both DC and AC biases.}

\subsubsection{Time-dependent response to a DC bias}

In Fig \ref{fig:n=00003D5Chain} (b), we plot the cross-correlation
$C^{\left(\times\right)}\left(t,t\right)$ through a molecular wire
for for different values of the \emph{static} ($A_{\alpha}^{\left(1\right)}=0=A_{\alpha}^{\left(2\right)}$)
bias $V=V_{L}=-V_{R}$, with $\Gamma=0.5$, on-site energy $E=1$
and the hopping parameter is set to $\lambda=0.1$. In Fig. \ref{fig:n=00003D5Chain}
(b), we observe the occurrence of a `kick' in the cross-correlation
signal beginning at a resonance time of about $t_{res}\simeq20$,
before the signal decays towards zero. This resonance is extremely
small (shown in the inset to Fig. \ref{fig:n=00003D5Chain} (b)) when
$V<E$, as in this case the on-site energy of the chain lies outside
the bias window $\left[-V,V\right]$. The magnitude of the resonance
sharply increases when $E$ crosses into the bias window at $V=1$,
before saturating at a maximum value at around $V=2$, which can be
seen from the fact that the $V=5$ (green) curve sits almost exactly
on top of the $V=2$ (blue) one. This resonance is transient, we associate
it with the relaxation time taken for the system to reach its steady
state.

\begin{figure}
\subfloat[$N_{s}=3$]{\includegraphics[scale=0.5]{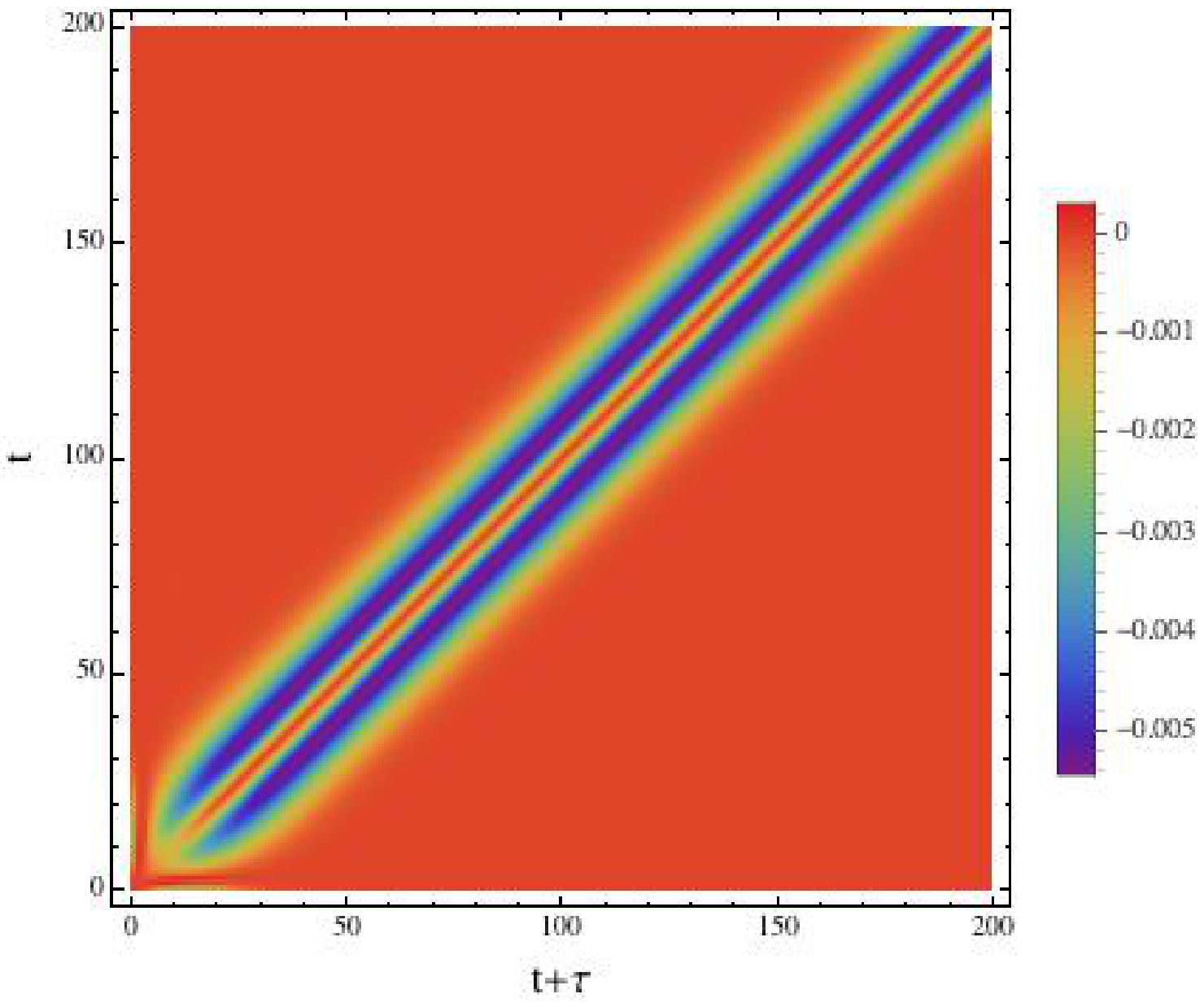}}\subfloat[$N_{s}=4$]{\includegraphics[scale=0.5]{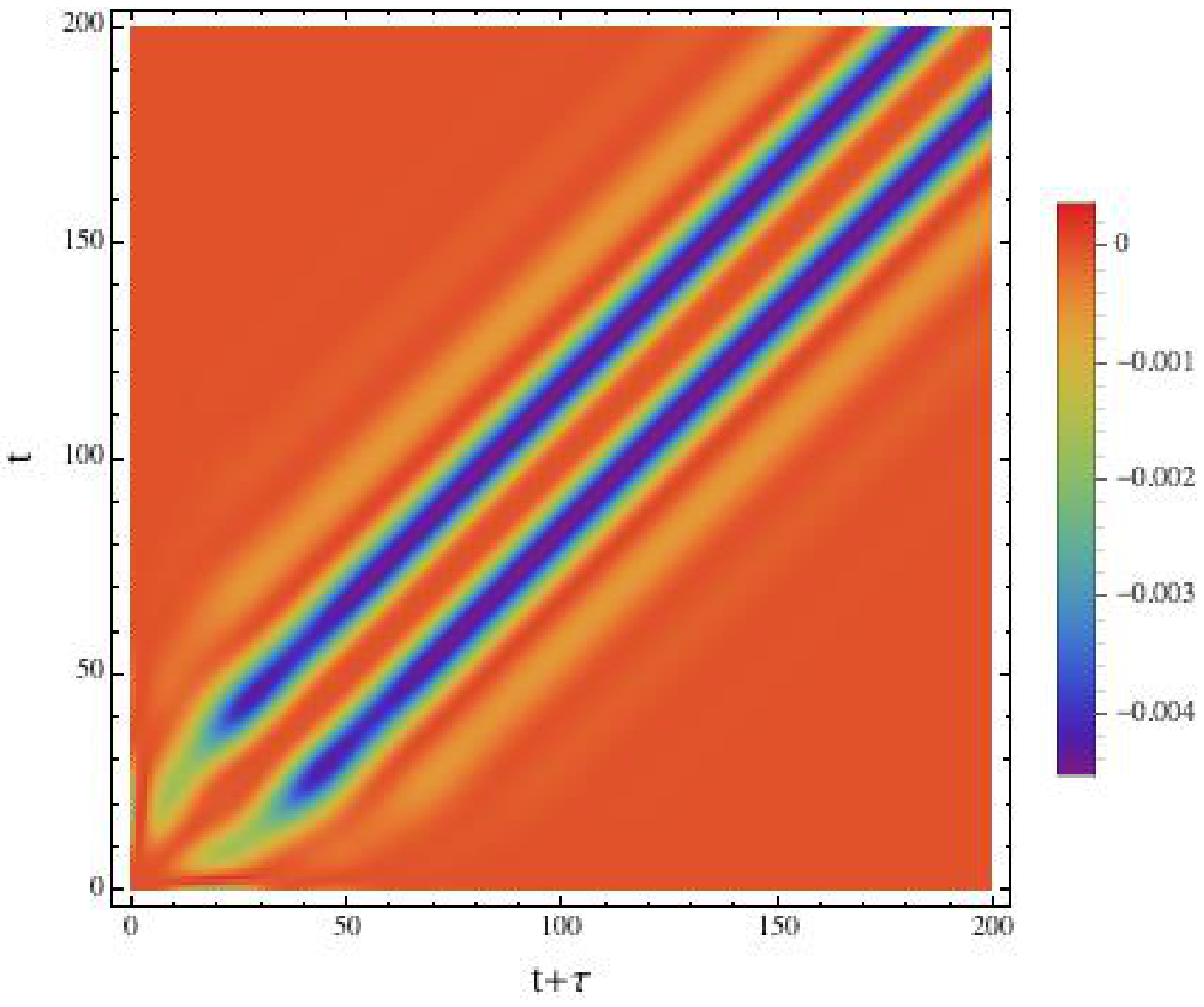}}

\subfloat[$N_{s}=5$]{\includegraphics[scale=0.5]{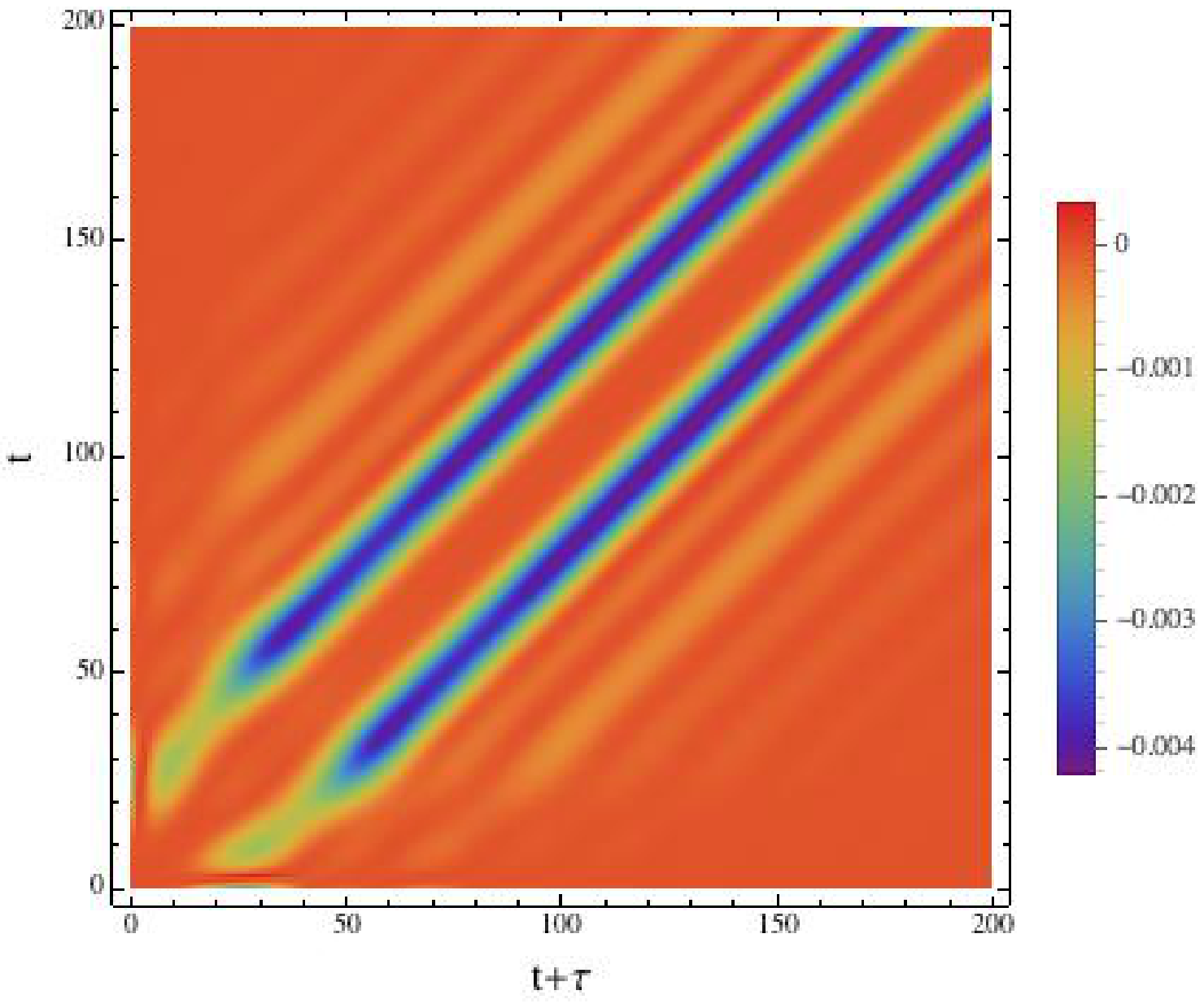}}\subfloat[$N_{s}=6$]{\includegraphics[scale=0.5]{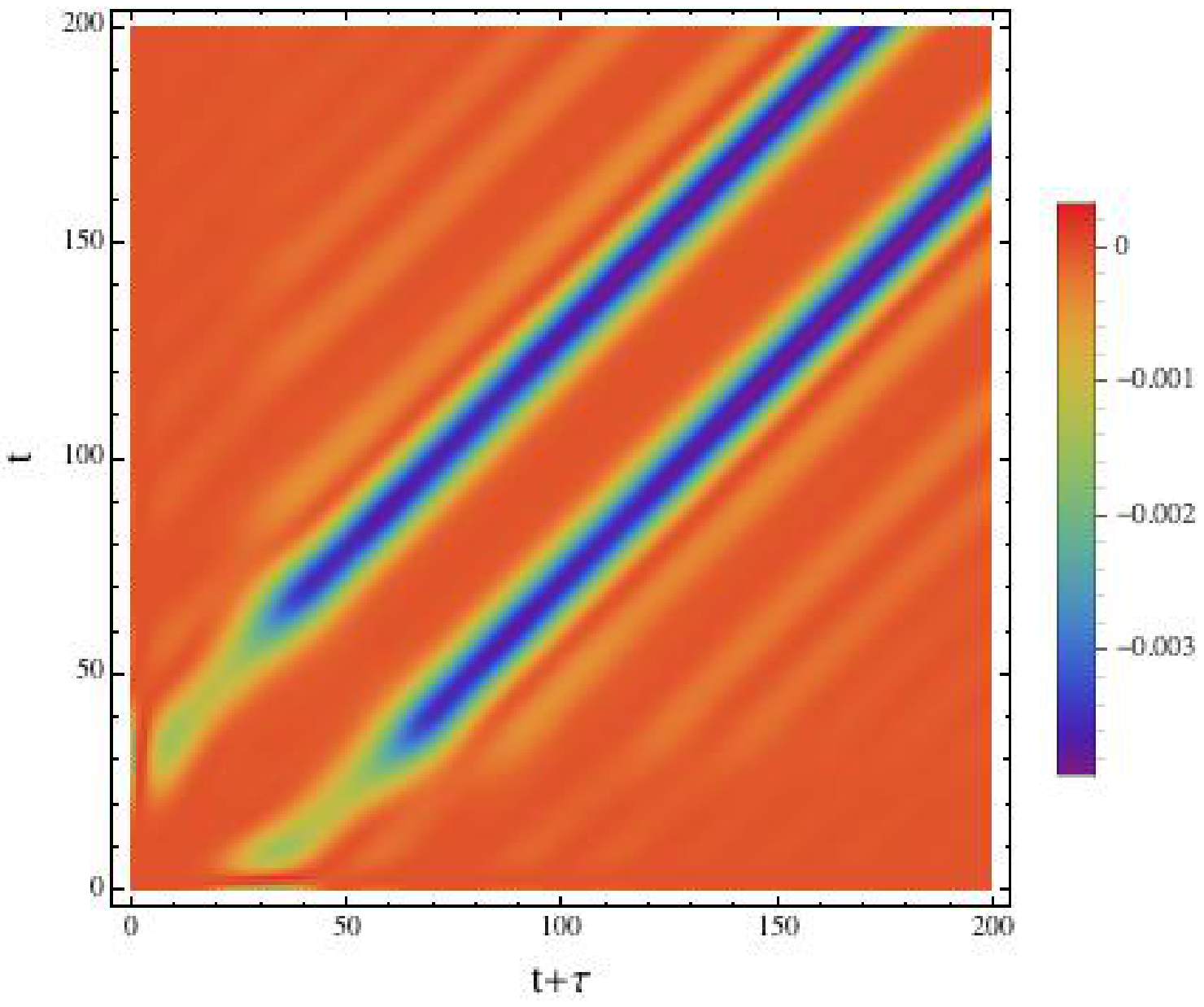}}\caption{\label{fig:wirebasecalculations} Plots of $\textrm{Re}\left[C^{\left(\times\right)}\left(t+\tau,t\right)\right]$
for $V=V_{L}=-V_{R}=5$, $\lambda=0.1$, $E=1$, $\Gamma=0.5$ where
the number of sites is varied so that (a) $N_{s}=3$, (b) $N_{s}=4$,
(c) $N_{s}=5$ and (d) $N_{s}=6$. }
\end{figure}

In Fig \ref{fig:wirebasecalculations} we exhibit the contour plot
of $\textrm{Re}\left[C^{\left(\times\right)}\left(t+\tau,t\right)\right]$
for the $V=5$ case, and for different numbers of atomic sites $N_{s}=3,4,5,6$.
Unlike the single-site case in Fig. \ref{fig:CLRdotsincontour}, and
for instance the contour plots in Ref. \cite{Yang2014}, the magnitude
of the cross-correlation is \textit{not} in general maximized along
the $\tau=0$ diagonal. Instead, we see a very strong `ripple' spreading
out from the diagonal for all values of $N_{s}$ with a maximum magnitude
at a value of $\tau_{max}=t_{1}-t_{2}$ \textcolor{black}{satisfying
the relation $\max\left|\textrm{Re}\left[C^{\left(\times\right)}\left(t+\tau,t\right)\right]\right|=\textrm{Re}\left[C^{\left(\times\right)}\left(t\pm\tau_{max},t\right)\right]$.
It appears }from the plots \ref{fig:wirebasecalculations} (a)-(d)
that $\tau_{max}$ increases linearly wi\textcolor{black}{th increasing
$N_{s}$. In the $N_{s}=5$ case $\tau_{max}\simeq20$, i.e. it is
roughly equal to the resonance time $t_{res}$ in $C^{\left(\times\right)}\left(t,t\right)$,
so we expect that that the two time scales $\tau_{max}$ and $t_{res}$
may }be physically related. 

The fact that $\tau_{max}$ increases with $N_{s}$ implies that its
position is due to the finite size of the molecular wire and its intrinsic
properties. To understand this heuristically, one may consider the
Schr{\"o}dinger equation for a wire of $N_{s}$ sites with a spacing of
size $1$ (i.e. of length $N_{s}-1$), energy $E$ and inter-site
coupling $\lambda$ . This leads to a dispersion $\varepsilon\left(k\right)=E+2\lambda\cos\left(k\right)$,
and therefore the traversal time for an electron of unit mass to pass
through the wire is approximated by (note that $\hbar=1$): 
\begin{equation}
\tau_{traversal}\approx\frac{N_{s}-1}{\partial_{k}\epsilon(k)}=\frac{N_{s}-1}{2\lambda\sin\left(k\right)}\label{eq:transittime}
\end{equation}
Whereas this expression neglects the presence of the leads and cannot
be taken as anything other than a rule of thumb, it indicates that
we may investigate the interplay of $\lambda$ and $N_{s}$ should
we wish to understand the effects on the dynamics of a finite system
size. 

\begin{figure}
\subfloat[]{\includegraphics[scale=0.5]{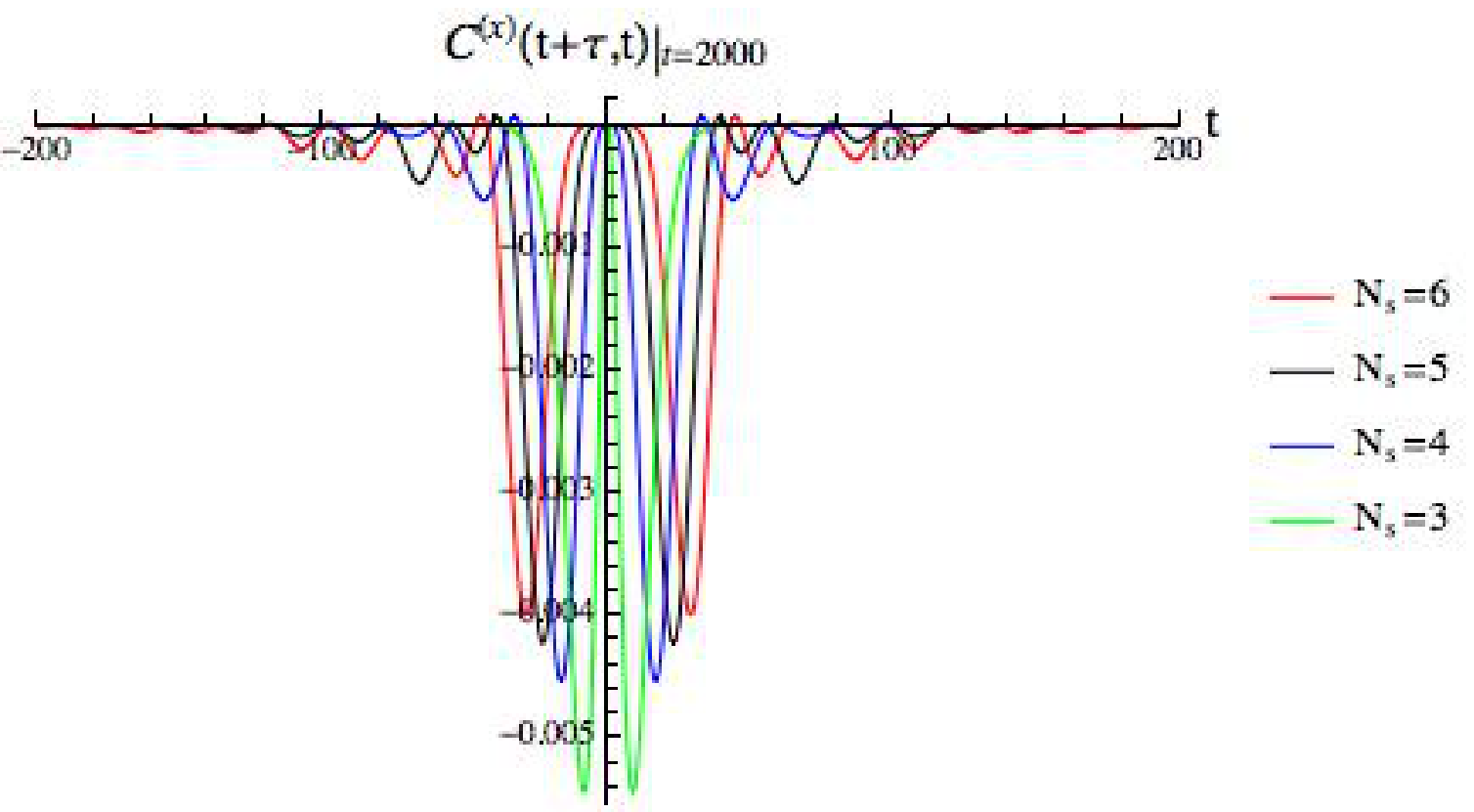}}\subfloat[]{\includegraphics[scale=0.5]{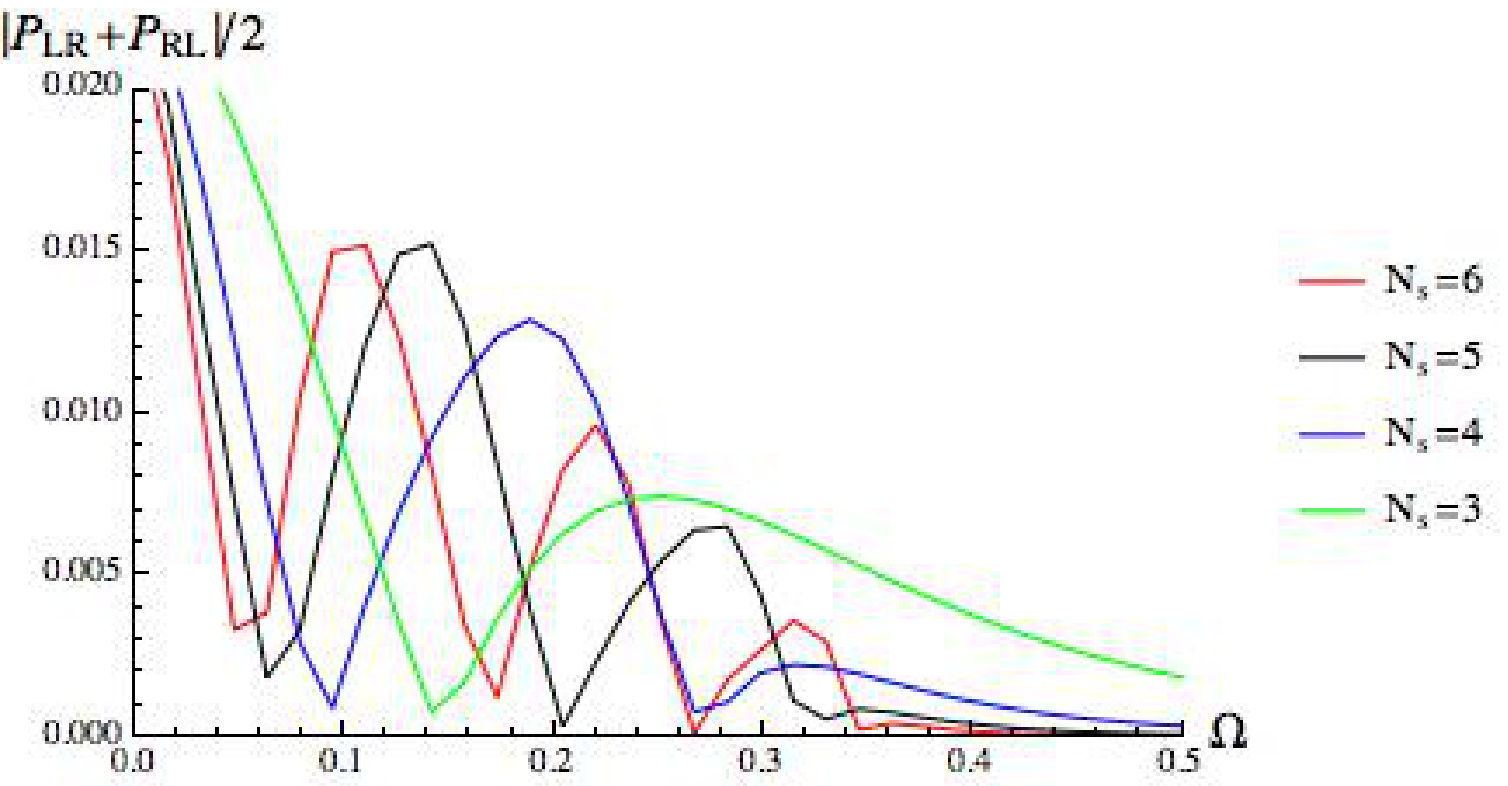}

}

\subfloat[]{\includegraphics[scale=0.5]{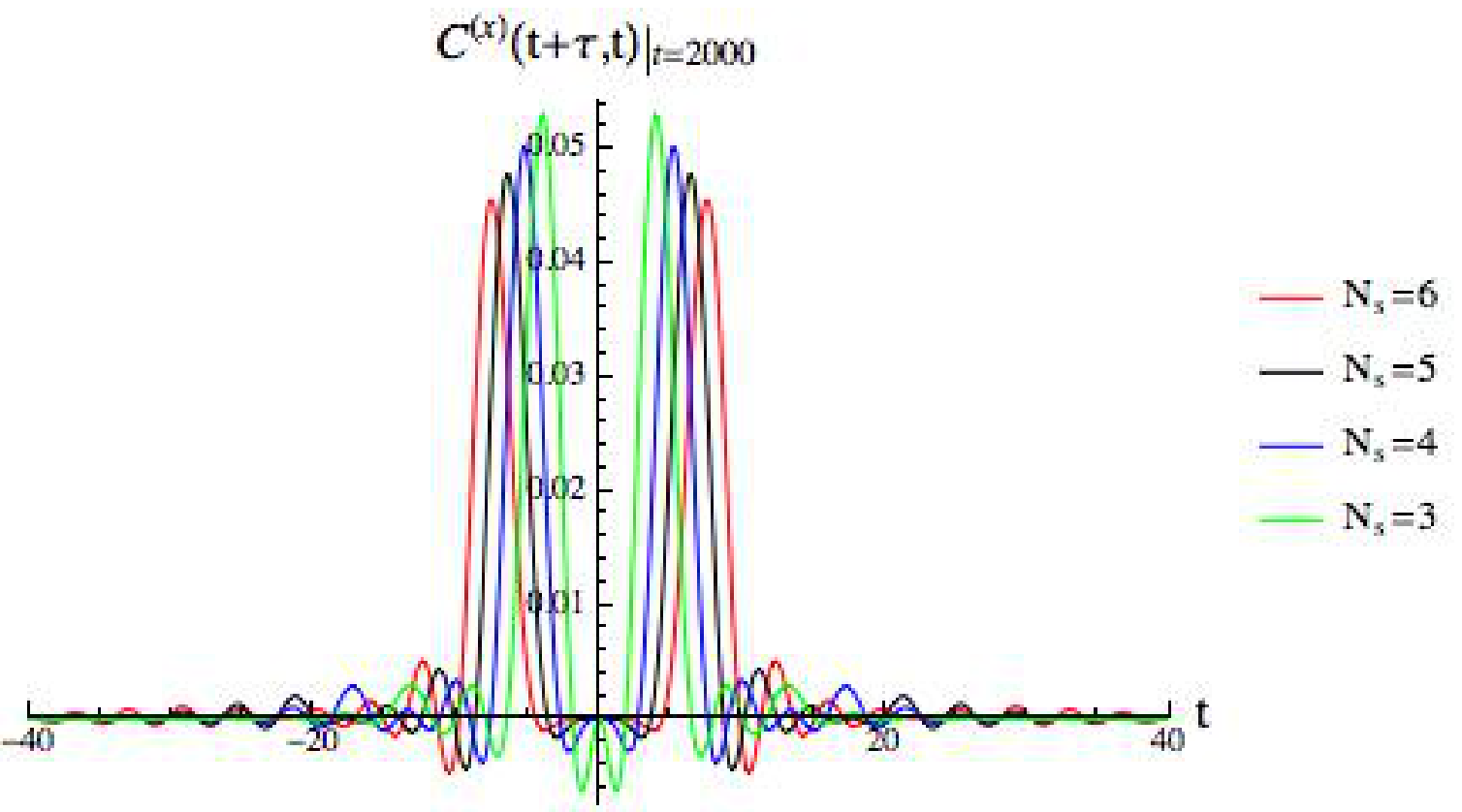}}\subfloat[]{\includegraphics[scale=0.5]{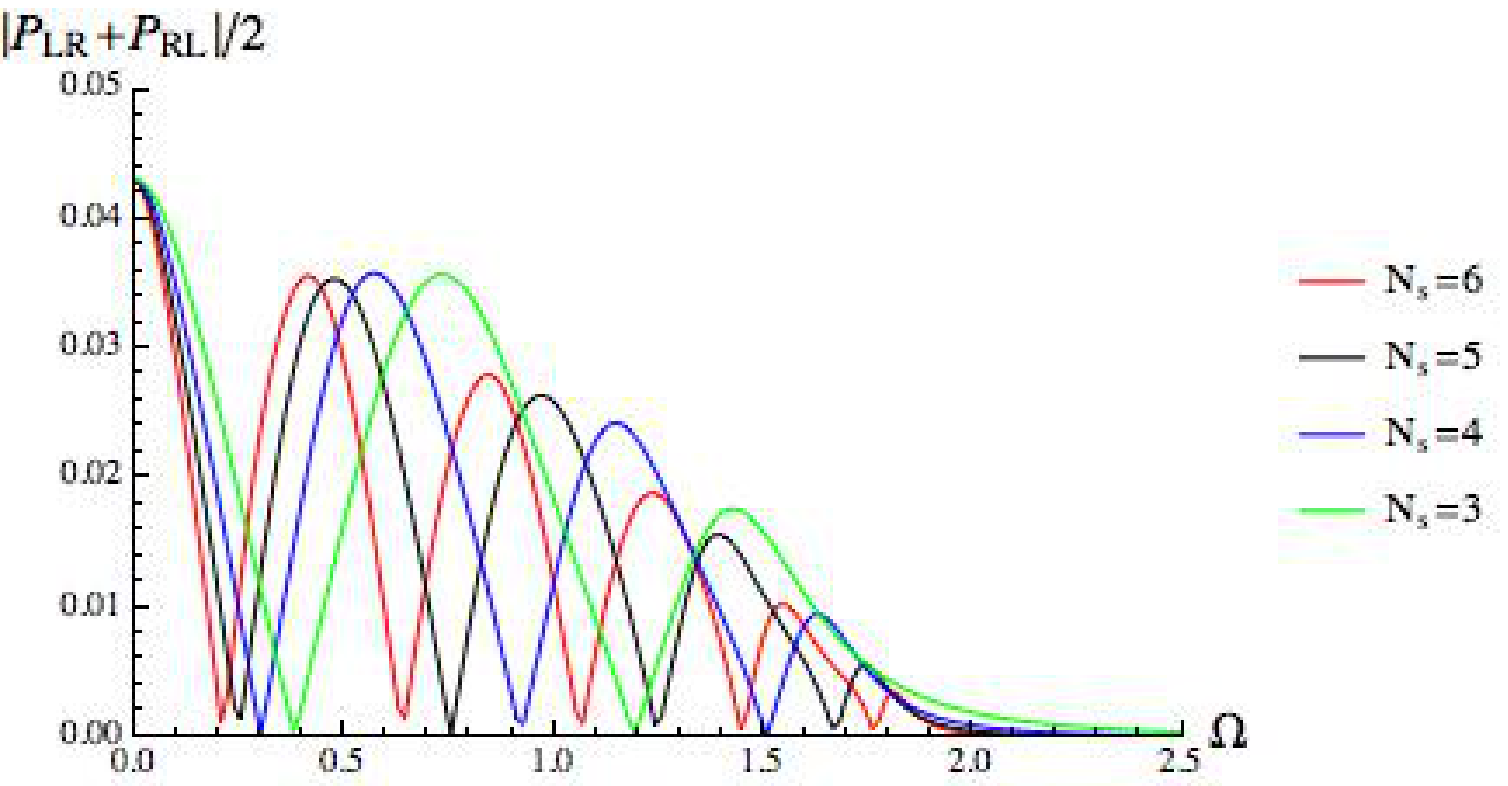}}\caption{\label{fig:lambdacompareconstV}(a) Cross-section of $\textrm{Re}\left[C^{\left(\times\right)}\left(t+\tau,t\right)|_{t=2000}\right]$
for the relative time range $\tau\in[-200,200]$, with $\lambda=0.1$.
(b) Plot of the low frequency end of the average symmetrized power
spectrum of cross-correlations, $\left(P_{LR}\left(\Omega\right)+P_{RL}\left(\Omega\right)\right)/2$,
obtained as the numerical Fourier transform of the signal in (a).
(c) Cross-section of $\textrm{Re}\left[C^{\left(\times\right)}\left(t+\tau,t\right)|_{t=2000}\right]$
for the relative time range $\tau\in[-40,40]$, with $\lambda=0.5$.
(d) The low frequency region of $\left(P_{LR}\left(\Omega\right)+P_{RL}\left(\Omega\right)\right)/2$
obtained from (c). We use the parameters $V=V_{L}=-V_{R}=5$, $E=1$,
$\Gamma=0.5$ throughout. }
\end{figure}

In Figs. \ref{fig:lambdacompareconstV} (b) and (d), we show the results
of calculations of the absolute value of the Fourier transform of
$\underset{t_{0}\rightarrow-\infty}{\lim}\textrm{Re}\left[C^{\left(\times\right)}\left(t+\tau,t\right)\right]$
with respect to $\tau$. This is done for each value of $N_{s}$ in
Fig. \ref{fig:wirebasecalculations} by fixing $t=2000$ and evaluating
a diagonal time slice of each plot shown there for $\tau\in\left[-200,200\right]$.
These time slices are shown in Fig. \ref{fig:lambdacompare} (a) ($\lambda=0.1$)
and (c) ($\lambda=0.5$). From Eqs. (\ref{eq:PSomega}) and (\ref{eq:longPS}),
the quantity plotted in Figs. \ref{fig:lambdacompareconstV} (b) and
(d) satisfies the following identity:

\begin{equation}
\mathcal{F}\left[\textrm{Re}\left[C^{\left(\times\right)}\left(\tau+t,t\right)\right];\Omega\right]=\frac{P_{LR}\left(\Omega\right)+P_{RL}\left(\Omega\right)}{2}\label{eq:symmetricID}
\end{equation}
We are therefore simply plotting the absolute value of the average
symmetrized cross-correlations in Figs. \ref{fig:lambdacompareconstV}
(b) and (d). In Fig. \ref{fig:lambdacompareconstV} (b), we plot this
for $\lambda=0.1$, and observe oscillating resonant frequencies at
values of $n\Omega_{N_{s}}$ for some intrinsic frequency $\Omega_{N_{s}}$
that depends on the length of the wire. We find that $\Omega_{N_{s}}$
decreases with increasing wire length. For example, the main $N_{s}=5$
resonance occurs at $\Omega_{5}\simeq0.15$, corresponding to a time
of $2\pi/\Omega_{5}\simeq40\simeq2\tau_{max}$, i.e. the distance
between peaks on Fig. \ref{fig:lambdacompareconstV} (a) and \ref{fig:wirebasecalculations}
(b). This is to be expected from the heuristic relation (\ref{eq:transittime})
and the contour plots of Fig. \ref{fig:wirebasecalculations}. When
we increase the intersite coupling to $\lambda=0.5$ in Fig. \ref{fig:lambdacompareconstV}
(d), we find that the position of the main resonance, for each value
of $N_{s}$, shifts by a factor of roughly $5$, so that these peaks
can be attributed to wire traversal events. We also see that the higher
frequency modes occurring at multiples of $\Omega_{N_{s}}$ are stronger
and more numerous in the $\lambda=0.5$ case than for $\lambda=0.1$.
These modes correspond to the subsidiary `ripples' seen to emanate
from the main resonances in Fig. \ref{fig:wirebasecalculations}.
Physically, these ripples are due to internally reflected electrons,
or `circular currents' that contribute weakly to the cross-lead correlations
in each lead when compared with the main influence of electrons propagating
directly from the other lead. 

\subsubsection{Time-dependent response to an AC bias}

\begin{figure}
\subfloat[]{\includegraphics[scale=0.5]{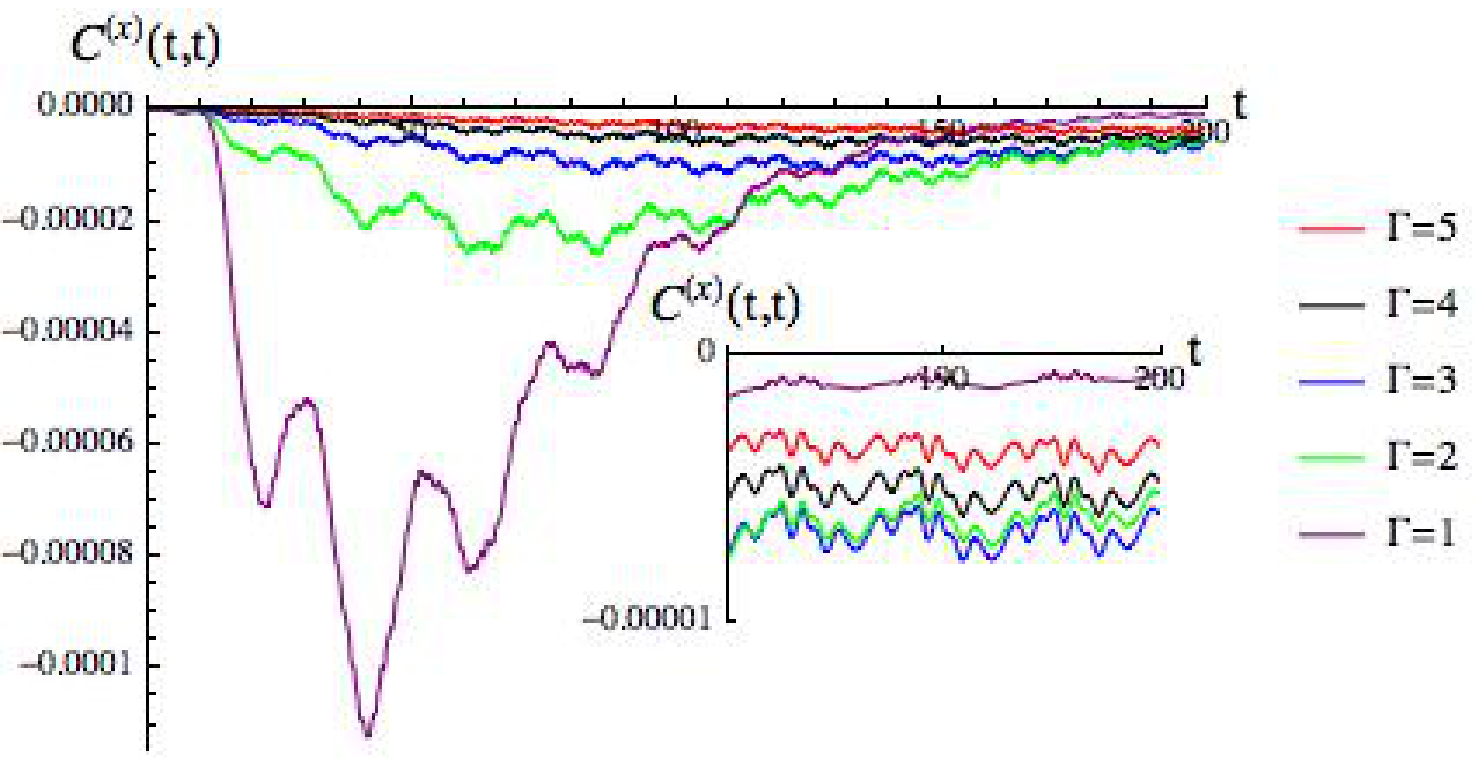}

}\subfloat[]{\includegraphics[scale=0.5]{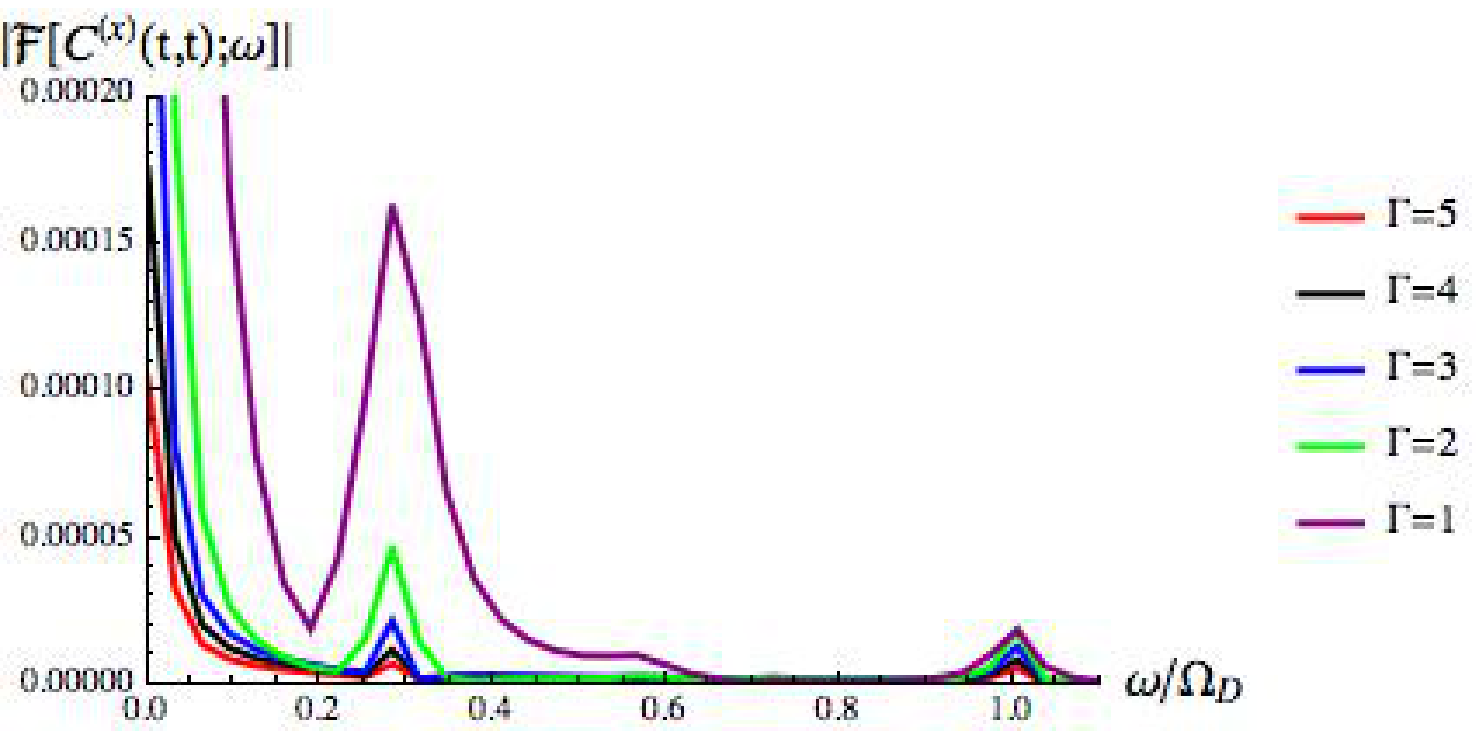}}

\subfloat[]{\includegraphics[scale=0.5]{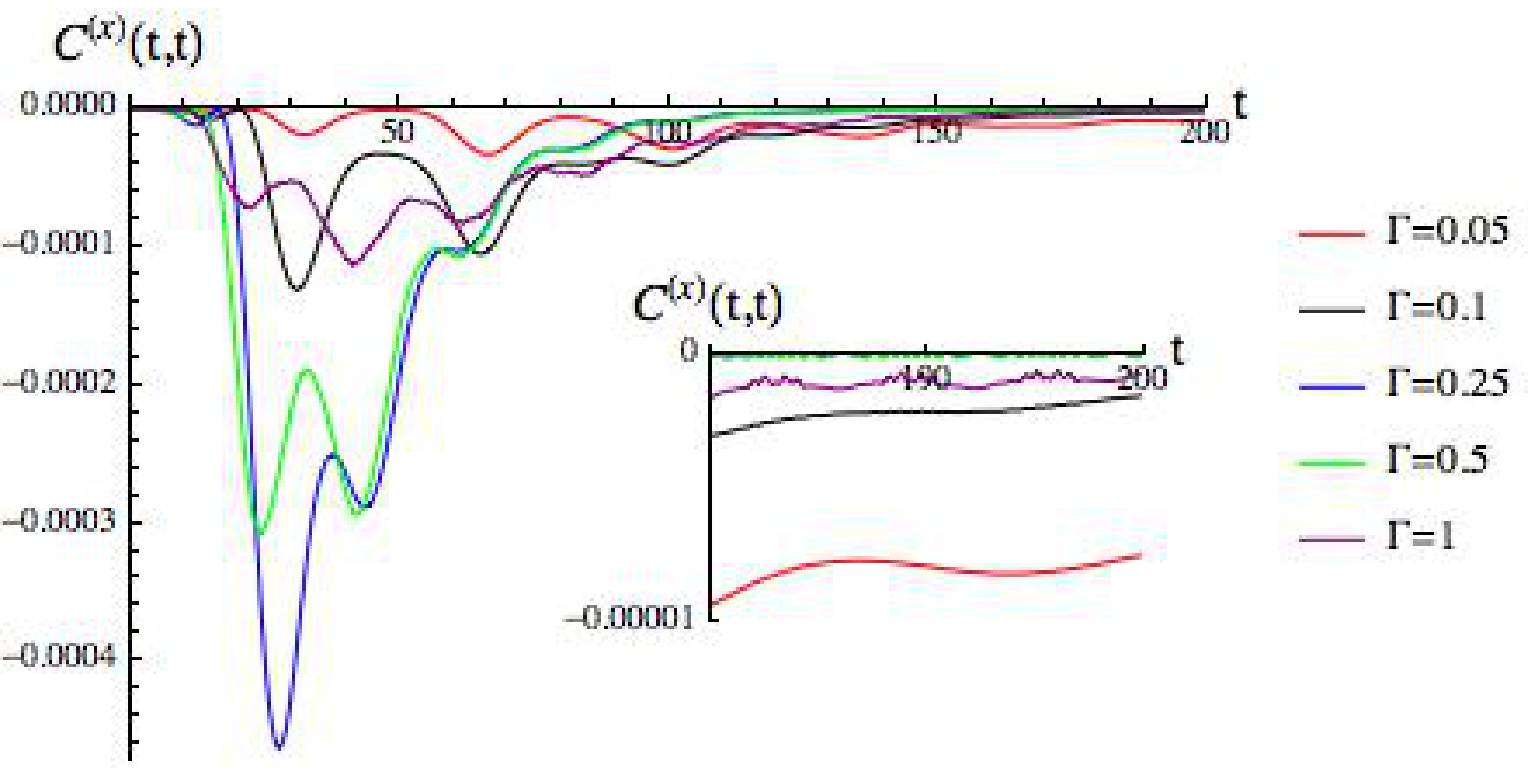}}\subfloat{\includegraphics[scale=0.5]{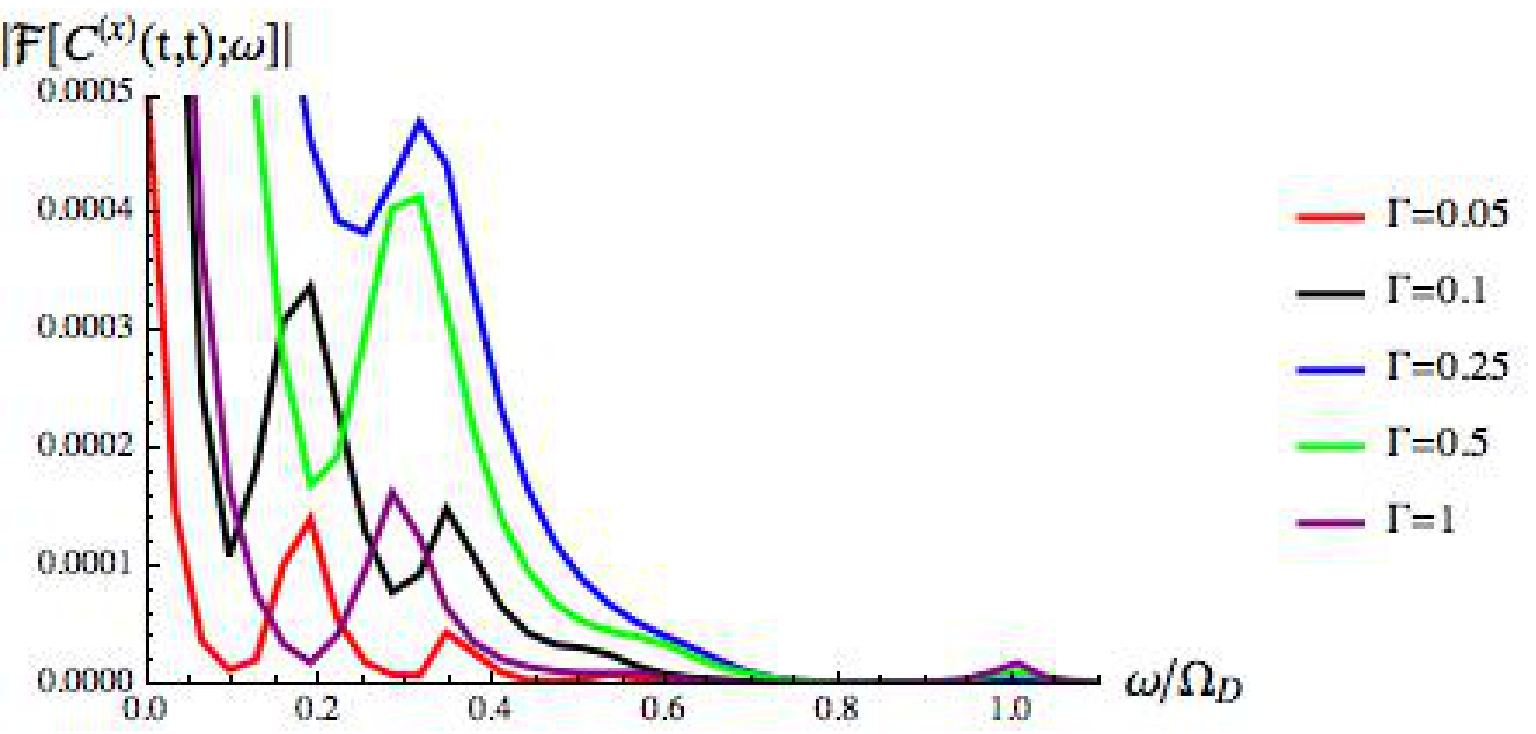}}\caption{\label{fig:wiregamcompare}(a) Plot of $C^{\left(\times\right)}\left(t,t\right)$
for $\Gamma\in\left[1,5\right]$. (b) Plot of the low-frequency end
of the absolute value of the Fourier transform $\mathcal{F}\left[C^{\left(\times\right)}\left(t,t\right);\omega\right]$
for the same parameters as (a), in units of the fundamental driving
frequency $\Omega_{D}$. (c) Plot of $C^{\left(\times\right)}\left(t,t\right)$
for $\Gamma\in\left[0.05,1\right]$. (d) Plot of the low-frequency
end of $\left|\mathcal{F}\left[C^{\left(\times\right)}\left(t,t\right);\omega\right]\right|$
for the same parameters as (c). Parameters chosen are $V_{L}=5$,
$V_{R}=5$, $A_{L}^{\left(1\right)}=4=A_{R}^{\left(1\right)}$, $A_{L}^{\left(2\right)}=0=A_{R}^{\left(2\right)}$,
$\Omega_{L}=1=\Omega_{R}\equiv\Omega_{D}$, $\phi_{L}=0$, $\phi_{R}=-\pi/2$,
$N_{s}=5$, $\lambda=0.1$.}
\end{figure}

To understand how the timescale of the resonance occurring in the
case of the perpendicular wire combines with an AC field, we will
now compute the cross-correlations for the same type of driving that
was studied in Ref. \cite{Ridley}, where long transients were observed
due to the relative sparsity of the level width matrix for a wire
in the configuration of Fig. \ref{fig:n=00003D5Chain} (a). Specifically,
we employ the bias (\ref{eq:biasdouble}) $V_{L}=5$, $V_{R}=5$,
$A_{L}^{\left(1\right)}=4=A_{R}^{\left(1\right)}$, $A_{L}^{\left(2\right)}=0=A_{R}^{\left(2\right)}$,
$\Omega_{L}=1=\Omega_{R}\equiv\Omega_{D}$, with the only difference
between the leads coming from a symmetry-breaking phase: $\phi_{L}=0$,
$\phi_{R}=-\pi/2$. In Fig. \ref{fig:wiregamcompare} (a) we plot
the $\tau=0$ cross-correlation $C^{\left(\times\right)}\left(t,t\right)$
for different values of the end-site level width parameter $\Gamma$
in a five-site wire. Similarly to Fig. \ref{fig:wirebasecalculations}
(a), we observe a resonance occurring in the absolute value of $C^{\left(\times\right)}\left(t,t\right)$
for the perpendicular five-site wire at most values of $\Gamma$,
and the frequency of this resonance, given in Fig. \ref{fig:wirebasecalculations}
(b), does not appear to be related to the driving frequency $\Omega_{D}$
as it is unchanged from its position of $t_{res}\simeq20$ in the
static bias case considered above in Fig. \ref{fig:n=00003D5Chain}
(b). After the resonance, the cross-correlation decays to a signal
with a smaller amplitude, whilst retaining a complex periodic `ringing'
signal, as shown in the inset to Fig. \ref{fig:wiregamcompare} (a).
In Fig. \ref{fig:wiregamcompare} (b), we plot the absolute value
of the Fourier transform of the signal in Fig. \ref{fig:wiregamcompare}
(a) with respect to the \textit{measurement time}, i.e. we compute
$\left|\mathcal{F}\left[C^{\left(\times\right)}\left(t,t\right);\omega\right]\right|$.
In addition to the peak at $\omega=\Omega_{D}$ corresponding to the
regular PAT process seen in Fig. \ref{fig:CLRdotsin}, we observe
an additional peak at a much lower frequency that does not occur in
the single level case, occurring at $\omega_{res}\simeq0.3\Omega_{D}$.
This frequency should be distinguished from the resonance in the steady
state correlations, $\Omega_{N_{s}}$: it corresponds to the `kick'
that the \textit{diagonal ($\tau=0$)} cross-correlation receives
at $t_{res}\simeq20$, via the relation $\omega_{res}=2\pi/t_{res}$.
This peak becomes increasingly dominant as $\Gamma$ is decreased
from $5$ to $1$. In Fig. \ref{fig:wiregamcompare} (c) we plot $C^{\left(\times\right)}\left(t,t\right)$
for $\Gamma\in\left[0.05,1\right]$. It is seen that the resonance
at $\omega_{res}\simeq0.3\Omega_{D}$ continues to grow as $\Gamma$
decreases, before saturating in the region of $\Gamma=0.25$, whereupon
the resonance decays into a less singular form. This is reflected
in the frequency spectrum of this signal, shown in Fig. \ref{fig:wiregamcompare}
(d), which shows how the resonance continues to grow before reaching
the saturation value of $\Gamma$ and splitting into two smaller resonant
peaks as $\Gamma$ tends to $0$. 

\begin{figure}
\subfloat[]{\includegraphics[scale=0.5]{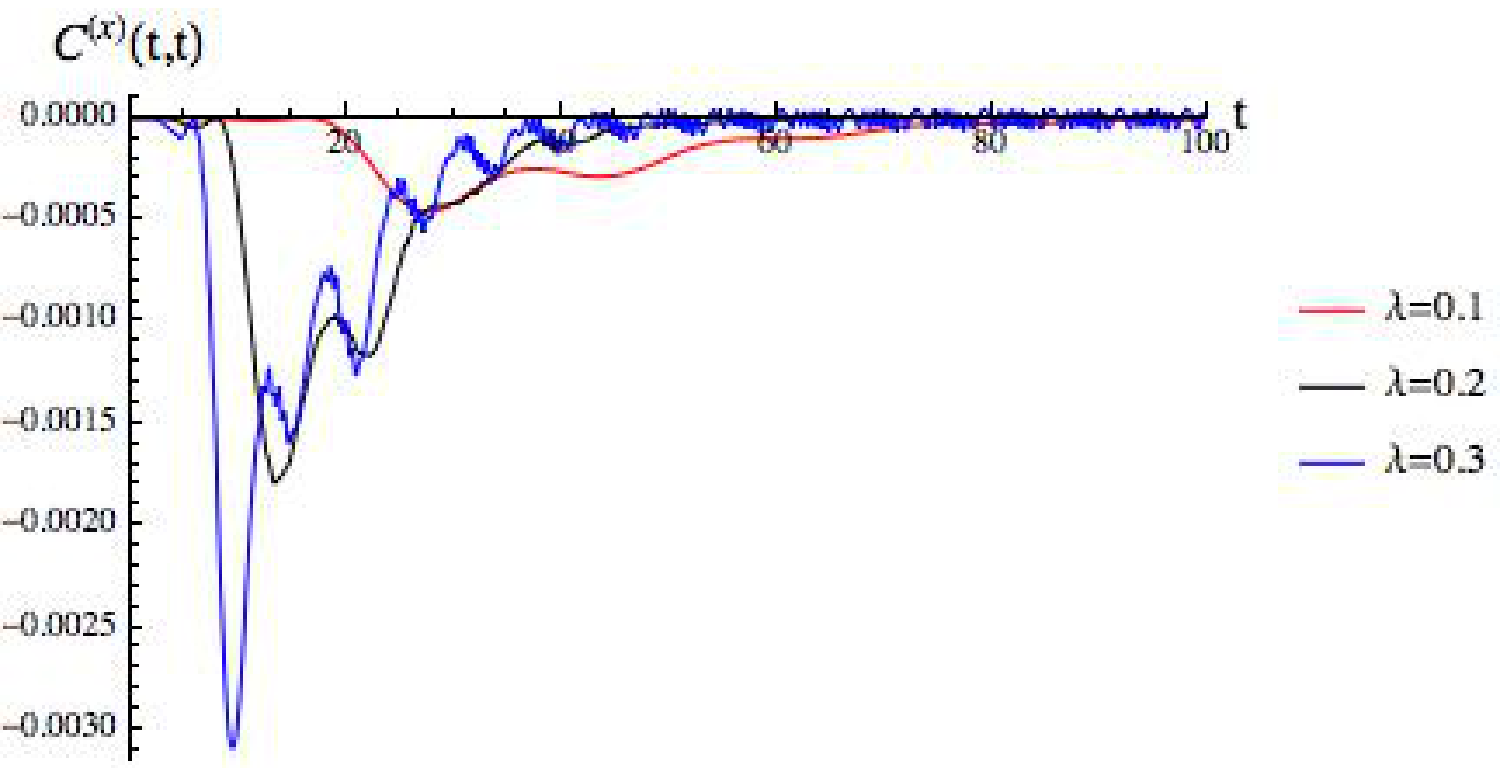}}\subfloat[]{\includegraphics[scale=0.5]{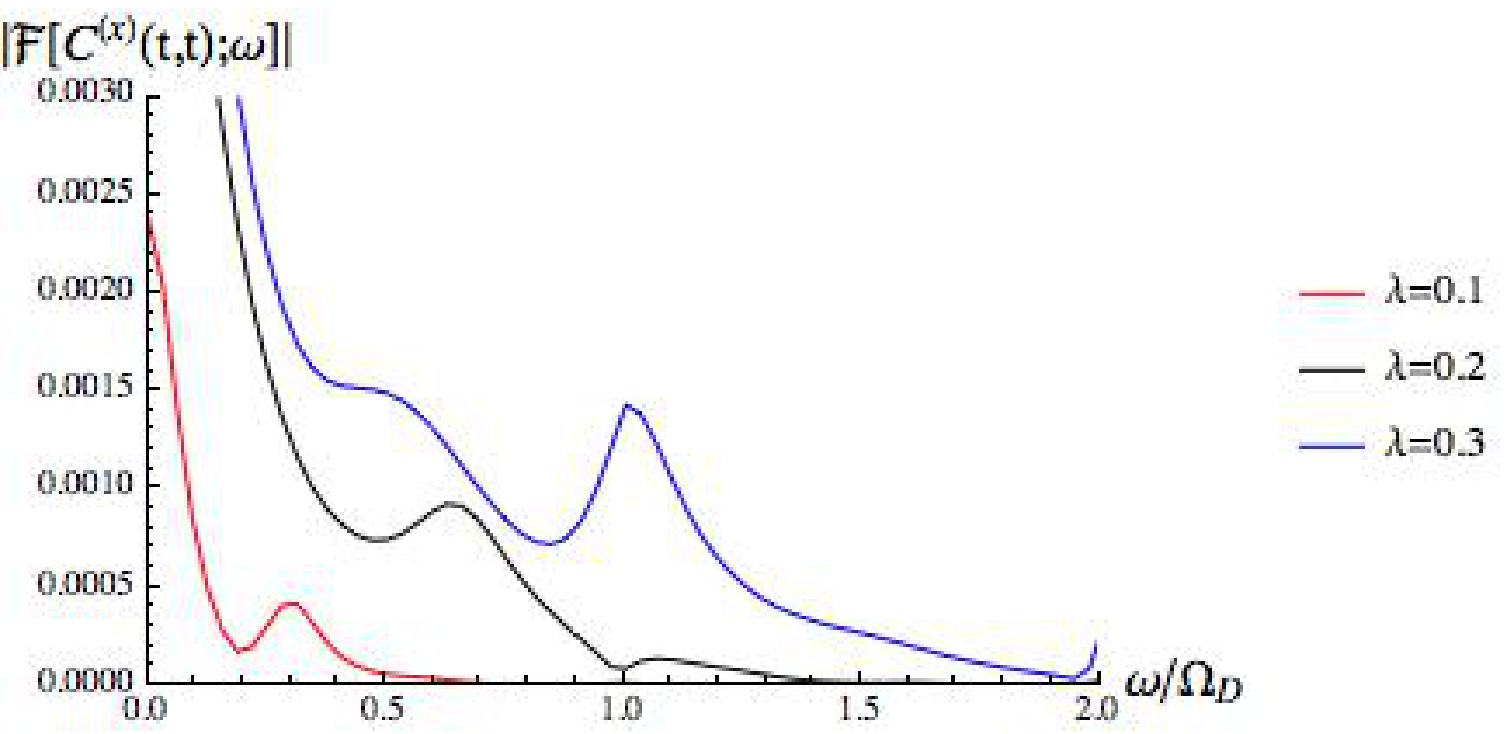}

}

\subfloat[]{\includegraphics[scale=0.5]{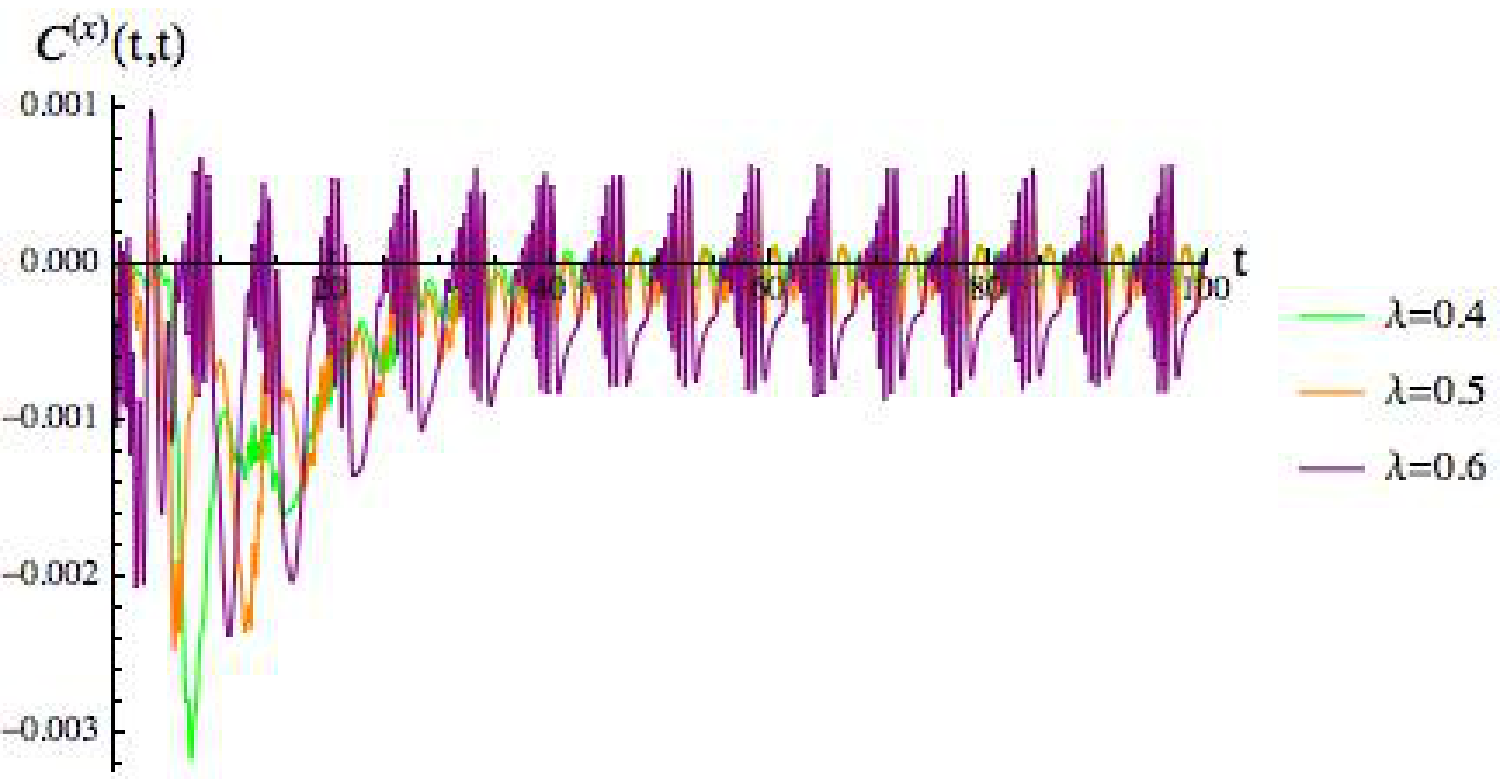}}\subfloat[]{\includegraphics[scale=0.5]{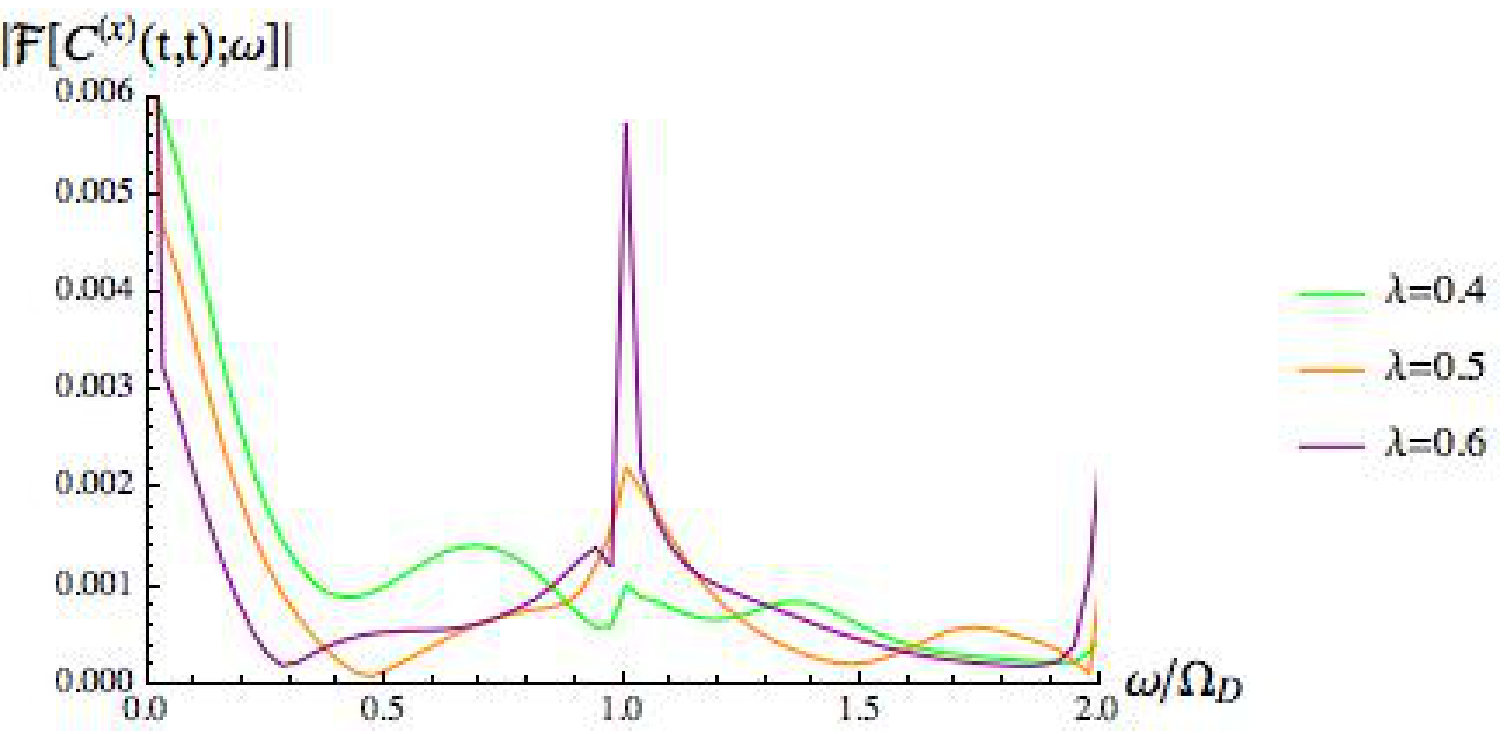}}\caption{\label{fig:lambdacompare}(a) Plot of $C^{\left(\times\right)}\left(t,t\right)$
for $\lambda\in\left[0.1,0.3\right]$. (b) Plot of the low-frequency
end of the absolute value of the Fourier transform $\mathcal{F}\left[C^{\left(\times\right)}\left(t,t\right);\omega\right]$
for the same parameters as in (a), in units of the fundamental driving
frequency $\Omega_{D}$. (c) Plot of $C^{\left(\times\right)}\left(t,t\right)$
for $\lambda\in\left[0.4,0.6\right]$. (d) Plot of the low-frequency
end of $\left|\mathcal{F}\left[C^{\left(\times\right)}\left(t,t\right);\omega\right]\right|$
for the same parameters as in (c). Parameters chosen are $V_{L}=5$,
$V_{R}=5$, $A_{L}^{\left(1\right)}=4=A_{R}^{\left(1\right)}$, $A_{L}^{\left(2\right)}=0=A_{R}^{\left(2\right)}$,
$\Omega_{L}=1=\Omega_{R}\equiv\Omega_{D}$, $\phi_{L}=0$, $\phi_{R}=-\pi/2$,
$N_{s}=5$.}
\end{figure}

The frequency of the resonance in Figs. \ref{fig:wiregamcompare}
(b) and (d) is located at about $0.3-0.35\Omega_{D}$ regardless of
whether the bias is AC or DC, and seems to be only moderately affected
by changes in $\Gamma$. We therefore suspect that it is due to the
finite size and intrinsic properties of the wire. In Fig. \ref{fig:lambdacompare},
we present calculations of the cross-correlation in the same system,
this time varying the hopping parameter $\lambda$ and keeping the
coupling parameter fixed at $\Gamma=0.5$. Fig. \ref{fig:lambdacompare}
(a) shows that, as $\lambda$ is doubled from $0.1$ to $0.2$, the
time at which the resonance kicks in is approximately halved, before
being scaled down by a factor of $\sim2/3$ as $\lambda$ is further
increased to $0.3$. This is reflected in the Fourier Transform of
these signals shown in Fig. \ref{fig:lambdacompare} (b), which show
that the position of the resonant frequency increases linearly with
increasing $\lambda$, as expected from the heuristic relation (\ref{eq:transittime}).
As we continue to increase the coupling parameter, the duration of
the transient resonance becomes shorter until it approaches the time
of $2\pi$, i.e. the time period of the fundamental driving frequency
$\omega_{res}=\Omega_{D}$ at around $\lambda=0.3$. For values of
the coupling greater than this, we see indeed in Fig. \ref{fig:lambdacompare}
(c) that that timescale of the transient becomes smaller than the
timescale associated with the driving. The frequency spectrum of cross-correlations,
shown in Fig. \ref{fig:lambdacompare} (d), contains peaks at $\omega_{res}\simeq1.4\Omega_{D}$
(when $\lambda=0.4$) and $\omega_{res}\simeq1.75\Omega_{D}$ (when
$\lambda=0.5$), corresponding to a continuation of the linear dependence
of the transient resonant frequency on $\lambda$. However, when we
cross into the regime of $\lambda/\Gamma>1$ (purple line), this resonance
has been submerged beneath the growing resonances at integer values
of $\omega/\Omega_{D}$, a fact which is reflected in the strongly
oscillating signal of the purple line in Fig. \ref{fig:lambdacompare}
(c). These resonances continue into the high-frequency part of the
spectrum beyond the narrow window exhibited here, and are due to PAT
processes. 

\subsubsection{Discussion of the results for the molecular wire}

The calculations presented here point to a rather clear physical interpretation
of the transient behaviour of cross-correlations in extended systems.
There are three factors which compete to determine how long electrons
take to cross the nanojunction: the strength of the end-site coupling
$\Gamma$, the length of the wire $N_{s}-1$ and the internal hopping
parameter $\lambda$. When $\Gamma/\lambda>1$, the molecular wire
is more resistant to propagating electrons than the molecule-electrode
interface. Although the large value of $\Gamma$ tends to reduce the
lifetime of molecular modes, the small value of $\lambda$ makes it
difficult for tunneling between sites to occur. This means that the
time taken for electrons to traverse the molecular region is significantly
longer than the timescale of the external driving field, and so the
currents in each lead, $I_{L}\left(t\right)$ and $I_{R}\left(t\right)$
become more strongly correlated at about the time taken for electrons
to propagate between the leads following the switch-on. This explains
the resonant `kick' at $t_{res}\simeq20$ in Figs. \ref{fig:wiregamcompare}
(a) and (c), which remains at this time so long as $\Gamma>\lambda$.
However, as $\Gamma$ is decreased to the weak coupling regime of
$\Gamma/\lambda<1$, we enter a regime in which the lifetime of modes
on the molecular sites exceeds that taken to tunnel between sites
and so the traversal time is increased. This explains the splitting
of the single frequency peak in Fig. \ref{fig:wiregamcompare} (d)
into two smaller peaks. 

The position of the resonance in frequency space is determined by
$\lambda$ and $N_{s}$. Physically, as the coupling between molecular
sites is increased, it becomes increasingly easier for electrons to
traverse the wire. In the DC case we saw that increasing the value
of $\lambda$ increases the strength of reflected currents within
the molecular wire. In the AC case, we found that in the regime of
$\lambda\sim\Gamma\sim\Omega_{D}$, the time taken for electrons to
pass from lead $L$ to lead $R$ is smaller than the rate at which
photon-assisted electrons\textcolor{red}{{} }tunnel from the lead onto
the molecule. In this regime, the electrons can travel between leads
in less time than it takes for the signal driving them to undergo
an appreciable change. 

Once the resonance in $C^{\left(\times\right)}\left(t,t\right)$ has
died out, a steady state is achieved which is oscillatory in the case
of an AC bias, and stationary in the case of a DC bias. In the latter
case, we saw the emergence of a regime in which each lead at time
$t$ felt the influence of the other most strongly at a time shifted
by $\tau_{max}$, which was roughly equal to the time $t_{res}$,
and which changed with $N_{s}$ and $\lambda$ in the same way as
$t_{res}$. This suggests the existence of a time delay for information
to propagate between the $L$ and $R$ leads that shows up in the
low frequency power spectrum, and in the equal time cross-correlations.
The question of how to define the traversal time for electrons tunneling
across a nanostructure has been the subject of much debate, with a
variety of different definitions proposed, mainly based on the rate
of change of wavepacket phase with respect to momentum \cite{Buttiker1982,Collins1987}.
The results presented here point to a new method of determining this
time for large molecular structures, namely, by identifying the low
frequency resonances in the transient of $C^{\left(\times\right)}\left(t,t\right)$,
or in the steady-state Fourier transform of $C^{\left(\times\right)}\left(t+\tau,t\right)$
with respect to the time difference $\tau$.

\section{Conclusions}

In this paper we have presented a formalism for the calculation of
the time-dependent quantum current correlations in nanojunctions which
can be used to study the transient current cross-correlations following
a partition-free bias switch-on process. The switched on bias may
have any time-dependence: our approach is not restricted to periodic
or constant biases. Moreover, our formalism applies to any molecular
structure to which the WBLA applies and will be very useful for transport
calculations on large molecular structures. Importantly, it perfectly
reproduces the steady-state quantum noise formulas obtained previously
in Refs. \cite{Buttiker1992,Yang1992,Yang2014} under the appropriate
limits. 

We then presented calculations of the cross-lead current correlations
both in the full two-time plane and for the equal time ($\tau=0$)
case. Whereas in the single level case, the magnitude of cross-correlations
was maximized for a correlation delay time $\tau_{max}=0$, $\tau_{max}$
was increased significantly with an increase in the number $N_{s}$
of atoms in a wire and with decreasing intersite hopping strength
$\lambda$, so we naturally interpret it as the traversal time for
electronic information to cross the nanojunction. In addition, a resonance
was observed in the $\tau=0$ cross-correlation at a time $t_{res}$
that could be orders of magnitude greater than the time taken for
electrons to tunnel onto the molecule. We found that $t_{res}$ was
independent of the particular bias chosen but scaled linearly with
$N_{s}$ and $1/\lambda$. This points once again to a signature of
the traversal time in the cross-correlations. Therefore we anticipate
that our method can be used to determine electron traversal times
in a rather more precise way than that offered by heuristic arguments.
This will be useful for functional device applications, for example
in determining the maximum operating frequencies in extended molecules
used as switches or frequency sensors in real circuits. 

We emphasize that we have only begun to explore the parameter space
and system size that is now accessible within the biharmonic bias
model of Eq. (\ref{eq:biasdouble}). The calculations presented in
this paper were intended to complement numerical work done in Refs.
\cite{Ridley2015,Ridley} and therefore only used a single harmonic
driving term. They were also applied to very simple model systems,
although our formulas are general and may be applied efficiently to
a comparatively larger systems, such as CNT and GNR. In forthcoming
work we will use the method described in the Numerics section to estimate
the traversal time for these kind of structures. We will also show
how the formulas presented in Appendix \ref{sec:FormulasBiharmonic}
can be used to achieve AC-DC rectification, or charge pumping, by
including the second harmonic in Eq. (\ref{eq:biasdouble}) and manipulating
the TR symmetry-breaking phase $\phi_{\alpha}$ appearing in the first
harmonic.

\paragraph{Acknowledgements}

Michael Ridley was supported through a studentship in the Centre for
Doctoral Training on Theory and Simulation of Materials at Imperial
College funded by the Engineering and Physical Sciences Research Council
under grant number EP/G036888/1.

\appendix

\section{\label{sec:Appendix-B} - Dyson Equations for Matrix Blocks of the
Green's Function}

\subsection{Lead-Molecule Coupling Terms}

In this section, the first and fourth terms in (\ref{eq: 1}) will
be evaluated. First one utilises the fact that the Dyson equations
for the full $\alpha-C$ and $C-\alpha$ Green's functions blocks
are given by:

\begin{equation}
\mathbf{G}_{\alpha C}\left(z_{1},z_{2}\right)=\int_{\gamma}d\bar{z}\mathbf{g}_{\alpha\alpha}\left(z_{1},\bar{z}\right)\mathbf{h}_{\alpha C}\left(\bar{z}\right)\mathbf{G}_{CC}\left(\bar{z},z_{2}\right)\label{eq: 2}
\end{equation}

\begin{equation}
\mathbf{G}_{C\alpha}\left(z_{1},z_{2}\right)=\int_{\gamma}d\bar{z}\mathbf{G}_{CC}\left(z_{1},\bar{z}\right)\mathbf{h}_{C\alpha}\left(\bar{z}\right)\mathbf{g}_{\alpha\alpha}\left(\bar{z},z_{2}\right)\label{eq: 3}
\end{equation}
We now introduce the notation `$\left(z_{1}^{-},z_{2}^{+}\right)$'
to denote that the first argument is always on the upper branch $C_{-}$
of the Konstantinov-Perel' contour, and the second argument always
on the lower branch $C_{+}$, and therefore that the second argument
is always `later' on the contour than the first. `$\left(z_{1}^{+},z_{2}^{-}\right)$'
denotes the opposite ordering of contour positions, and retaining
this ordering is necessary to obtain the correct initial conditions
when the limit $t_{1},t_{2}\rightarrow t_{0}$ is taken. Next we apply
the Langreth rules \cite{Langreth1972,Langreth1976} to Eqs. (\ref{eq: 2})
and (\ref{eq: 3}) to give:

\begin{equation}
\mathbf{G}_{\alpha C}^{>}\left(t_{1},t_{2}\right)=\left[\mathbf{g}_{\alpha\alpha}^{>}\mathbf{h}_{\alpha C}\cdot\mathbf{G}_{CC}^{a}+\mathbf{g}_{\alpha\alpha}^{r}\mathbf{h}_{\alpha C}\cdot\mathbf{G}_{CC}^{>}+\mathbf{g}_{\alpha\alpha}^{\urcorner}\mathbf{h}_{\alpha C}\star\mathbf{G}_{CC}^{\ulcorner}\right]_{\left(t_{1}^{+},t_{2}^{-}\right)}
\end{equation}

\begin{equation}
\mathbf{G}_{\beta C}^{<}\left(t_{2},t_{1}\right)=\left[\mathbf{g}_{\beta\beta}^{<}\mathbf{h}_{\beta C}\cdot\mathbf{G}_{CC}^{a}+\mathbf{g}_{\beta\beta}^{r}\mathbf{h}_{\beta C}\cdot\mathbf{G}_{CC}^{<}+\mathbf{g}_{\beta\beta}^{\urcorner}\mathbf{h}_{\beta C}\star\mathbf{G}_{CC}^{\ulcorner}\right]_{\left(t_{2}^{-},t_{1}^{+}\right)}
\end{equation}

\begin{equation}
\mathbf{G}_{C\beta}^{>}\left(t_{1},t_{2}\right)=\left[\mathbf{G}_{CC}^{>}\cdot\mathbf{h}_{C\beta}\mathbf{g}_{\beta\beta}^{a}+\mathbf{G}_{CC}^{r}\cdot\mathbf{h}_{C\beta}\mathbf{g}_{\beta\beta}^{>}+\mathbf{G}_{CC}^{\urcorner}\star\mathbf{h}_{C\beta}\mathbf{g}_{\beta\beta}^{\ulcorner}\right]_{\left(t_{1}^{+},t_{2}^{-}\right)}
\end{equation}

\begin{equation}
\mathbf{G}_{C\alpha}^{<}\left(t_{2},t_{1}\right)=\left[\mathbf{G}_{CC}^{<}\cdot\mathbf{h}_{C\alpha}\mathbf{g}_{\alpha\alpha}^{a}+\mathbf{G}_{CC}^{r}\cdot\mathbf{h}_{C\alpha}\mathbf{g}_{\alpha\alpha}^{<}+\mathbf{G}_{CC}^{\urcorner}\star\mathbf{h}_{C\alpha}\mathbf{g}_{\alpha\alpha}^{\ulcorner}\right]_{\left(t_{2}^{-},t_{1}^{+}\right)}
\end{equation}
These expressions are combined with the definition of the embedding
self-energy to give:

\[
\mathbf{h}_{C\alpha}\mathbf{G}_{\alpha C}^{>}\left(t_{1},t_{2}\right)\mathbf{h}_{C\beta}\mathbf{G}_{\beta C}^{<}\left(t_{2},t_{1}\right)+\mathbf{G}_{C\beta}^{>}\left(t_{1},t_{2}\right)\mathbf{h}_{\beta C}\mathbf{G}_{C\alpha}^{<}\left(t_{2},t_{1}\right)\mathbf{h}_{\alpha C}=
\]

\[
\left(\mathbf{\Sigma}_{\alpha}^{>}\cdot\mathbf{G}_{CC}^{a}+\mathbf{\Sigma}_{\alpha}^{r}\cdot\mathbf{G}_{CC}^{>}+\mathbf{\Sigma}_{\alpha}^{\urcorner}\star\mathbf{G}_{CC}^{\ulcorner}\right)_{\left(t_{1}^{+},t_{2}^{-}\right)}\left(\mathbf{\Sigma}_{\beta}^{<}\cdot\mathbf{G}_{CC}^{a}+\mathbf{\Sigma}_{\beta}^{r}\cdot\mathbf{G}_{CC}^{<}+\mathbf{\Sigma}_{\beta}^{\urcorner}\star\mathbf{G}_{CC}^{\ulcorner}\right)_{\left(t_{2}^{-},t_{1}^{+}\right)}
\]

\begin{equation}
+\left(\mathbf{G}_{CC}^{>}\cdot\mathbf{\Sigma}_{\beta}^{a}+\mathbf{G}_{CC}^{r}\cdot\mathbf{\Sigma}_{\beta}^{>}+\mathbf{G}_{CC}^{\urcorner}\star\mathbf{\Sigma}_{\beta}^{\ulcorner}\right)_{\left(t_{1}^{+},t_{2}^{-}\right)}\left(\mathbf{G}_{CC}^{<}\cdot\mathbf{\Sigma}_{\alpha}^{a}+\mathbf{G}_{CC}^{r}\cdot\mathbf{\Sigma}_{\alpha}^{<}+\mathbf{G}_{CC}^{\urcorner}\star\mathbf{\Sigma}_{\alpha}^{\ulcorner}\right)_{\left(t_{2}^{-},t_{1}^{+}\right)}\label{eq:p1}
\end{equation}

\subsection{Lead-Lead and Molecule-Molecule Terms}

In this section the second and third terms in (\ref{eq: 1}) will
be evaluated. The EoM for the full GF is projected onto the $\alpha\beta$
region:

\begin{equation}
\left[i\frac{d}{dz_{1}}-\mathbf{h}_{\alpha\alpha}\left(z_{1}\right)\right]\mathbf{G}_{\alpha\beta}\left(z_{1},z_{2}\right)=\mathbf{I}_{\alpha}\delta_{\alpha\beta}\delta\left(z_{1},z_{2}\right)+\mathbf{h}_{\alpha C}\left(z_{1}\right)\mathbf{G}_{C\beta}\left(z_{1},z_{2}\right)
\end{equation}

\begin{equation}
\mathbf{G}_{\alpha\beta}\left(z_{1},z_{2}\right)\left[-i\overleftarrow{\frac{d}{dz_{2}}}-\mathbf{h}_{\beta\beta}\left(z_{2}\right)\right]=\mathbf{I}_{\alpha}\delta_{\alpha\beta}\delta\left(z_{1},z_{2}\right)+\mathbf{G}_{\alpha C}\left(z_{1},z_{2}\right)\mathbf{h}_{C\beta}\left(z_{2}\right)
\end{equation}
We insert the fomulae (\ref{eq: 2}) and (\ref{eq: 3}) into these
EoM to get:

\begin{equation}
\left[i\frac{d}{dz_{1}}-\mathbf{h}_{\alpha\alpha}\left(z_{1}\right)\right]\mathbf{G}_{\alpha\beta}\left(z_{1},z_{2}\right)=\mathbf{I}_{\alpha}\delta_{\alpha\beta}\delta\left(z_{1},z_{2}\right)+\int_{\gamma}d\bar{z}\,\mathbf{h}_{\alpha C}\left(z_{1}\right)\mathbf{G}_{CC}\left(z_{1},\bar{z}\right)\mathbf{h}_{C\beta}\left(\bar{z}\right)\mathbf{g}_{\beta\beta}\left(\bar{z},z_{2}\right)
\end{equation}

\begin{equation}
\mathbf{G}_{\alpha\beta}\left(z_{1},z_{2}\right)\left[-i\overleftarrow{\frac{d}{dz_{2}}}-\mathbf{h}_{\beta\beta}\left(z_{2}\right)\right]=\mathbf{I}_{\alpha}\delta_{\alpha\beta}\delta\left(z_{1},z_{2}\right)+\int_{\gamma}d\bar{z}\,\mathbf{g}_{\alpha\alpha}\left(z_{1},\bar{z}\right)\mathbf{h}_{\alpha C}\left(\bar{z}\right)\mathbf{G}_{CC}\left(\bar{z},z_{2}\right)\mathbf{h}_{C\beta}\left(z_{2}\right)
\end{equation}
Introducing the GF of the bare leads,

\begin{equation}
\left[i\frac{d}{dz_{1}}-\mathbf{h}_{\alpha\alpha}\left(z_{1}\right)\right]\mathbf{g}_{\alpha\alpha}\left(z_{1},z_{2}\right)=\mathbf{I}_{\alpha}\delta\left(z_{1},z_{2}\right)
\end{equation}
we can extract the desired Dyson equation:

\begin{equation}
\mathbf{G}_{\alpha\beta}\left(z_{1},z_{2}\right)=\mathbf{g}_{\alpha\alpha}\left(z_{1},z_{2}\right)\delta_{\alpha\beta}+\int_{\gamma}d\bar{z}d\bar{z}'\mathbf{g}_{\alpha\alpha}\left(z_{1},\bar{z}\right)\mathbf{h}_{\alpha C}\left(\bar{z}\right)\mathbf{G}_{CC}\left(\bar{z},\bar{z}'\right)\mathbf{h}_{C\beta}\left(\bar{z}'\right)\mathbf{g}_{\beta\beta}\left(\bar{z}',z_{2}\right)
\end{equation}
Once more applying the Langreth rules, the greater and lesser GFs
can then be found:

\[
\mathbf{G}_{\alpha\beta}^{>}\left(t_{1},t_{2}\right)=\mathbf{g}_{\alpha\alpha}^{>}\left(t_{1},t_{2}\right)\delta_{\alpha\beta}+\left[\left(\mathbf{g}_{\alpha\alpha}^{>}\mathbf{h}_{\alpha C}\cdot\mathbf{G}_{CC}^{a}+\mathbf{g}_{\alpha\alpha}^{r}\mathbf{h}_{\alpha C}\oslash\mathbf{G}_{CC}^{>}+\mathbf{g}_{\alpha\alpha}^{\urcorner}\mathbf{h}_{\alpha C}\star\mathbf{G}_{CC}^{\ulcorner}\right)\cdot\mathbf{h}_{C\beta}\mathbf{g}_{\beta\beta}^{a}\right.
\]

\begin{equation}
\left.+\mathbf{g}_{\alpha\alpha}^{r}\mathbf{h}_{\alpha C}\cdot\mathbf{G}_{CC}^{r}\cdot\mathbf{h}_{C\beta}\mathbf{g}_{\beta\beta}^{>}+\left(\mathbf{g}_{\alpha\alpha}^{r}\mathbf{h}_{\alpha C}\cdot\mathbf{G}_{CC}^{\urcorner}+\mathbf{g}_{\alpha\alpha}^{\urcorner}\mathbf{h}_{\alpha C}\star\mathbf{G}_{CC}^{M}\right)\star\mathbf{h}_{C\beta}\mathbf{g}_{\beta\beta}^{\ulcorner}\right]_{\left(t_{1}^{+},t_{2}^{-}\right)}
\end{equation}

\[
\mathbf{G}_{\beta\alpha}^{<}\left(t_{2},t_{1}\right)=\mathbf{g}_{\alpha\alpha}^{<}\left(t_{2},t_{1}\right)\delta_{\alpha\beta}+\left[\left(\mathbf{g}_{\beta\beta}^{<}\mathbf{h}_{\beta C}\cdot\mathbf{G}_{CC}^{a}+\mathbf{g}_{\beta\beta}^{r}\mathbf{h}_{\beta C}\cdot\mathbf{G}_{CC}^{<}+\mathbf{g}_{\beta\beta}^{\urcorner}\mathbf{h}_{\beta C}\star\mathbf{G}_{CC}^{\ulcorner}\right)\cdot\mathbf{h}_{C\alpha}\mathbf{g}_{\alpha\alpha}^{a}\right.
\]

\begin{equation}
\left.+\mathbf{g}_{\beta\beta}^{r}\mathbf{h}_{\beta C}\cdot\mathbf{G}_{CC}^{r}\cdot\mathbf{h}_{C\alpha}\mathbf{g}_{\alpha\alpha}^{<}+\left(\mathbf{g}_{\beta\beta}^{r}\mathbf{h}_{\beta C}\cdot\mathbf{G}_{CC}^{\urcorner}+\mathbf{g}_{\beta\beta}^{\urcorner}\mathbf{h}_{\beta C}\star\mathbf{G}_{CC}^{M}\right)\star\mathbf{h}_{C\alpha}\mathbf{g}_{\alpha\alpha}^{\ulcorner}\right]_{\left(t_{2}^{-},t_{1}^{+}\right)}
\end{equation}
We are thus able to write the second and third terms in the correlation
function (\ref{eq: 1}) in terms of self energy and GF components:

\[
\mathbf{h}_{C\alpha}\mathbf{G}_{\alpha\beta}^{>}\left(t_{1},t_{2}\right)\mathbf{h}_{\beta C}=\mathbf{\Sigma}_{\alpha}^{>}\left(t_{1},t_{2}\right)\delta_{\alpha\beta}+\left[\left(\mathbf{\Sigma}_{\alpha}^{>}\cdot\mathbf{G}_{CC}^{a}+\mathbf{\Sigma}_{\alpha}^{r}\cdot\mathbf{G}_{CC}^{>}+\mathbf{\Sigma}_{\alpha}^{\urcorner}\star\mathbf{G}_{CC}^{\ulcorner}\right)\cdot\mathbf{\Sigma}_{\beta}^{a}\right.
\]

\begin{equation}
\left.+\mathbf{\Sigma}_{\alpha}^{r}\cdot\mathbf{G}_{CC}^{r}\cdot\mathbf{\Sigma}_{\beta}^{>}+\left(\mathbf{\Sigma}_{\alpha}^{r}\cdot\mathbf{G}_{CC}^{\urcorner}+\mathbf{\Sigma}_{\alpha}^{\urcorner}\star\mathbf{G}_{CC}^{M}\right)\star\mathbf{\Sigma}_{\beta}^{\ulcorner}\right]_{\left(t_{1}^{+},t_{2}^{-}\right)}\label{eq:p2a}
\end{equation}

\[
\mathbf{h}_{C\beta}\mathbf{G}_{\beta\alpha}^{<}\left(t_{2},t_{1}\right)\mathbf{h}_{\alpha C}=\mathbf{\Sigma}_{\alpha}^{<}\left(t_{2},t_{1}\right)\delta_{\alpha\beta}+\left[\left(\mathbf{\Sigma}_{\beta}^{<}\cdot\mathbf{G}_{CC}^{a}+\mathbf{\Sigma}_{\beta}^{r}\cdot\mathbf{G}_{CC}^{<}+\mathbf{\Sigma}_{\beta}^{\urcorner}\star\mathbf{G}_{CC}^{\ulcorner}\right)\cdot\mathbf{\Sigma}_{\alpha}^{a}\right.
\]

\begin{equation}
\left.+\mathbf{\Sigma}_{\beta}^{r}\cdot\mathbf{G}_{CC}^{r}\cdot\mathbf{\Sigma}_{\alpha}^{<}+\left(\mathbf{\Sigma}_{\beta}^{r}\cdot\mathbf{G}_{CC}^{\urcorner}+\mathbf{\Sigma}_{\beta}^{\urcorner}\star\mathbf{G}_{CC}^{M}\right)\star\mathbf{\Sigma}_{\alpha}^{\ulcorner}\right]_{\left(t_{2}^{-},t_{1}^{+}\right)}\label{eq:p2b}
\end{equation}
To simplify these expressions we use an identity \cite{Tuovinen2013,Stefanucci2013}:

\begin{eqnarray}
\left(\mathbf{\Sigma}_{\alpha}^{\urcorner}\star\mathbf{G}_{CC}^{M}\star\mathbf{\Sigma}_{\beta}^{\ulcorner}\right)_{\left(t_{1},t_{2}\right)} & = & 0
\end{eqnarray}
This enables us to neglect the terms in the correlation function arising
from a double convolution on $C_{M}$ in Eqs. (\ref{eq:p2a}) and
(\ref{eq:p2b}), so that in conjunction with Eq. (\ref{eq:p1}) one
finally obtains the correlation function in Eq. (\ref{eq:power1}).

\section{\label{sec:Appendix-A}- Green's Functions and Self-Energies for
the Time-Dependent Model Hamiltonian}

We have previously obtained all Green's Functions and self-energy
components for the switch-on process described by the Hamiltonian
in Eq. (\ref{eq:Hamiltonian}) \cite{Ridley2016}. We list these below
for expediency:

\begin{equation}
\mathbf{G}_{CC}^{r}\left(t_{1},t_{2}\right)=-i\theta\left(t_{1}-t_{2}\right)e^{-i\mathbf{\widetilde{h}}{}_{CC}^{eff}\left(t_{1}-t_{2}\right)}e^{-i\varphi_{C}\left(t_{1},t_{2}\right)}
\end{equation}

\begin{equation}
\mathbf{G}_{CC}^{a}\left(t_{1},t_{2}\right)=i\theta\left(t_{2}-t_{1}\right)e^{-i\left(\mathbf{\widetilde{h}}{}_{CC}^{eff}\right)^{\dagger}\left(t_{1}-t_{2}\right)}e^{-i\varphi_{C}\left(t_{1},t_{2}\right)}
\end{equation}

\begin{equation}
\mathbf{G}_{CC}^{M}\left(\tau_{1},\tau_{2}\right)=\frac{i}{\beta}\underset{q=-\infty}{\overset{\infty}{\sum}}e^{-\omega_{q}\left(\tau_{1}-\tau_{2}\right)}\begin{cases}
\begin{array}{c}
\left(\omega_{q}-\mathbf{h}_{CC}^{eff}+\mu\right)^{-1},\, Im\left(\omega_{q}\right)>0\\
\left(\omega_{q}-\left(\mathbf{h}_{CC}^{eff}\right)^{\dagger}+\mu\right)^{-1},\, Im\left(\omega_{q}\right)<0
\end{array}\end{cases}
\end{equation}

\[
\mathbf{G}_{CC}^{\urcorner}\left(t_{1},\tau_{2}\right)=e^{-i\mathbf{\widetilde{h}}^{eff}\left(t_{1}-t_{0}\right)}e^{-i\varphi_{C}\left(t_{1},t_{0}\right)}\left[\mathbf{G}^{M}\left(0^{+},\tau_{2}\right)-i\intop_{t_{0}}^{t_{1}}d\bar{t}e^{i\mathbf{\widetilde{h}}^{eff}\left(\bar{t}-t_{0}\right)}e^{i\varphi_{C}\left(\bar{t},t_{0}\right)}\left[\mathbf{\Sigma}^{\urcorner}\star\mathbf{G}^{M}\right]_{\left(\bar{t},\tau_{2}\right)}\right]
\]

\begin{equation}
\mathbf{G}_{CC}^{\ulcorner}\left(\tau_{1},t_{2}\right)=\left[\mathbf{G}^{M}\left(\tau_{1},0^{+}\right)+i\intop_{t_{0}}^{t_{2}}d\bar{t}\left[\mathbf{G}^{M}\star\mathbf{\Sigma}^{\ulcorner}\right]_{\left(\tau_{1},\bar{t}\right)}e^{-i\left(\mathbf{\widetilde{h}}^{eff}\right)^{\dagger}\left(\bar{t}-t_{0}\right)}e^{-i\varphi_{C}\left(\bar{t},t_{0}\right)}\right]e^{i\left(\mathbf{\widetilde{h}}^{eff}\right)^{\dagger}\left(t_{2}-t_{0}\right)}e^{i\varphi_{C}\left(t_{2},t_{0}\right)}
\end{equation}

\begin{equation}
\mathbf{\Sigma}_{\alpha}^{r}\left(t_{1},t_{2}\right)=-\frac{i\mathbf{\Gamma}_{\alpha}}{2}\delta\left(t_{1}-t_{2}\right)\label{eq:retardedself}
\end{equation}

\begin{equation}
\mathbf{\Sigma}_{\alpha}^{a}\left(t_{1},t_{2}\right)=\frac{i\mathbf{\Gamma}_{\alpha}}{2}\delta\left(t_{1}-t_{2}\right)\label{eq:advancedself}
\end{equation}

\begin{equation}
\mathbf{\Sigma}_{\alpha}^{M}\left(\tau_{1},\tau_{2}\right)=\frac{i}{\beta}\underset{q=-\infty}{\overset{\infty}{\sum}}e^{-\omega_{q}\left(\tau_{1}-\tau_{2}\right)}\frac{i\mathbf{\Gamma}_{\alpha}}{2}\begin{cases}
\begin{array}{c}
-1,\, Im\left(\omega_{q}\right)>0\\
+1,\, Im\left(\omega_{q}\right)<0
\end{array}\end{cases}
\end{equation}

\begin{equation}
\mathbf{\Sigma}_{\alpha}^{\urcorner}\left(t_{1},\tau_{2}\right)=\frac{i\mathbf{\Gamma}_{\alpha}}{\beta}\underset{q=-\infty}{\overset{\infty}{\sum}}e^{-i\psi_{\alpha}\left(t_{1},t_{0}\right)}e^{\omega_{q}\tau_{2}}\int\frac{d\omega}{2\pi}\frac{e^{-i\omega\left(t_{1}-t_{0}\right)}}{\omega_{q}-\omega+\mu_{\alpha}}
\end{equation}

\begin{equation}
\mathbf{\Sigma}_{\alpha}^{\ulcorner}\left(\tau_{1},t_{2}\right)=\frac{i\mathbf{\Gamma}_{\alpha}}{\beta}\underset{q=-\infty}{\overset{\infty}{\sum}}e^{-\omega_{q}\tau_{1}}e^{i\psi_{\alpha}\left(t_{2},t_{0}\right)}\int\frac{d\omega}{2\pi}\frac{e^{i\omega\left(t_{2}-t_{0}\right)}}{\omega_{q}-\omega+\mu_{\alpha}}
\end{equation}

\begin{equation}
\mathbf{\Sigma}_{\alpha}^{\lessgtr}\left(t_{1},t_{2}\right)=\pm i\mathbf{\Gamma}_{\alpha}e^{-i\psi_{\alpha}\left(t_{1},t_{2}\right)}\int\frac{d\omega}{2\pi}f\left(\pm\left(\omega-\mu\right)\right)e^{-i\omega\left(t_{1}-t_{2}\right)}\label{eq:lessgreatself}
\end{equation}

\section{\label{sec:Appendix-C} - Steady-State Results}

In the limits of long time and static bias, the general WBLA formula
for the two-time correlation given in Eq. (\ref{eq:corrcompact})
may be mapped to the frequency domain as a summation over five terms:
\begin{equation}
C_{\alpha\beta}\left(\Omega\right)=\underset{i=1}{\overset{5}{\sum}}C_{\alpha\beta}^{\left(i\right)}\left(\Omega\right)\label{eq:cnonsym}
\end{equation}
The first term is nonzero only when $\alpha=\beta$

\[
C_{\alpha\beta}^{\left(1\right)}\left(\Omega\right)=\underset{t_{0}\rightarrow-\infty,V_{\alpha}\left(t\right)\rightarrow V_{\alpha}}{\lim}\int d\tau e^{i\Omega\tau}\delta_{\alpha\beta}4q^{2}\mathrm{Tr}_{C}\left[\mathbf{\Sigma}_{\alpha}^{>}\left(t_{1},t_{2}\right)\mathbf{G}_{CC}^{<}\left(t_{2},t_{1}\right)+\mathbf{G}_{CC}^{>}\left(t_{1},t_{2}\right)\mathbf{\Sigma}_{\alpha}^{<}\left(t_{2},t_{1}\right)\right]
\]

\begin{equation}
=\delta_{\alpha\beta}4q^{2}\underset{\gamma}{\sum}\int\frac{d\omega}{2\pi}\left[\left(1-f_{\alpha}\left(\omega+\Omega-\mu\right)\right)f_{\gamma}\left(\omega-\mu\right)+\left(1-f_{\gamma}\left(\omega-\mu\right)\right)f_{\alpha}\left(\omega-\Omega-\mu\right)\right]T_{\gamma\alpha}\left(\omega\right)\label{eq:C1}
\end{equation}
where we have introduced the transmission probability 

\begin{equation}
T_{\gamma\alpha}\left(\omega\right)\equiv\mbox{Tr}_{C}\left[\mathbf{\Gamma}_{\alpha}\mathbf{G}^{r}\left(\omega\right)\mathbf{\Gamma}_{\gamma}\mathbf{G}^{a}\left(\omega\right)\right]\label{eq:transmission}
\end{equation}
Following the interpretative scheme of Ref. \cite{Zamoum2015}, we
identify the physical origin of this term in processes involving the
excitation and propagation of a quasiparticle electron-hole pair,
one of which is excited by an energy of $\hbar\Omega$ with respect
to the other. The other terms in $C_{\alpha\beta}\left(\Omega\right)$
occur in higher orders of the level width and involve more complicated
electron-hole energy transfer processes:

\[
C_{\alpha\beta}^{\left(2\right)}\left(\Omega\right)=\underset{t_{0}\rightarrow-\infty,V_{\alpha}\left(t\right)\rightarrow V_{\alpha}}{\lim}\int d\tau e^{i\Omega\tau}4q^{2}\mathrm{Tr}_{C}\left[\mathbf{\Gamma}_{\alpha}\mathbf{G}_{CC}^{>}\left(t+\tau,t\right)\mathbf{\Gamma}_{\beta}\mathbf{G}_{CC}^{<}\left(t,t+\tau\right)\right]
\]

\begin{equation}
=4q^{2}\underset{\gamma,\gamma'}{\sum}\int\frac{d\omega}{2\pi}\left(1-f_{\gamma}\left(\omega-\mu\right)\right)f_{\gamma'}\left(\omega-\Omega-\mu\right)\textrm{Tr}_{C}\left[\mathbf{T}_{CC}^{\left(\alpha\gamma\right)}\left(\omega\right)\mathbf{T}_{CC}^{\dagger\left(\beta\gamma\right)}\left(\omega\right)\mathbf{T}_{CC}^{\left(\beta\gamma'\right)}\left(\omega-\Omega\right)\mathbf{T}_{CC}^{\dagger\left(\alpha\gamma'\right)}\left(\omega-\Omega\right)\right]\label{eq:C2}
\end{equation}

\[
C_{\alpha\beta}^{\left(3\right)}\left(\Omega\right)=\underset{t_{0}\rightarrow-\infty,V_{\alpha}\left(t\right)\rightarrow V_{\alpha}}{\lim}\int d\tau e^{i\Omega\tau}i4q^{2}\mathrm{Tr}_{C}\left[\mathbf{G}_{CC}^{>}\left(t+\tau,t\right)\left[\mathbf{\Lambda}_{\beta}^{+}\left(t,t+\tau\right)\mathbf{\Gamma}_{\alpha}+\mathbf{\Gamma}_{\beta}\left(\mathbf{\Lambda}_{\alpha}^{+}\right)^{\dagger}\left(t+\tau,t\right)\right]\right]
\]

\[
=i4q^{2}\underset{\gamma}{\sum}\int\frac{d\omega}{2\pi}\mathrm{Tr}_{C}\left[\left(1-f_{\gamma}\left(\omega-\mu\right)\right)f_{\beta}\left(\omega-\Omega-\mu\right)\mathbf{A}_{\gamma}\left(\omega\right)\mathbf{\Gamma}_{\beta}\mathbf{G}_{CC}^{a}\left(\omega-\Omega\right)\mathbf{\Gamma}_{\alpha}\right.
\]

\begin{equation}
\left.-\left(1-f_{\gamma}\left(\omega-\mu\right)\right)f_{\alpha}\left(\omega-\Omega-\mu\right)\mathbf{A}_{\gamma}\left(\omega\right)\mathbf{\Gamma}_{\beta}\mathbf{G}_{CC}^{r}\left(\omega-\Omega\right)\mathbf{\Gamma}_{\alpha}\right]\label{eq:C3}
\end{equation}

\[
C_{\alpha\beta}^{\left(4\right)}\left(\Omega\right)=\underset{t_{0}\rightarrow-\infty,V_{\alpha}\left(t\right)\rightarrow V_{\alpha}}{\lim}\int d\tau e^{i\Omega\tau}i4q^{2}\mathrm{Tr}_{C}\left[\left[\mathbf{\Lambda}_{\alpha}^{-}\left(t+\tau,t\right)\mathbf{\Gamma}_{\beta}+\mathbf{\Gamma}_{\alpha}\left(\mathbf{\Lambda}_{\beta}^{-}\right)^{\dagger}\left(t,t+\tau\right)\right]\mathbf{G}_{CC}^{<}\left(t,t+\tau\right)\right]
\]

\[
=i4q^{2}\underset{\gamma}{\sum}\int\frac{d\omega}{2\pi}\mathrm{Tr}_{C}\left[\left(1-f_{\alpha}\left(\omega-\mu\right)\right)f_{\gamma}\left(\omega-\Omega-\mu\right)\mathbf{\Gamma}_{\alpha}\mathbf{G}_{CC}^{a}\left(\omega\right)\mathbf{\Gamma}_{\beta}\mathbf{A}_{\gamma}\left(\omega-\Omega\right)\right.
\]

\begin{equation}
\left.-\left(1-f_{\beta}\left(\omega-\mu\right)\right)f_{\gamma}\left(\omega-\Omega-\mu\right)\mathbf{\Gamma}_{\alpha}\mathbf{G}_{CC}^{r}\left(\omega\right)\mathbf{\Gamma}_{\beta}\mathbf{A}_{\gamma}\left(\omega-\Omega\right)\right]\label{eq:C3-1}
\end{equation}

\[
C_{\alpha\beta}^{\left(5\right)}\left(\Omega\right)=\underset{t_{0}\rightarrow-\infty,V_{\alpha}\left(t\right)\rightarrow V_{\alpha}}{\lim}-\int d\tau e^{i\Omega\tau}i4q^{2}\mathrm{Tr}_{C}\left[\mathbf{\Lambda}_{\beta}^{+}\left(t,t+\tau\right)\mathbf{\Lambda}_{\alpha}^{-}\left(t+\tau,t\right)+\left(\mathbf{\Lambda}_{\alpha}^{+}\right)^{\dagger}\left(t+\tau,t\right)\left(\mathbf{\Lambda}_{\beta}^{-}\right)^{\dagger}\left(t,t+\tau\right)\right]
\]

\[
=-4q^{2}\underset{\gamma}{\sum}\int\frac{d\omega}{2\pi}\mathrm{Tr}_{C}\left[f_{\beta}\left(\omega-\Omega-\mu\right)\left(1-f_{\alpha}\left(\omega-\mu\right)\right)\mathbf{\Gamma}_{\alpha}\mathbf{G}_{CC}^{a}\left(\omega\right)\mathbf{\Gamma}_{\beta}\mathbf{G}_{CC}^{a}\left(\omega-\Omega\right)\right.
\]

\begin{equation}
\left.+f_{\alpha}\left(\omega-\Omega-\mu\right)\left(1-f_{\beta}\left(\omega-\mu\right)\right)\mathbf{\Gamma}_{\alpha}\mathbf{G}_{CC}^{r}\left(\omega\right)\mathbf{\Gamma}_{\beta}\mathbf{G}_{CC}^{r}\left(\omega-\Omega\right)\right]\label{eq:C3-1-1}
\end{equation}
These formulas are then substituted into Eq. (\ref{eq:cnonsym}) to
get (\ref{eq:CautoZhang}) after some lengthy algebra.

\section{\label{sec:Appendix-D} - Formulas for a fast Numerical Implementation}

In this appendix, we provide exact formulas for all terms appearing
in Eq. (\ref{eq:corrcompact}). Our method is based on the fact that
we can expand the Fermi function into a series expansion whose terms
possess a simple pole structure \cite{Ozaki2007}:

\begin{equation}
f\left(x\right)=\frac{1}{e^{\beta x}+1}=\frac{1}{2}-\underset{N_{p}\rightarrow\infty}{\lim}\underset{l=1}{\overset{N_{p}}{\sum}}\eta_{l}\left(\frac{1}{\beta x+i\zeta_{l}}+\frac{1}{\beta x-i\zeta_{l}}\right)\label{eq:PADE}
\end{equation}
When the parameter values are $\eta_{l}=1$ and $\zeta_{l}=\pi\left(2l-1\right)$,
this is referred to as the Matsubara expansion, but one can also improve
the convergence of this series for finite $N_{p}$ by expressing the
Fermi function as a finite continued fraction, and then poles of the
Fermi function can be found as the solution to an eigenproblem for
a tridiagonal matrix \cite{Croy2009b,Hu2010,Hu2011}. From the Matsubara
expansion, one can write the lesser/greater self energies as follows:

\begin{equation}
\mathbf{\Sigma}_{\alpha}^{\lessgtr}\left(t_{1},t_{2}\right)=\pm i\frac{\mathbf{\Gamma}_{\alpha}}{2}\delta\left(t_{1}-t_{2}\right)-\mathbf{\Gamma}_{\alpha}e^{-i\psi_{\alpha}\left(t_{1},t_{2}\right)}e^{-i\mu\left(t_{1}-t_{2}\right)}\mbox{cosech}\left.\left(\frac{\pi}{\beta}\left(t_{1}-t_{2}\right)\right)\right|_{t_{1}\neq t_{2}}\label{eq:selflessgreat}
\end{equation}
where we define $\mbox{cosech}\left.\left(\frac{\pi}{\beta}\left(t_{1}-t_{2}\right)\right)\right|_{t_{1}\neq t_{2}}$
such that it is equal to zero when $t_{1}=t_{2}$. In practice this
function is implemented using the Pad� parameters as in Ref. \cite{Ridley}

\begin{equation}
\mbox{cosech}\left.\left(\frac{\pi}{\beta}\left(t_{1}-t_{2}\right)\right)\right|_{t_{1}\neq t_{2}}\simeq2\underset{l=1}{\overset{N_{p}}{\sum}}\left[\theta\left(t_{1}-t_{2}\right)e^{-\frac{\zeta_{l}}{\beta}\left(t_{1}-t_{2}\right)}-\theta\left(t_{2}-t_{1}\right)e^{-\frac{\zeta_{l}}{\beta}\left(t_{2}-t_{1}\right)}\right]\label{eq:cosech}
\end{equation}
where the step function is defined by the midpoint convention:
\begin{equation}
\theta\left(x\right)=\begin{cases}
\begin{array}{c}
1,\, x>0\\
\frac{1}{2},\, x=0\\
0,\, x<0
\end{array}\end{cases}\label{eq:theta}
\end{equation}
This evaluation in Eq. (\ref{eq:cosech}) is extremely precise at
large $t_{1}-t_{2}$, but diverges less rapidly than the true $\mbox{cosech}$
at $t_{1}\sim t_{2}$, thus avoiding numerical errors in the integration.
We remark that the delta function in the first term of Eq. (\ref{eq:selflessgreat})
is the reason for the divergence in the current autocorrelation at
$t_{1}=t_{2}$ in the WBLA.

The effective Hamiltonian $\mathbf{h}_{CC}^{eff}$ is non-Hermitian.
We introduce the left and right eigenvectors of this, which are known
to share the same eigenvalues \cite{Tuovinen2014}: 

\begin{eqnarray}
\mathbf{h}_{CC}^{eff}\left|\varphi_{j}^{R}\right\rangle  & = & \bar{\varepsilon}_{j}\left|\varphi_{j}^{R}\right\rangle \,\,\mbox{and}\,\,\left\langle \varphi_{j}^{L}\right|\mathbf{h}_{CC}^{eff}=\bar{\varepsilon}_{j}\left\langle \varphi_{j}^{L}\right|\label{eq:leftrighteigenproblem}
\end{eqnarray}
By inserting the expansion in Eq. (\ref{eq:PADE}) and removing all
frequency integrals, it is possible to evaluate exactly the $\mathbf{\Lambda}^{\pm}$
matrices defined in Eqs. (\ref{eq:amatrix}) and (\ref{eq:lambda2})
in terms of the so-called \textit{Hurwitz-Lerch Transcendent}\textcolor{black}{{}
$\Phi$ \cite{Lerch1887}:}

\begin{equation}
\Phi\left(z,s,a\right)\equiv\underset{n=0}{\overset{\infty}{\sum}}\frac{z^{n}}{\left(n+a\right)^{s}}\label{eq:HLT}
\end{equation}
This arises from integrals over terms of the form $e^{i\omega\tau}/\left(\omega-z\right)$,
where $z$ is a complex valued pole. Thus, we derive expressions for
the $\mathbf{\Lambda}^{\pm}$ matrices in the left/right eigenbasis
and with all frequency integrals removed, and with the partitioning
parameter $\xi_{p}$ defined in Eq. (\ref{eq:partitioningparameter})
introduced in order to distinguish between the partitioned and partition-free
switch-on processes:

\[
\mathbf{\Lambda}_{\beta}^{+}\left(t_{2},t_{1}\right)=\underset{j}{\sum}\frac{\mathbf{\Gamma}_{\beta}\left|\varphi_{j}^{L}\right\rangle \left\langle \varphi_{j}^{R}\right|}{\left\langle \varphi_{j}^{R}\right|\left.\varphi_{j}^{L}\right\rangle }\left[-\frac{i}{2\beta}\underset{t_{0}}{\overset{t_{1}}{\int}}d\tau e^{i\bar{\varepsilon}_{j}^{*}\left(t_{1}-\tau\right)}e^{-i\mu\left(t_{2}-\tau\right)}e^{-i\psi_{\beta}\left(t_{2},\tau\right)}\mbox{cosech}\left.\left(\frac{\pi}{\beta}\left(t_{2}-\tau\right)\right)\right|_{t_{2}\neq\tau}\right.
\]

\begin{equation}
\left.-\theta\left(t_{1}-t_{2}\right)\frac{e^{i\bar{\varepsilon}_{j}^{*}\left(t_{1}-t_{2}\right)}}{2}-\xi_{p}\frac{i}{2\pi}e^{i\bar{\varepsilon}_{j}^{*}\left(t_{1}-t_{0}\right)}e^{-i\mu\left(t_{2}-t_{0}\right)}e^{-i\psi_{\beta}\left(t_{2},t_{0}\right)}\bar{\Phi}\left(\beta,t_{2}-t_{0},\bar{\varepsilon}_{j}^{*}-\mu\right)\right]\label{eq:amatrix-1}
\end{equation}

\[
\left(\mathbf{\Lambda}_{\alpha}^{+}\right)^{\dagger}\left(t_{1},t_{2}\right)=\underset{j}{\sum}\frac{\left|\varphi_{j}^{R}\right\rangle \left\langle \varphi_{j}^{L}\right|\mathbf{\Gamma}_{\alpha}}{\left\langle \varphi_{j}^{L}\right|\left.\varphi_{j}^{R}\right\rangle }\left[\frac{i}{2\beta}\underset{t_{0}}{\overset{t_{2}}{\int}}d\tau e^{-i\bar{\varepsilon}_{j}\left(t_{2}-\tau\right)}e^{i\mu\left(t_{1}-\tau\right)}e^{i\psi_{\alpha}\left(t_{1},\tau\right)}\mbox{cosech}\left.\left(\frac{\pi}{\beta}\left(t_{1}-\tau\right)\right)\right|_{t_{1}\neq\tau}\right.
\]

\begin{equation}
\left.-\theta\left(t_{2}-t_{1}\right)\frac{e^{-i\bar{\varepsilon}_{j}\left(t_{2}-t_{1}\right)}}{2}+\xi_{p}\frac{i}{2\pi}e^{-i\bar{\varepsilon}_{j}\left(t_{2}-t_{0}\right)}e^{i\mu\left(t_{1}-t_{0}\right)}e^{i\psi_{\alpha}\left(t_{1},t_{0}\right)}\bar{\Phi}\left(\beta,t_{1}-t_{0},-\left(\bar{\varepsilon}_{j}-\mu\right)\right)\right]
\end{equation}

\[
\mathbf{\Lambda}_{\alpha}^{-}\left(t_{1},t_{2}\right)=\underset{j}{\sum}\frac{\mathbf{\Gamma}_{\alpha}\left|\varphi_{j}^{L}\right\rangle \left\langle \varphi_{j}^{R}\right|}{\left\langle \varphi_{j}^{R}\right|\left.\varphi_{j}^{L}\right\rangle }\left[-\frac{i}{2\beta}\underset{t_{0}}{\overset{t_{2}}{\int}}d\tau e^{i\bar{\varepsilon}_{j}^{*}\left(t_{2}-\tau\right)}e^{-i\mu\left(t_{1}-\tau\right)}e^{-i\psi_{\alpha}\left(t_{1},\tau\right)}\mbox{cosech}\left.\left(\frac{\pi}{\beta}\left(t_{1}-\tau\right)\right)\right|_{t_{1}\neq\tau}\right.
\]

\begin{equation}
\left.+\theta\left(t_{2}-t_{1}\right)\frac{e^{i\bar{\varepsilon}_{j}^{*}\left(t_{2}-t_{1}\right)}}{2}-\xi_{p}\frac{i}{2\pi}e^{i\bar{\varepsilon}_{j}^{*}\left(t_{2}-t_{0}\right)}e^{-i\mu\left(t_{1}-t_{0}\right)}e^{-i\psi_{\alpha}\left(t_{1},t_{0}\right)}\bar{\Phi}\left(\beta,t_{1}-t_{0},\bar{\varepsilon}_{j}^{*}-\mu\right)\right]\label{eq:bmatrix-1}
\end{equation}

\[
\left(\mathbf{\Lambda}_{\beta}^{-}\right)^{\dagger}\left(t_{2},t_{1}\right)=\underset{j}{\sum}\frac{\left|\varphi_{j}^{R}\right\rangle \left\langle \varphi_{j}^{L}\right|\mathbf{\Gamma}_{\beta}}{\left\langle \varphi_{j}^{L}\right|\left.\varphi_{j}^{R}\right\rangle }\left[\frac{i}{2\beta}\underset{t_{0}}{\overset{t_{1}}{\int}}d\tau e^{-i\bar{\varepsilon}_{j}\left(t_{1}-\tau\right)}e^{i\mu\left(t_{2}-\tau\right)}e^{i\psi_{\beta}\left(t_{2},\tau\right)}\mbox{cosech}\left.\left(\frac{\pi}{\beta}\left(t_{2}-\tau\right)\right)\right|_{t_{2}\neq\tau}\right.
\]

\begin{equation}
\left.+\theta\left(t_{1}-t_{2}\right)\frac{e^{-i\bar{\varepsilon}_{j}\left(t_{1}-t_{2}\right)}}{2}+\xi_{p}\frac{i}{2\pi}e^{-i\bar{\varepsilon}_{j}\left(t_{1}-t_{0}\right)}e^{i\mu\left(t_{2}-t_{0}\right)}e^{i\psi_{\beta}\left(t_{2},t_{0}\right)}\bar{\Phi}\left(\beta,t_{2}-t_{0},-\left(\bar{\varepsilon}_{j}-\mu\right)\right)\right]
\end{equation}
Here we have defined the following compact object in terms of the
Hurwitz-Lerch Transcendent: 
\begin{equation}
\bar{\Phi}\left(\beta,\tau,z\right)\equiv\exp\left(-\frac{\pi}{\beta}\tau\right)\Phi\left(e^{-\frac{2\pi\tau}{\beta}},1,\frac{1}{2}+\frac{\beta z}{2i\pi}\right)\label{eq:phigen-1}
\end{equation}

In addition, the lesser and greater Green's functions can be put into
a convenient form for the numerical evaluation:

\begin{eqnarray*}
\mathbf{G}_{CC}^{\gtrless}\left(t_{1},t_{2}\right) & = & \frac{1}{2\pi}\underset{\gamma,k,j}{\sum}\frac{\left|\varphi_{j}^{R}\right\rangle \left\langle \varphi_{j}^{L}\right|\mathbf{\Gamma}_{\gamma}\left|\varphi_{k}^{L}\right\rangle \left\langle \varphi_{k}^{R}\right|}{\left\langle \varphi_{j}^{L}\mid\varphi_{j}^{R}\right\rangle \left\langle \varphi_{k}^{R}\mid\varphi_{k}^{L}\right\rangle }e^{-i\varphi_{c}\left(t_{1},t_{2}\right)}e^{-i\bar{\varepsilon}_{j}\left(t_{1}-t_{0}\right)}e^{i\bar{\varepsilon}_{k}^{*}\left(t_{2}-t_{0}\right)}
\end{eqnarray*}

\[
\times\left\{ \frac{i}{\bar{\varepsilon}_{k}^{*}-\bar{\varepsilon}_{j}}\left[\Psi\left(\frac{1}{2}+\frac{\beta}{2i\pi}\left(\bar{\varepsilon}_{k}^{*}-\mu\right)\right)-\Psi\left(\frac{1}{2}-\frac{\beta}{2i\pi}\left(\bar{\varepsilon}_{j}-\mu\right)\right)\right]\right.
\]

\[
\pm\frac{\pi}{\bar{\varepsilon}_{k}^{*}-\bar{\varepsilon_{j}}}\left[\theta\left(t_{1}-t_{2}\right)e^{i\left(\bar{\varepsilon}_{j}-\bar{\varepsilon}_{k}^{*}\right)\left(t_{2}-t_{0}\right)}+\theta\left(t_{2}-t_{1}\right)e^{i\left(\bar{\varepsilon}_{j}-\bar{\varepsilon}_{k}^{*}\right)\left(t_{1}-t_{0}\right)}\right]
\]
\[
-\xi_{p}\left(\underset{t_{0}}{\overset{t_{1}}{\int}}d\tau e^{i\left(\bar{\varepsilon}_{j}-\mu\right)\left(\tau-t_{0}\right)}e^{i\left(\varphi_{c}-\psi_{\gamma}\right)\left(\tau,t_{0}\right)}\bar{\Phi}\left(\beta,\tau-t_{0},\bar{\varepsilon}_{k}^{*}-\mu\right)-c.c._{j\leftrightarrow k,t_{1}\leftrightarrow t_{2}}\right)
\]

\[
-\frac{2\pi}{\beta}\left[\theta\left(t_{1}-t_{2}\right)I\left(t_{2},\beta,\mu,\bar{\varepsilon}_{j},\bar{\varepsilon}_{k}^{*}\right)+\theta\left(t_{2}-t_{1}\right)I\left(t_{1},\beta,\mu,\bar{\varepsilon}_{j},\bar{\varepsilon}_{k}^{*}\right)\right]
\]

\begin{equation}
\left.-\frac{2\pi}{\beta}\underset{l}{\sum}\eta_{l}\left[\theta\left(t_{1}-t_{2}\right)\underset{t_{2}}{\overset{t_{1}}{\int}}d\tau\underset{t_{0}}{\overset{t_{2}}{\int}}d\bar{\tau}e^{i\left(\bar{\varepsilon}_{j}-\mu+i\frac{\zeta_{l}}{\beta}\right)\left(\tau-t_{0}\right)}e^{-i\left(\bar{\varepsilon}_{k}^{*}-\mu+i\frac{\zeta_{l}}{\beta}\right)\left(\bar{\tau}-t_{0}\right)}e^{i\left(\varphi_{c}-\psi_{\gamma}\right)\left(\tau,\bar{\tau}\right)}-c.c._{j\leftrightarrow k,t_{1}\leftrightarrow t_{2}}\right]\right\} \label{eq:ggreatlessexpanded-1}
\end{equation}
where $c.c._{j\leftrightarrow k,t_{1}\leftrightarrow t_{2}}$ denotes
the complex conjugation of the preceding term with both the $j$ and
$k$ indices and the times $t_{1}$ and $t_{2}$ exchanged, and we
have defined the function:

\begin{equation}
I\left(t,\beta,\mu,\bar{\varepsilon}_{j},\bar{\varepsilon}_{k}^{*}\right)=\underset{t_{0}}{\overset{t}{\int}}d\tau\underset{t_{0}}{\overset{t}{\int}}d\bar{\tau}e^{i\left(\bar{\varepsilon}_{j}-\mu\right)\left(\tau-t_{0}\right)}e^{-i\left(\bar{\varepsilon}_{k}^{*}-\mu\right)\left(\bar{\tau}-t_{0}\right)}e^{i\left(\varphi_{c}-\psi_{\gamma}\right)\left(\tau,\bar{\tau}\right)}\mbox{cosech}\left.\left(\frac{\pi}{\beta}\left(\tau-\bar{\tau}\right)\right)\right|_{\tau\neq\bar{\tau}}\label{eq:Ifunction}
\end{equation}
Here we have introduced the \textit{digamma function} $\Psi$, defined
as the logarithmic derivative of the complex gamma function, $\Psi\left(z\right)\equiv\frac{d\ln\Gamma\left(z\right)}{dz}$
\cite{Jauho1994}. Note that we can get $\mathbf{G}^{<}$ directly
from $\mathbf{G}^{>}$ on each summation cycle via the following useful
property:

\begin{equation}
\mathbf{G}^{>}\left(t_{1},t_{2}\right)-\mathbf{G}^{<}\left(t_{1},t_{2}\right)=-ie^{-i\varphi_{C}\left(t_{1},t_{2}\right)}\underset{j}{\sum}\left[\frac{\left|\varphi_{j}^{R}\right\rangle \left\langle \varphi_{j}^{L}\right|}{\left\langle \varphi_{j}^{L}\mid\varphi_{j}^{R}\right\rangle }e^{-i\bar{\varepsilon}_{j}\left(t_{1}-t_{2}\right)}\theta\left(t_{1}-t_{2}\right)+\frac{\left|\varphi_{j}^{L}\right\rangle \left\langle \varphi_{j}^{R}\right|}{\left\langle \varphi_{j}^{R}\mid\varphi_{j}^{L}\right\rangle }e^{i\bar{\varepsilon}_{j}^{*}\left(t_{2}-t_{1}\right)}\theta\left(t_{2}-t_{1}\right)\right]\label{eq:lessgreatrelation}
\end{equation}
This means a single nested loop of calculations in the two-time plane
is sufficient to calculate both Green's functions. We then use the
fact that $\mathbf{G}^{<}\left(t_{1},t_{2}\right)=-\mathbf{G}^{<}\left(t_{2},t_{1}\right)^{\dagger}$
to get the time-reversed GFs, thus further reducing the calculation
time by a half.

\section{Formulas for the Green's Functions and Lambda Matrices in the Biharmonic
Model \label{sec:FormulasBiharmonic}}

When we substitute Eq. (\ref{eq:BESSELphase}) into Eq. (\ref{eq:ggreatlessexpanded-1}),
we obtain the following result for the greater and lesser Green's
functions:

\[
\mathbf{G}^{\gtrless}\left(t_{1},t_{2}\right)=\frac{1}{2\pi}\underset{\gamma,k,j}{\sum}\frac{\left|\varphi_{j}^{R}\right\rangle \left\langle \varphi_{j}^{L}\right|\mathbf{\Gamma}_{\gamma}\left|\varphi_{k}^{L}\right\rangle \left\langle \varphi_{k}^{R}\right|}{\left\langle \varphi_{j}^{L}\mid\varphi_{j}^{R}\right\rangle \left\langle \varphi_{k}^{R}\mid\varphi_{k}^{L}\right\rangle }\left\{ \pm\frac{\pi}{\bar{\varepsilon}_{k}^{*}-\bar{\varepsilon_{j}}}\left[\theta\left(t_{1}-t_{2}\right)e^{-i\bar{\varepsilon}_{j}\left(t_{1}-t_{2}\right)}+\theta\left(t_{2}-t_{1}\right)e^{i\bar{\varepsilon}_{k}^{*}\left(t_{2}-t_{1}\right)}\right]\right.
\]

\[
+\frac{ie^{-i\bar{\varepsilon}_{j}\left(t_{1}-t_{0}\right)}e^{i\bar{\varepsilon}_{k}^{*}\left(t_{2}-t_{0}\right)}}{\bar{\varepsilon}_{k}^{*}-\bar{\varepsilon}_{j}}\left[\Psi\left(\frac{1}{2}+\frac{\beta}{2i\pi}\left(\bar{\varepsilon}_{k}^{*}-\mu\right)\right)-\Psi\left(\frac{1}{2}-\frac{\beta}{2i\pi}\left(\bar{\varepsilon}_{j}-\mu\right)\right)\right]
\]

\[
+\xi_{p}i\underset{r,s}{\sum}J_{r}\left(\frac{A_{\gamma}^{\left(1\right)}}{p_{1}\Omega_{\gamma}}\right)J_{s}\left(\frac{A_{\gamma}^{\left(2\right)}}{p_{2}\Omega_{\gamma}}\right)\left[\frac{e^{-ir\phi_{\gamma}}e^{i\frac{A_{\gamma}^{\left(1\right)}}{p_{1}\Omega_{\gamma}}\sin\phi_{\gamma}}}{\bar{\varepsilon}_{j}-\bar{\varepsilon}_{k}^{*}-V_{\gamma}-\Omega_{\gamma}\left(p_{1}r+p_{2}s\right)}\right.
\]

\[
\times\left[e^{-i\bar{\varepsilon}_{j}\left(t_{1}-t_{0}\right)}e^{i\bar{\varepsilon}_{k}^{*}\left(t_{2}-t_{0}\right)}\left[\Psi\left(\frac{1}{2}+\frac{\beta}{2\pi i}\left(\bar{\varepsilon}_{k}^{*}-\mu\right)\right)-\Psi\left(\frac{1}{2}+\frac{\beta}{2\pi i}\left(\bar{\varepsilon}_{j}-\mu-V_{\gamma}-\Omega_{\gamma}\left(p_{1}r+p_{2}s\right)\right)\right)\right]\right.
\]

\[
\left.+e^{i\bar{\varepsilon}_{k}^{*}\left(t_{2}-t_{0}\right)}e^{-i\left(\mu+V_{\gamma}+\Omega_{\gamma}\left(p_{1}r+p_{2}s\right)\right)\left(t_{1}-t_{0}\right)}\left[\bar{\Phi}\left(t_{1}-t_{0},\beta,\bar{\varepsilon}_{k}^{*}-\mu\right)-\bar{\Phi}\left(t_{1}-t_{0},\beta,\bar{\varepsilon}_{j}-\mu-V_{\gamma}-\Omega_{\gamma}\left(p_{1}r+p_{2}s\right)\right)\right]\right]
\]

\[
+\frac{e^{ir\phi_{\gamma}}e^{-i\frac{A_{\gamma}^{\left(1\right)}}{p_{1}\Omega_{\gamma}}\sin\phi_{\gamma}}}{\bar{\varepsilon}_{k}^{*}-\bar{\varepsilon}_{j}-V_{\gamma}-\Omega_{\gamma}\left(p_{1}r+p_{2}s\right)}\left[e^{-i\bar{\varepsilon}_{j}\left(t_{1}-t_{0}\right)}e^{i\bar{\varepsilon}_{k}^{*}\left(t_{2}-t_{0}\right)}\left[\Psi\left(\frac{1}{2}-\frac{\beta}{2\pi i}\left(\bar{\varepsilon}_{j}-\mu\right)\right)-\Psi\left(\frac{1}{2}-\frac{\beta}{2\pi i}\left(\bar{\varepsilon}_{k}^{*}-\mu-V_{\gamma}-\Omega_{\gamma}\left(p_{1}r+p_{2}s\right)\right)\right)\right]\right.
\]

\[
\left.\left.+e^{-i\bar{\varepsilon}_{j}\left(t_{1}-t_{0}\right)}e^{i\left(\mu+V_{\gamma}+\Omega_{\gamma}\left(p_{1}r+p_{2}s\right)\right)\left(t_{2}-t_{0}\right)}\left[\bar{\Phi}\left(t_{2}-t_{0},\beta,-\left(\bar{\varepsilon}_{j}-\mu\right)\right)-\bar{\Phi}\left(t_{2}-t_{0},\beta,-\left(\bar{\varepsilon}_{k}^{*}-\mu-V_{\gamma}-\Omega_{\gamma}\left(p_{1}r+p_{2}s\right)\right)\right)\right]\right]\right]
\]

\[
+i\underset{r,r',s,s'}{\sum}J_{r}\left(\frac{A_{\gamma}^{\left(1\right)}}{p_{1}\Omega_{\gamma}}\right)J_{r'}\left(\frac{A_{\gamma}^{\left(1\right)}}{p_{1}\Omega_{\gamma}}\right)J_{s}\left(\frac{A_{\gamma}^{\left(2\right)}}{p_{2}\Omega_{\gamma}}\right)J_{s'}\left(\frac{A_{\gamma}^{\left(2\right)}}{p_{2}\Omega_{\gamma}}\right)\frac{e^{-i\left(r-r'\right)\phi_{\gamma}}}{\bar{\varepsilon}_{j}-\bar{\varepsilon}_{k}^{*}-\Omega_{\gamma}\left(p_{1}\left(r-r'\right)+p_{2}(s-s')\right)}
\]

\[
\times\left[e^{-i\bar{\varepsilon}_{j}\left(t_{1}-t_{0}\right)}e^{i\bar{\varepsilon}_{k}^{*}\left(t_{2}-t_{0}\right)}\left[\Psi\left(\frac{1}{2}+\frac{\beta}{2\pi i}\left(\bar{\varepsilon}_{j}-\mu-V_{\gamma}-\Omega_{\gamma}\left(p_{1}r+p_{2}s\right)\right)\right)-\Psi\left(\frac{1}{2}-\frac{\beta}{2\pi i}\left(\bar{\varepsilon}_{k}^{*}-\mu-V_{\gamma}-\Omega_{\gamma}\left(p_{1}r'+p_{2}s'\right)\right)\right)\right]\right.
\]

\[
+e^{i\bar{\varepsilon}_{k}^{*}\left(t_{2}-t_{0}\right)}e^{-i\left(\mu+V_{\gamma}+\Omega_{\gamma}\left(p_{1}r+p_{2}s\right)\right)\left(t_{1}-t_{0}\right)}\left[\bar{\Phi}\left(t_{1}-t_{0},\beta,\bar{\varepsilon}_{j}-\mu-V_{\gamma}-\Omega_{\gamma}\left(p_{1}r+p_{2}s\right)\right)\right.
\]

\[
\left.-\bar{\Phi}\left(t_{1}-t_{0},\beta,\bar{\varepsilon}_{k}^{*}-\mu-V_{\gamma}-\Omega_{\gamma}\left(p_{1}r'+p_{2}s'\right)\right)\right]
\]

\[
+e^{-i\bar{\varepsilon}_{j}\left(t_{1}-t_{0}\right)}e^{i\left(\mu+V_{\gamma}+\Omega_{\gamma}\left(p_{1}r'+p_{2}s'\right)\right)\left(t_{2}-t_{0}\right)}\left[\bar{\Phi}\left(t_{2}-t_{0},\beta,-\left(\bar{\varepsilon}_{j}-\mu-V_{\gamma}-\Omega_{\gamma}\left(p_{1}r+p_{2}s\right)\right)\right)\right.
\]

\[
\left.-\bar{\Phi}\left(t_{2}-t_{0},\beta,-\left(\bar{\varepsilon}_{k}^{*}-\mu-V_{\gamma}-\Omega_{\gamma}\left(p_{1}r'+p_{2}s'\right)\right)\right)\right]
\]

\[
+\theta\left(t_{1}-t_{2}\right)\left[e^{-i\Omega_{\gamma}\left(p_{1}r+p_{2}s\right)\left(t_{1}-t_{0}\right)}e^{i\Omega_{\gamma}\left(p_{1}r'+p_{2}s'\right)\left(t_{2}-t_{0}\right)}e^{-i\left(\mu+V_{\gamma}\right)\left(t_{1}-t_{2}\right)}\right.
\]

\[
\times\left[\bar{\Phi}\left(t_{1}-t_{2},\beta,\bar{\varepsilon}_{k}^{*}-\mu-V_{\gamma}-\Omega_{\gamma}\left(p_{1}r'+p_{2}s'\right)\right)-\bar{\Phi}\left(t_{1}-t_{2},\beta,\bar{\varepsilon}_{j}-\mu-V_{\gamma}-\Omega_{\gamma}\left(p_{1}r+p_{2}s\right)\right)\right]
\]

\[
+e^{-i\bar{\varepsilon}_{j}\left(t_{1}-t_{2}\right)}e^{-i\Omega_{\gamma}\left(p_{1}\left(r-r'\right)+p_{2}\left(s-s'\right)\right)\left(t_{2}-t_{0}\right)}\left[\Psi\left(\frac{1}{2}-\frac{\beta}{2\pi i}\left(\bar{\varepsilon}_{j}-\mu-V_{\gamma}-\Omega_{\gamma}\left(p_{1}r+p_{2}s\right)\right)\right)\right.
\]

\[
\left.\left.-\Psi\left(\frac{1}{2}+\frac{\beta}{2\pi i}\left(\bar{\varepsilon}_{j}-\mu-V_{\gamma}-\Omega_{\gamma}\left(p_{1}r+p_{2}s\right)\right)\right)\right]\right]
\]

\[
+\theta\left(t_{2}-t_{1}\right)\left[e^{-i\Omega_{\gamma}\left(p_{1}r+p_{2}s\right)\left(t_{1}-t_{0}\right)}e^{i\Omega_{\gamma}\left(p_{1}r'+p_{2}s'\right)\left(t_{2}-t_{0}\right)}e^{-i\left(\mu+V_{\gamma}\right)\left(t_{1}-t_{2}\right)}\right.
\]

\[
\times\left[\bar{\Phi}\left(t_{2}-t_{1},\beta,-\left(\bar{\varepsilon}_{k}^{*}-\mu-V_{\gamma}-\Omega_{\gamma}\left(p_{1}r'+p_{2}s'\right)\right)\right)-\bar{\Phi}\left(t_{2}-t_{1},\beta,-\left(\bar{\varepsilon}_{j}-\mu-V_{\gamma}-\Omega_{\gamma}\left(p_{1}r+p_{2}s\right)\right)\right)\right]
\]

\[
+e^{i\bar{\varepsilon}_{k}^{*}\left(t_{2}-t_{1}\right)}e^{-i\Omega_{\gamma}\left(p_{1}\left(r-r'\right)+p_{2}\left(s-s'\right)\right)\left(t_{1}-t_{0}\right)}\left[\Psi\left(\frac{1}{2}-\frac{\beta}{2\pi i}\left(\bar{\varepsilon}_{k}^{*}-\mu-V_{\gamma}-\Omega_{\gamma}\left(p_{1}r+p_{2}s\right)\right)\right)\right.
\]

\begin{equation}
\left.\left.\left.\left.-\Psi\left(\frac{1}{2}+\frac{\beta}{2\pi i}\left(\bar{\varepsilon}_{k}^{*}-\mu-V_{\gamma}-\Omega_{\gamma}\left(p_{1}r+p_{2}s\right)\right)\right)\right]\right]\right]\right\} \label{eq:GFgreatlessBESS}
\end{equation}
The formula for the current in lead $\alpha$ can be similarly derived
by inserting Eq. (\ref{eq:BESSELphase}) into Eq. (22) of Ref. \cite{Ridley};
we defer its publication to a forthcoming paper on the quantum pump.
To evaluate the two-time current correlation function in Eq. (\ref{eq:corrcompact})
for the biharmonic driving model, it is necessary to evaluate the
integral appearing in the expression (\ref{eq:amatrix-1}) for $\mathbf{\Lambda}_{\beta}^{+}\left(t_{2},t_{1}\right)$.
Expanding the integrand using Eq. (\ref{eq:BESSELphase}), we obtain:

\[
\underset{t_{0}}{\overset{t_{1}}{\int}}d\tau e^{i\bar{\varepsilon}_{j}^{*}\left(t_{1}-\tau\right)}e^{-i\mu\left(t_{2}-\tau\right)}e^{-i\psi_{\beta}\left(t_{2},\tau\right)}\mbox{cosech}\left.\left(\frac{\pi}{\beta}\left(t_{2}-\tau\right)\right)\right|_{t_{2}\neq\tau}
\]

\[
=\frac{\beta}{\pi}e^{i\bar{\varepsilon}_{j}^{*}\left(t_{1}-t_{0}\right)}e^{-i\left(\mu+V_{\beta}\right)\left(t_{2}-t_{0}\right)}e^{-i\frac{A_{\beta}^{\left(1\right)}}{p_{1}\Omega_{\beta}}\sin\left(p_{1}\Omega_{\beta}\left(t_{2}-t_{0}\right)+\phi_{\beta}\right)}e^{-i\frac{A_{\beta}^{\left(2\right)}}{p_{2}\Omega_{\beta}}\sin\left(p_{2}\Omega_{\beta}\left(t_{2}-t_{0}\right)\right)}\underset{r,s}{\sum}J_{r}\left(\frac{A_{\beta}^{\left(1\right)}}{p_{1}\Omega_{\beta}}\right)J_{s}\left(\frac{A_{\beta}^{\left(2\right)}}{p_{2}\Omega_{\beta}}\right)e^{ir\phi_{\beta}}
\]

\[
\times\left\{ \theta\left(t_{1}-t_{2}\right)e^{-i\left(\bar{\varepsilon}_{j}^{*}-\mu-V_{\beta}-\Omega_{\beta}\left(p_{1}r+p_{2}s\right)\right)\left(t_{2}-t_{0}\right)}\left[\Psi\left(\frac{1}{2}-\frac{\beta}{2\pi i}\left(\bar{\varepsilon}_{j}^{*}-\mu-V_{\beta}-\Omega_{\beta}\left(p_{1}r+p_{2}s\right)\right)\right)\right.\right.
\]

\[
\left.-\Psi\left(\frac{1}{2}+\frac{\beta}{2\pi i}\left(\bar{\varepsilon}_{j}^{*}-\mu-V_{\beta}-\Omega_{\beta}\left(p_{1}r+p_{2}s\right)\right)\right)\right]-\bar{\Phi}\left(t_{2}-t_{0},\beta,\bar{\varepsilon}_{j}^{*}-\mu-V_{\beta}-\Omega_{\beta}\left(p_{1}r+p_{2}s\right)\right)
\]

\[
+e^{-i\left(\bar{\varepsilon}_{j}^{*}-\mu-V_{\beta}-\Omega_{\beta}\left(p_{1}r+p_{2}s\right)\right)\left(t_{1}-t_{0}\right)}\left[\theta\left(t_{1}-t_{2}\right)\bar{\Phi}\left(t_{1}-t_{2},\beta,-\left(\bar{\varepsilon}_{j}^{*}-\mu-V_{\beta}-\Omega_{\beta}\left(p_{1}r+p_{2}s\right)\right)\right)\right.
\]

\begin{equation}
\left.\left.+\theta\left(t_{2}-t_{1}\right)\bar{\Phi}\left(t_{2}-t_{1},\beta,\bar{\varepsilon}_{j}^{*}-\mu-V_{\beta}-\Omega_{\beta}\left(p_{1}r+p_{2}s\right)\right)\right]\right\} \label{eq:lambdaint-1}
\end{equation}
The integral in $\left(\mathbf{\Lambda}_{\beta}^{-}\right)^{\dagger}\left(t_{2},t_{1}\right)$
is obtained as the complex conjugate of Eq. (\ref{eq:lambdaint-1}),
the integral in $\mathbf{\Lambda}_{\alpha}^{-}\left(t_{1},t_{2}\right)$
is obtained by exchanging indices $\alpha\leftrightarrow\beta$ and
times $t_{1}\leftrightarrow t_{2}$, and the integral in $\left(\mathbf{\Lambda}_{\alpha}^{+}\right)^{\dagger}\left(t_{1},t_{2}\right)$
is obtained as the complex conjugate of the latter expression. Thus,
one obtains for the $\mathbf{\Lambda}^{\pm}$ matrices:

\[
\mathbf{\Lambda}_{\beta}^{+}\left(t_{2},t_{1}\right)=\underset{j}{\sum}\frac{\mathbf{\Gamma}_{\beta}\left|\varphi_{j}^{L}\right\rangle \left\langle \varphi_{j}^{R}\right|}{\left\langle \varphi_{j}^{R}\right|\left.\varphi_{j}^{L}\right\rangle }\left[-\frac{i}{2\pi}e^{i\bar{\varepsilon}_{j}^{*}\left(t_{1}-t_{0}\right)}e^{-i\left(\mu+V_{\beta}\right)\left(t_{2}-t_{0}\right)}e^{-i\frac{A_{\beta}^{\left(1\right)}}{p_{1}\Omega_{\beta}}\sin\left(p_{1}\Omega_{\beta}\left(t_{2}-t_{0}\right)+\phi_{\beta}\right)}e^{-i\frac{A_{\beta}^{\left(2\right)}}{p_{2}\Omega_{\beta}}\sin\left(p_{2}\Omega_{\beta}\left(t_{2}-t_{0}\right)\right)}\right.
\]

\[
\times\left[\xi_{p}e^{i\frac{A_{\beta}^{\left(1\right)}}{p_{1}\Omega_{\beta}}\sin\left(\phi_{\beta}\right)}\bar{\Phi}\left(\beta,t_{2}-t_{0},\bar{\varepsilon}_{j}^{*}-\mu\right)+\underset{r,s}{\sum}J_{r}\left(\frac{A_{\beta}^{\left(1\right)}}{p_{1}\Omega_{\beta}}\right)J_{s}\left(\frac{A_{\beta}^{\left(2\right)}}{p_{2}\Omega_{\beta}}\right)e^{ir\phi_{\beta}}\left\{ \theta\left(t_{1}-t_{2}\right)e^{-i\left(\bar{\varepsilon}_{j}^{*}-\mu-V_{\beta}-\Omega_{\beta}\left(p_{1}r+p_{2}s\right)\right)\left(t_{2}-t_{0}\right)}\right.\right.
\]

\[
\times\left[\Psi\left(\frac{1}{2}-\frac{\beta}{2\pi i}\left(\bar{\varepsilon}_{j}^{*}-\mu-V_{\beta}-\Omega_{\beta}\left(p_{1}r+p_{2}s\right)\right)\right)-\Psi\left(\frac{1}{2}+\frac{\beta}{2\pi i}\left(\bar{\varepsilon}_{j}^{*}-\mu-V_{\beta}-\Omega_{\beta}\left(p_{1}r+p_{2}s\right)\right)\right)\right]
\]

\[
-\bar{\Phi}\left(t_{2}-t_{0},\beta,\bar{\varepsilon}_{j}^{*}-\mu-V_{\beta}-\Omega_{\beta}\left(p_{1}r+p_{2}s\right)\right)
\]

\[
+e^{-i\left(\bar{\varepsilon}_{j}^{*}-\mu-V_{\beta}-\Omega_{\beta}\left(p_{1}r+p_{2}s\right)\right)\left(t_{1}-t_{0}\right)}\left[\theta\left(t_{1}-t_{2}\right)\bar{\Phi}\left(t_{1}-t_{2},\beta,-\left(\bar{\varepsilon}_{j}^{*}-\mu-V_{\beta}-\Omega_{\beta}\left(p_{1}r+p_{2}s\right)\right)\right)\right.
\]

\begin{equation}
\left.\left.\left.\left.+\theta\left(t_{2}-t_{1}\right)\bar{\Phi}\left(t_{2}-t_{1},\beta,\bar{\varepsilon}_{j}^{*}-\mu-V_{\beta}-\Omega_{\beta}\left(p_{1}r+p_{2}s\right)\right)\right]\right\} \right]-\theta\left(t_{1}-t_{2}\right)\frac{e^{i\bar{\varepsilon}_{j}^{*}\left(t_{1}-t_{2}\right)}}{2}\right]\label{eq:lambda1BESS-1}
\end{equation}

\[
\mathbf{\Lambda}_{\alpha}^{-}\left(t_{1},t_{2}\right)=\underset{j}{\sum}\frac{\mathbf{\Gamma}_{\alpha}\left|\varphi_{j}^{L}\right\rangle \left\langle \varphi_{j}^{R}\right|}{\left\langle \varphi_{j}^{R}\right|\left.\varphi_{j}^{L}\right\rangle }\left[-\frac{i}{2\pi}e^{i\bar{\varepsilon}_{j}^{*}\left(t_{2}-t_{0}\right)}e^{-i\left(\mu+V_{\alpha}\right)\left(t_{1}-t_{0}\right)}e^{-i\frac{A_{\alpha}^{\left(1\right)}}{p_{1}\Omega_{\alpha}}\sin\left(p_{1}\Omega_{\alpha}\left(t_{1}-t_{0}\right)+\phi_{\alpha}\right)}e^{-i\frac{A_{\alpha}^{\left(2\right)}}{p_{2}\Omega_{\alpha}}\sin\left(p_{2}\Omega_{\alpha}\left(t_{1}-t_{0}\right)\right)}\right.
\]

\[
\times\left[\xi_{p}e^{i\frac{A_{\alpha}^{\left(1\right)}}{p_{1}\Omega_{\alpha}}\sin\left(\phi_{\alpha}\right)}\bar{\Phi}\left(\beta,t_{1}-t_{0},\bar{\varepsilon}_{j}^{*}-\mu\right)+\underset{r,s}{\sum}J_{r}\left(\frac{A_{\alpha}^{\left(1\right)}}{p_{1}\Omega_{\alpha}}\right)J_{s}\left(\frac{A_{\alpha}^{\left(2\right)}}{p_{2}\Omega_{\alpha}}\right)e^{ir\phi_{\alpha}}\left\{ \theta\left(t_{2}-t_{1}\right)e^{-i\left(\bar{\varepsilon}_{j}^{*}-\mu-V_{\alpha}-\Omega_{\alpha}\left(p_{1}r+p_{2}s\right)\right)\left(t_{1}-t_{0}\right)}\right.\right.
\]

\[
\times\left[\Psi\left(\frac{1}{2}-\frac{\beta}{2\pi i}\left(\bar{\varepsilon}_{j}^{*}-\mu-V_{\alpha}-\Omega_{\alpha}\left(p_{1}r+p_{2}s\right)\right)\right)-\Psi\left(\frac{1}{2}+\frac{\beta}{2\pi i}\left(\bar{\varepsilon}_{j}^{*}-\mu-V_{\alpha}-\Omega_{\alpha}\left(p_{1}r+p_{2}s\right)\right)\right)\right]
\]

\[
-\bar{\Phi}\left(t_{1}-t_{0},\beta,\bar{\varepsilon}_{j}^{*}-\mu-V_{\alpha}-\Omega_{\alpha}\left(p_{1}r+p_{2}s\right)\right)
\]

\[
+e^{-i\left(\bar{\varepsilon}_{j}^{*}-\mu-V_{\alpha}-\Omega_{\alpha}\left(p_{1}r+p_{2}s\right)\right)\left(t_{2}-t_{0}\right)}\left[\theta\left(t_{1}-t_{2}\right)\bar{\Phi}\left(t_{1}-t_{2},\beta,\bar{\varepsilon}_{j}^{*}-\mu-V_{\alpha}-\Omega_{\alpha}\left(p_{1}r+p_{2}s\right)\right)\right.
\]

\begin{equation}
\left.\left.\left.\left.+\theta\left(t_{2}-t_{1}\right)\bar{\Phi}\left(t_{2}-t_{1},\beta,-\left(\bar{\varepsilon}_{j}^{*}-\mu-V_{\alpha}-\Omega_{\alpha}\left(p_{1}r+p_{2}s\right)\right)\right)\right]\right\} \right]+\theta\left(t_{2}-t_{1}\right)\frac{e^{i\bar{\varepsilon}_{j}^{*}\left(t_{2}-t_{1}\right)}}{2}\right]\label{eq:lambda2BESS-1}
\end{equation}
The matrices $\left(\mathbf{\Lambda}_{\beta}^{-}\right)^{\dagger}\left(t_{2},t_{1}\right)$
and $\left(\mathbf{\Lambda}_{\alpha}^{+}\right)^{\dagger}\left(t_{1},t_{2}\right)$
are then obtained via complex conjugation and exchange of the lead
indices $\alpha\leftrightarrow\beta$ and times $t_{1}\leftrightarrow t_{2}$
in Eqs. (\ref{eq:lambda1BESS-1}) and (\ref{eq:lambda2BESS-1}), respectively. 

\bibliographystyle{unsrt}

\begin{thebibliography}{100}

\bibitem{Bergren2016}
Adam~Johan Bergren, Lucas Zeer-Wanklyn, Mitchell Semple, Nikola Pekas, Bryan
  Szeto, and Richard~L McCreery.
\newblock {Musical molecules : the molecular junction as an active component in
  audio distortion circuits}.
\newblock {\em Journal of Physics: Condensed Matter}, 28:094011, 2016.

\bibitem{Shalf2015}
J.~M. Shalf and R.~Leland.
\newblock {Computing beyond Moore's Law}.
\newblock {\em Computer}, 48(12):14--23, 2015.

\bibitem{Selzer2013}
Yoram Selzer and Uri Peskin.
\newblock {Transient dynamics in molecular junctions: Picosecond resolution
  from dc measurements by a laser pulse pair sequence excitation}.
\newblock {\em Journal of Physical Chemistry C}, 117(43):22369--22376, 2013.

\bibitem{Aviram1974}
Arieh Aviram and Mark~A Ratner.
\newblock {Molecular Rectifiers}.
\newblock {\em Chem. Phys. Lett.}, 29(2):277--283, 1974.

\bibitem{Blum2003}
Amy~Szuchmacher Blum, John~C. Yang, Ranganathan Shashidhar, and Banahalli
  Ratna.
\newblock {Comparing the conductivity of molecular wires with the scanning
  tunneling microscope}.
\newblock {\em Applied Physics Letters}, 82(19):3322--3324, 2003.

\bibitem{Wohlgamuth2013}
Chris~H. Wohlgamuth, Marc~A. McWilliams, and Jason~D. Slinker.
\newblock {DNA as a molecular wire: Distance and sequence dependence}.
\newblock {\em Analytical Chemistry}, 85(18):8634--8640, 2013.

\bibitem{Rokhinson2000a}
L~P Rokhinson, L~J Guo, S~Y Chou, and D~C Tsui.
\newblock {Double-dot charge transport in Si single electron/hole transistors}.
\newblock {\em Applied Physics Letters}, 76(12):4, 2000.

\bibitem{Swager1998}
Timothy~M Swager.
\newblock {The Molecular Wire Approach to Sensory Signal Amplification}.
\newblock {\em Accounts of Chemical Research}, 31(5):201--207, 1998.

\bibitem{Iniguez-de-la-Torre2010}
I~I{\~{n}}iguez-de-la Torre, T~Gonz{\'{a}}lez, D~Pardo, C~Gard{\`{e}}s,
  Y~Roelens, S~Bollaert, a~Curutchet, C~Gaquiere, and J~Mateos.
\newblock {Three-terminal junctions operating as mixers, frequency doublers and
  detectors: a broad-band frequency numerical and experimental study at room
  temperature}.
\newblock {\em Semiconductor Science and Technology}, 25(12):125013, 2010.

\bibitem{JanvanderMolen2010}
Sense {Jan van der Molen} and Peter Liljeroth.
\newblock {Charge transport through molecular switches}.
\newblock {\em Journal of Physics: Condensed Matter}, 22(13):133001--133030,
  2010.

\bibitem{Liu2013}
Wei Liu, Sergey~N Filimonov, Javier Carrasco, and Alexandre Tkatchenko.
\newblock {Molecular switches from benzene derivatives adsorbed on metal
  surfaces}.
\newblock {\em Nature Communications}, 4:3569, 2013.

\bibitem{Drexler1995}
H.~Drexler, J.~S. Scott, S.~J. Allen, K.~L. Campman, and A.~C. Gossard.
\newblock {Photon-assisted tunneling in a resonant tunneling diode: Stimulated
  emission and absorption in the THz range}.
\newblock {\em Applied Physics Letters}, 67(1995):2816, 1995.

\bibitem{Covington2000}
M~Covington, Mw~Keller, Rl~Kautz, and Jm~Martinis.
\newblock {Photon-assisted tunneling in electron pumps}.
\newblock {\em Physical Review Letters}, 84(22):5192--5, 2000.

\bibitem{Li2004}
Shengdong Li, Zhen Yu, Sheng-Feng Yen, W.~C. Tang, and Peter~J. Burke.
\newblock {Carbon Nanotube Transistor Operation at 2.6 GHz}.
\newblock {\em Nano Letters}, 4(4):753--756, 2004.

\bibitem{Burke2004}
Peter~J. Burke.
\newblock {Carbon Nanotube Devices for GHz to THz Applications}.
\newblock {\em Proceedings of SPIE}, 5593:52--61, 2004.

\bibitem{Chaste2008}
J~Chaste, L~Lechner, P~Morfin, G~Feve, T~Kontos, J~M. Berroir, D~C Glattli,
  H~Happy, P~Hakonen, and B~Placais.
\newblock {Single Carbon Nanotube Transistor at GHz Frequency}.
\newblock {\em Nano Lett.}, 8(2):525--528, 2008.

\bibitem{Zhang2014}
Weidong Zhang, Phi {H. Q. Pham}, Elliott~R. Brown, and Peter {J. Burke}.
\newblock {AC conductivity parameters of graphene derived from THz etalon
  transmittance.}
\newblock {\em Nanoscale}, 6(22):13895--9, 2014.

\bibitem{Ratner2013}
Mark Ratner.
\newblock {A brief history of molecular electronics}.
\newblock {\em Nature Nanotechnology}, 8(June), 2013.

\bibitem{Blanter2000}
Ya.~M. Blanter and M.~B{\"{u}}ttiker.
\newblock {Shot noise in mesoscopic conductors}.
\newblock {\em Physics Reports}, 336(1-2):1--166, 2000.

\bibitem{Reznikov1995}
M.~Reznikov, M.~Heiblum, Hadas Shtrikman, and D.~Mahalu.
\newblock {Temporal correlation of electrons: Suppression of shot noise in a
  ballistic quantum point contact}.
\newblock {\em Physical Review Letters}, 75(18):3340--3343, 1995.

\bibitem{Kumar1996}
a.~Kumar, L.~Saminadayar, D.~Glattli, Y.~Jin, and B.~Etienne.
\newblock {Experimental Test of the Quantum Shot Noise Reduction Theory}.
\newblock {\em Physical Review Letters}, 76(15):2778--2781, 1996.

\bibitem{Saminadayar1997}
L.~Saminadayar, D.~Glattli, Y.~Jin, and B.~Etienne.
\newblock {Observation of the e/3 Fractionally Charged Laughlin Quasiparticle}.
\newblock {\em Physical Review Letters}, 79(13):2526--2529, 1997.

\bibitem{Djukic2006}
D~Djukic and J~M van Ruitenbeek.
\newblock {Shot noise measurements on a single molecule}.
\newblock {\em Nano Letters}, 6(4):789--793, 2006.

\bibitem{Ochoa2015}
Maicol~A. Ochoa, Yoram Selzer, Uri Peskin, and Michael Galperin.
\newblock {Pump-probe noise spectroscopy of molecular junctions}.
\newblock {\em Journal of Physical Chemistry Letters}, 6(3):470--476, 2015.

\bibitem{Parmentier2016}
F~D Parmentier, L~N Serkovic-Loli, P~Roulleau, and D~C Glattli.
\newblock {Photon-Assisted Shot Noise in Graphene in the Terahertz Range}.
\newblock {\em Physical Review Letters}, 227401(June):1--5, 2016.

\bibitem{Nyquist1928}
H.~Nyquist.
\newblock {Thermal agitation of electric charge in conductors}.
\newblock {\em Physical Review}, 32(1):110--113, 1928.

\bibitem{Johnson1928}
J.~B. Johnson.
\newblock {Thermal agitation of electricity in conductors}.
\newblock {\em Physical Review}, 32(1928), 1928.

\bibitem{Aguado2000}
Ram{\'{o}}n Aguado and Leo~P. Kouwenhoven.
\newblock {Double Quantum Dots as Detectors of High-Frequency Quantum Noise in
  Mesoscopic Conductors}.
\newblock {\em Physical Review Letters}, 84(9):1986--1989, 2000.

\bibitem{Gavish2000}
U.~Gavish, Y.~Levinson, and Y.~Imry.
\newblock {Detection of quantum noise}.
\newblock {\em Physical Review B}, 62(16):R10637, 2000.

\bibitem{Billangeon2006}
P.~M. Billangeon, F.~Pierre, H.~Bouchiat, and R.~Deblock.
\newblock {Emission and absorption asymmetry in the quantum noise of a
  Josephson junction}.
\newblock {\em Physical Review Letters}, 96(13):1--4, 2006.

\bibitem{Zamoum2015}
R.~Zamoum, M.~Lavagna, and A.~Cr{\'{e}}pieux.
\newblock {Non-symmetrized noise in a quantum dot: interpretation in terms of
  photon emission and coherent superposition of scattering paths}.
\newblock {\em Physical Review B}, 93(23):235449, 2016.

\bibitem{Yang1992}
S.~R.~Eric Yang.
\newblock {Quantum Shot Noise Spectrum of a Point Contact}.
\newblock {\em Solid State Communications}, 81(5):375--378, 1992.

\bibitem{Buttiker1992}
M.~B{\"{u}}ttiker.
\newblock {Scattering theory of current and intensity noise correlations in
  conductors and wave guides}.
\newblock {\em Physical Review B}, 46(19):12485--12507, 1992.

\bibitem{Buttiker1992a}
M.~B{\"{u}}ttiker.
\newblock {Role of scattering amplitudes in frequency-dependent current
  fluctuations in small conductors}.
\newblock {\em Physical Review B}, 45(7):3807--3810, 1992.

\bibitem{Wu2010}
B.~H. Wu and C.~Timm.
\newblock {Noise spectra of ac-driven quantum dots: Floquet master-equation
  approach}.
\newblock {\em Physical Review B}, 81(7):075309, 2010.

\bibitem{Entin-Wohlman2007}
O.~Entin-Wohlman, Y.~Imry, S.~A. Gurvitz, and A.~Aharony.
\newblock {Steps and dips in the ac conductance and noise of mesoscopic
  structures}.
\newblock {\em Physical Review B}, 75(19):3--6, 2007.

\bibitem{Rothstein2009}
E.~A. Rothstein, O.~Entin-Wohlman, and A.~Aharony.
\newblock {Noise spectra of a biased quantum dot}.
\newblock {\em Physical Review B}, 79(7):1--7, 2009.

\bibitem{Gabdank2011}
N.~Gabdank, E.~A. Rothstein, O.~Entin-Wohlman, and A.~Aharony.
\newblock {Noise spectra of an interacting quantum dot}.
\newblock {\em Physical Review B}, 84(23), 2011.

\bibitem{Buttiker1990}
M.~B{\"{u}}ttiker.
\newblock {Scattering theory of thermal and excess noise in open conductors}.
\newblock {\em Physical Review Letters}, 65(23):2901--2904, 1990.

\bibitem{Landauer1993}
Rolf Landauer.
\newblock {Solid-state shot noise}.
\newblock {\em Physical Review B}, 47(24):16427--16432, 1993.

\bibitem{VandenBrom1999}
H.~E. van~den Brom and J.~M. van Ruitenbeek.
\newblock {Quantum suppression of shot noise in atom-size metallic contacts}.
\newblock {\em Phys. Rev. Lett.}, 82(7):1526--1529, 1999.

\bibitem{Dicarlo2008}
L.~Dicarlo, J.~R. Williams, Yiming Zhang, D.~T. McClure, and C.~M. Marcus.
\newblock {Shot noise in graphene}.
\newblock {\em Physical Review Letters}, 100(15):2--5, 2008.

\bibitem{Schoelkopf1997}
R.~Schoelkopf, P.~Burke, a.~Kozhevnikov, D.~Prober, and M.~Rooks.
\newblock {Frequency Dependence of Shot Noise in a Diffusive Mesoscopic
  Conductor}.
\newblock {\em Physical Review Letters}, 78(17):3370--3373, 1997.

\bibitem{Gabelli2008}
J.~Gabelli and B.~Reulet.
\newblock {Dynamics of quantum noise in a tunnel junction under ac excitation}.
\newblock {\em Physical Review Letters}, 100(2):2--5, 2008.

\bibitem{Moskalets2004}
M~Moskalets and M.~B{\"{u}}ttiker.
\newblock {Floquet scattering theory for current and heat noise in large
  amplitude adiabatic pumps}.
\newblock {\em Physical Review B}, 70(24):15, 2004.

\bibitem{Moskalets2007}
M.~Moskalets and M.~B{\"{u}}ttiker.
\newblock {Time-resolved noise of adiabatic quantum pumps}.
\newblock {\em Physical Review B}, 75(3):035315, 2007.

\bibitem{Moskalets2012}
Michael Moskalets.
\newblock {\em {Scattering Matrix Approach to Non-Stationary Quantum
  Transport}}.
\newblock Imperial College Press, 2012.

\bibitem{Dai2014}
Jiao~Hua Dai and Rui Zhu.
\newblock {Fano resonance in the nonadiabatic pumped shot noise of a
  time-dependent quantum well}.
\newblock {\em European Physical Journal B}, 87(12), 2014.

\bibitem{Zhu2015}
Rui Zhu, Jiao~Hua Dai, and Yong Guo.
\newblock {Fano resonance in the nonadiabatically pumped shot noise of a
  time-dependent quantum well in a two-dimensional electron gas and graphene}.
\newblock {\em Journal of Applied Physics}, 117(16):164306, 2015.

\bibitem{Camalet2004}
S{\'{e}}bastien Camalet, Sigmund Kohler, and Peter H{\"{a}}nggi.
\newblock {Shot-noise control in ac-driven nanoscale conductors}.
\newblock {\em Physical Review B}, 70(1):155326, 2004.

\bibitem{Lesovik1994}
G.~B. Lesovik and L.~S. Levitov.
\newblock {Noise in an ac Biased Junction: Nonstationary Aharonov-Bohm Effect}.
\newblock {\em Physical Review Letters}, 72(4):538--541, 1994.

\bibitem{Peskin2016}
Uri Peskin.
\newblock {Formulation of charge transport in molecular junctions with
  time-dependent molecule-leads coupling operators}.
\newblock {\em Fortschritte der Physik}, 9:1--9, 2016.

\bibitem{Vanevic2012}
Mihajlo Vanevi{\'{c}} and Wolfgang Belzig.
\newblock {Control of electron-hole pair generation by biharmonic voltage drive
  of a quantum point contact}.
\newblock {\em Physical Review B}, 86(24):241306, 2012.

\bibitem{Rychkov2005}
Valentin~S. Rychkov, Mikhail~L. Polianski, and Markus B{\"{u}}ttiker.
\newblock {Photon-assisted electron-hole shot noise in multiterminal
  conductors}.
\newblock {\em Physical Review B}, 72(15):155326, 2005.

\bibitem{Vanevic2008}
Mihajlo Vanevi{\'{c}}, Yuli~V. Nazarov, and Wolfgang Belzig.
\newblock {Elementary charge-transfer processes in mesoscopic conductors}.
\newblock {\em Physical Review B}, 78(24):245308, 2008.

\bibitem{Ivanov1997}
D.~A. Ivanov, H.~W. Lee, and L.~S. Levitov.
\newblock {Coherent states of alternating current}.
\newblock {\em Physical Review B}, 56(11):6839--6850, 1997.

\bibitem{Keeling2006}
J.~Keeling, I.~Klich, and L.~S. Levitov.
\newblock {Minimal excitation states of electrons in one-dimensional wires}.
\newblock {\em Physical Review Letters}, 97(11):15--18, 2006.

\bibitem{Dubois2013}
J~Dubois, T~Jullien, F~Portier, P~Roche, A~Cavanna, Y~Jin, W~Wegscheider,
  P~Roulleau, and D~C Glattli.
\newblock {Minimal-excitation states for electron quantum optics using
  levitons}.
\newblock {\em Nature}, 502(7473):659--663, 2013.

\bibitem{Gabelli2013}
Julien Gabelli and Bertrand Reulet.
\newblock {Shaping a time-dependent excitation to minimize the shot noise in a
  tunnel junction}.
\newblock {\em Physical Review B}, 87(7):1--7, 2013.

\bibitem{Pedersen1998}
Morten~Holm Pedersen and M.~B{\"{u}}ttiker.
\newblock {Scattering theory of photon-assisted electron transport.}
\newblock {\em Physical Review B}, 58(19):12993--13006, 1998.

\bibitem{Hammer2011}
Jan Hammer and Wolfgang Belzig.
\newblock {Quantum noise in ac-driven resonant-tunneling double-barrier
  structures: Photon-assisted tunneling versus electron antibunching}.
\newblock {\em Physical Review B}, 84(8):85419, 2011.

\bibitem{Strass2005}
M~Strass, P~H{\"{a}}nggi, and S~Kohler.
\newblock {Nonadiabatic Electron Pumping: Maximal Current with Minimal Noise}.
\newblock {\em Physical Review Letters}, 95(13):130601, 2005.

\bibitem{Riwar2013}
Roman~Pascal Riwar, Janine Splettstoesser, and J{\"{u}}rgen K{\"{o}}nig.
\newblock {Zero-frequency noise in adiabatically driven interacting quantum
  systems}.
\newblock {\em Physical Review B}, 87(19):195407, 2013.

\bibitem{Rocha2010}
Claudia~G Rocha, Luis E F~Foa Torres, and Gianaurelio Cuniberti.
\newblock {ac transport in graphene-based Fabry-P{\'{e}}rot devices}.
\newblock {\em Physical Review B}, 81(11):115435, 2010.

\bibitem{Konstantinov1961}
O.~V. Konstantinov and V.~I. Perel'.
\newblock {A diagram technique for evaluating transport quantities}.
\newblock {\em Soviet Physics JETP}, 12(1):142--149, 1961.

\bibitem{Kadanoff1962}
L.~P. Kadanoff and G.~Baym.
\newblock {\em {Quantum statistical mechanics: Green's function methods in
  equilibrium and nonequilibrium problems}}.
\newblock New York: Benjamin, 1962.

\bibitem{Keldysh1965}
L~V Keldysh.
\newblock {Diagram Technique for Nonequilibrium Processes}.
\newblock {\em Soviet Physics JETP}, 20(4):1018--1030, 1965.

\bibitem{Stefanucci2013}
G.~Stefanucci and R.~van Leeuwen.
\newblock {\em {Nonequilibrium Many-Body Theory of Quantum Systems: A Modern
  Introduction}}.
\newblock Cambridge University Press, 2013.

\bibitem{Arrachea2006}
Liliana Arrachea and Michael Moskalets.
\newblock {Relation between scattering-matrix and Keldysh formalisms for
  quantum transport driven by time-periodic fields}.
\newblock {\em Physical Review B}, 74(24):1--13, 2006.

\bibitem{DiVentra2008}
Massimiliano {Di Ventra}.
\newblock {\em {Electrical Transport in Nanoscale Systems}}.
\newblock Cambridge University Press, 2008.

\bibitem{Joho2012}
K.~Joho, S.~Maier, and A.~Komnik.
\newblock {Transient noise spectra in resonant tunneling setups: Exactly
  solvable models}.
\newblock {\em Physical Review B}, 86(15):1--14, 2012.

\bibitem{Caroli1971}
C~Caroli, R~Combescot, and P~Nozieres.
\newblock {Direct calculation of the tunneling current}.
\newblock {\em Sciences-New York}, 4(58):916--929, 1971.

\bibitem{Caroli1971a}
C~Caroli, R~Combescot, D~Lederer, P~Nozieres, and D~Saint-James.
\newblock {A direct calculation of the tunnelling current. II. Free electron
  description}.
\newblock {\em Journal of Physics C: Solid State Physics}, 4:2598--2610, 1971.

\bibitem{Wingreen1993}
Ned~S. Wingreen, Antti~Pekka Jauho, and Yigal Meir.
\newblock {Time-dependent transport through a mesoscopic structure}.
\newblock {\em Physical Review B}, 48(11):8487--8490, 1993.

\bibitem{Jauho1994}
Antti~Pekka Jauho, Ned~S. Wingreen, and Yigal Meir.
\newblock {Time-dependent transport in interacting and noninteracting
  resonant-tunneling systems}.
\newblock {\em Physical Review B}, 50(8):5528--5544, 1994.

\bibitem{Stefanucci2004}
Gianluca Stefanucci and Carl~Olof Almbladh.
\newblock {Time-dependent partition-free approach in resonant tunneling
  systems}.
\newblock {\em Physical Review B}, 69(19):195318, 2004.

\bibitem{Tuovinen2016}
Riku Tuovinen, Niko S{\"{a}}kkinen, Daniel Karlsson, Gianluca Stefanucci, and
  Robert van Leeuwen.
\newblock {Phononic heat transport in the transient regime: An analytic
  solution}.
\newblock {\em Physical Review B}, 93(21):214301, 2016.

\bibitem{Cini1980}
Michele Cini.
\newblock {Time-dependent approach to electron transport through junctions:
  General theory and simple applications}.
\newblock {\em Physical Review B}, 22(12):5887--5899, 1980.

\bibitem{Yang2015}
Pei-Yun Yang, Chuan-Yu Lin, and Wei-Min Zhang.
\newblock {Master equation approach to transient quantum transport
  incorporating initial correlations}.
\newblock {\em Physical Review B}, 92(16):165403, 2015.

\bibitem{Feng2008}
Zimin Feng, Joseph Maciejko, Jian Wang, and Hong Guo.
\newblock {Current fluctuations in the transient regime: An exact formulation
  for mesoscopic systems}.
\newblock {\em Physical Review B}, 77(7):1--11, 2008.

\bibitem{Yang2014}
Pei-Yun Yang, Chuan-Yu Lin, and Wei-Min Zhang.
\newblock {Transient current-current correlations and noise spectra}.
\newblock {\em Physical Review B}, 89(11):115411, 2014.

\bibitem{Perfetto2008}
Enrico Perfetto, Gianluca Stefanucci, and Michele Cini.
\newblock {Spin-flip scattering in time-dependent transport through a quantum
  dot: Enhanced spin-current and inverse tunneling magnetoresistance}.
\newblock {\em Physical Review B}, 78(15):155301(11), 2008.

\bibitem{Tuovinen2013}
Riku Tuovinen, Robert van Leeuwen, Enrico Perfetto, and Gianluca Stefanucci.
\newblock {Time-dependent Landauer-B{\"{u}}ttiker formula for transient
  dynamics}.
\newblock {\em Journal of Physics: Conference Series}, 427(1):012014, 2013.

\bibitem{Tuovinen2014}
Riku Tuovinen, Enrico Perfetto, Gianluca Stefanucci, and Robert {Van Leeuwen}.
\newblock {Time-dependent Landauer-B{\"{u}}ttiker formula: Application to
  transient dynamics in graphene nanoribbons}.
\newblock {\em Physical Review B}, 89(8):085131, 2014.

\bibitem{GomesdaRocha2015}
Claudia {Gomes da Rocha}, Riku Tuovinen, Robert van Leeuwen, and Pekka
  Koskinen.
\newblock {Curvature in graphene nanoribbons generates temporally and spatially
  focused electric currents}.
\newblock {\em Nanoscale}, 7(18):8627--8635, 2015.

\bibitem{Tuovinen2016b}
Riku Tuovinen.
\newblock {\em {Time-Dependent Quantum Transport in Nanosystems: A
  Nonequilibrium Green's Function Approach}}.
\newblock PhD thesis, University of Jyv{\"{a}}skyl{\"{a}}, 2016.

\bibitem{Ridley2015}
Michael Ridley, Angus Mackinnon, and Lev Kantorovich.
\newblock {Current through a multilead nanojunction in response to an arbitrary
  time-dependent bias}.
\newblock {\em Physical Review B}, 91(12):125433, 2015.

\bibitem{Ridley}
Michael Ridley, Angus MacKinnon, and Lev Kantorovich.
\newblock {Calculation of the current response in a nanojunction for an
  arbitrary time-dependent bias: application to the molecular wire}.
\newblock {\em Journal of Physics: Conference Series}, 696(1):012017, mar 2016.

\bibitem{Tuovinen2016a}
Riku Tuovinen, Robert van Leeuwen, Enrico Perfetto, and Gianluca Stefanucci.
\newblock {Time-dependent Landauer-B{\"{u}}ttiker formalism for superconducting
  junctions at arbitrary temperatures}.
\newblock {\em Journal of Physics: Conference Series}, 696(1):012016, 2016.

\bibitem{Ridley2016}
Michael Ridley, Angus MacKinnon, and Lev Kantorovich.
\newblock {Fluctuating-bias controlled electron transport in molecular
  junctions}.
\newblock {\em Physical Review B}, 93(20):205408, 2016.

\bibitem{Gaury2014}
Benoit Gaury and Xavier Waintal.
\newblock {Dynamical control of interference using voltage pulses in the
  quantum regime}.
\newblock {\em Nature Communications}, 5(May):1--11, 2014.

\bibitem{Langreth1972}
David Langreth and John Wilkins.
\newblock {Theory of Spin Resonance in Dilute Magnetic Alloys}.
\newblock {\em Physical Review B}, 6(9):3189--3227, 1972.

\bibitem{Langreth1976}
D.~C. Langreth.
\newblock {Linear and Nonlinear Response Theory with Applications}.
\newblock In J.~T. Devreese and E.~van Doren, editors, {\em NATO Advanced
  Studies Series B}, pages 3--32. Plenum, New York, 1976.

\bibitem{Engel2004}
Hans~Andreas Engel and Daniel Loss.
\newblock {Asymmetric quantum shot noise in quantum dots}.
\newblock {\em Physical Review Letters}, 93(13):1--4, 2004.

\bibitem{Verzijl2012}
C.~J~O Verzijl and J.~M. Thijssen.
\newblock {DFT-based molecular transport implementation in ADF/BAND}.
\newblock {\em Journal of Physical Chemistry C}, 116(46):24393--24412, 2012.

\bibitem{Verzijl2013}
C.~J~O Verzijl, J.~S. Seldenthuis, and J.~M. Thijssen.
\newblock {Applicability of the wide-band limit in DFT-based molecular
  transport calculations}.
\newblock {\em Journal of Chemical Physics}, 138(9):094102, 2013.

\bibitem{Zakka-Bajjani2007}
E.~Zakka-Bajjani, J.~S{\'{e}}gala, F.~Portier, P.~Roche, D.~C. Glattli,
  A.~Cavanna, and Y.~Jin.
\newblock {Experimental test of the high-frequency quantum shot noise theory in
  a quantum point contact}.
\newblock {\em Physical Review Letters}, 99(23):236803, 2007.

\bibitem{Dayem1962}
A.~H. Dayem and R.~J. Martin.
\newblock {Quantum Interaction of Microwave Radiation with Tunneling Between
  Superconductors}.
\newblock {\em Physical Review Letters}, 8(6):246--248, 1962.

\bibitem{Wang2003}
Baigeng Wang, Jian Wang, and Hong Guo.
\newblock {Current plateaus of nonadiabatic charge pump: Multiphoton assisted
  processes}.
\newblock {\em Physical Review B}, 68(15):155326, 2003.

\bibitem{Mujica1994}
V.~Mujica, M.~Kemp, and M.~A. Ratner.
\newblock {Electron conduction in molecular wires. I. A scattering formalism}.
\newblock {\em The Journal of Chemical Physics}, 101(8):6849, 1994.

\bibitem{Mujica1994a}
V~Mujica, M~Kemp, and M~A Ratner.
\newblock {Electron Conduction in Molecular Tunneling Microscopy Wires. II.
  Application}.
\newblock {\em Journal of Chemical Physics}, 101(8):6856--6864, 1994.

\bibitem{Lehmann2002}
J{\"{o}}rg Lehmann, Sigmund Kohler, Peter H{\"{a}}nggi, and Abraham Nitzan.
\newblock {Molecular Wires Acting as Coherent Quantum Ratchets}.
\newblock {\em Physical Review Letters}, 88(22):228305, 2002.

\bibitem{Chen2013}
Shuguang Chen, Hang Xie, Yu~Zhang, Xiaodong Cui, and Guanhua Chen.
\newblock {Quantum transport through an array of quantum dots.}
\newblock {\em Nanoscale}, 5(1):169--173, 2013.

\bibitem{Buttiker1982}
M.~B{\"{u}}ttiker and R.~Landauer.
\newblock {Traversal Time for Tunneling}.
\newblock {\em Physical Review Letters}, 49(23):1739--1742, 1982.

\bibitem{Collins1987}
S.~Collins, S.~Lowe, and J.~R. Barker.
\newblock {The quantum mechanical tunnelling time problem-revisited}.
\newblock {\em J. Phys. C: Solid State Phys.}, 20(36):6213, 1987.

\bibitem{Ozaki2007}
Taisuke Ozaki.
\newblock {Continued fraction representation of the Fermi-Dirac function for
  large-scale electronic structure calculations}.
\newblock {\em Physical Review B}, 75(3):1--9, 2007.

\bibitem{Croy2009b}
Alexander Croy and Ulf Saalmann.
\newblock {Propagation Scheme for Non-Equilibrium Dynamics of Electron
  Transport in Nanoscale Devices}.
\newblock {\em Physical Review B}, 80(December):245311, 2009.

\bibitem{Hu2010}
Jie Hu, Rui~Xue Xu, and Yijing Yan.
\newblock {Communication: Pad{\'{e}} spectrum decomposition of Fermi function
  and Bose function}.
\newblock {\em Journal of Chemical Physics}, 133(10):0--4, 2010.

\bibitem{Hu2011}
Jie Hu, Meng Luo, Feng Jiang, Rui~Xue Xu, and Yijing Yan.
\newblock {Pad{\'{e}} spectrum decompositions of quantum distribution functions
  and optimal hierarchical equations of motion construction for quantum open
  systems}.
\newblock {\em Journal of Chemical Physics}, 134(24):0--10, 2011.

\bibitem{Lerch1887}
M.~Lerch.
\newblock {Note sur la function}.
\newblock {\em Acta Math.}, 11(1-4):19--24, 1887.

\end{thebibliography}

\end{document}